\pdfoutput=1
\documentclass[journal]{IEEEtran}

\usepackage{graphicx}
\usepackage{amsmath}
\usepackage{stfloats}
\usepackage[table,xcdraw]{xcolor}
\usepackage{booktabs}
\usepackage{hyperref}
\usepackage{cleveref}
\usepackage{amssymb}
\usepackage{multirow}
\usepackage{tabularx}
\usepackage{balance}
\usepackage{soul}
\usepackage{float}
\usepackage{verbatim}
\usepackage{cite}
\usepackage{bm}
\usepackage{algorithm}
\usepackage{algpseudocode}
\usepackage{subfig}
\definecolor{refdarkred}{RGB}{128,0,0}
\hypersetup{
  colorlinks=true,
  linkcolor=refdarkred,
  citecolor=blue,
  urlcolor=blue
}

\begin{document}

\title{Closed-Loop Multi-Stage Construction for Intelligent Radio Maps}

\author{Chengxi~Li\IEEEmembership{},
       Minghui~Liwang,~\IEEEmembership{Senior Member,~IEEE},
       Wei~Gong,~\IEEEmembership{Member,~IEEE},
       Li~Li,~\IEEEmembership{Member,~IEEE},
       Baoxian~Zhang,~\IEEEmembership{Senior Member,~IEEE},
       Cheng~Li,~\IEEEmembership{Senior Member,~IEEE},
       and Jie~Chen,~\IEEEmembership{Fellow,~IEEE}
    \IEEEcompsocitemizethanks{
    \IEEEcompsocthanksitem This work was supported in part by the National Key Research and Development Program of China under grant No. 2022YFA1004700, Shanghai Municipal Science and Technology, China Major Project under grant 2021SHZDZX0100, the National Natural Science Foundation of China under grant Nos.~62403360, 72171172, 92367101, 62271424, 62088101, iF open-funds from Xinghuo Eco and China Institute of Communications, and the SFU University under Grants N001210 and N001230.

    \IEEEcompsocthanksitem C.~Li, W.~Gong, M.~Liwang, L.~Li, and J.~Chen are with the Department of Control Science and Engineering, Tongji University, Shanghai 201804, China, Shanghai Institute of Intelligent Science and Technology, Tongji University, Shanghai 201210, China, and also with the Shanghai Research Institute for Intelligent Autonomous Systems, Tongji University, Shanghai 201210, China (e-mail: \protect\href{mailto:chengxili@tongji.edu.cn}{chengxili@tongji.edu.cn}; \protect\href{mailto:weigong@tongji.edu.cn}{weigong@tongji.edu.cn};
    \protect\href{mailto:minghuiliwang@tongji.edu.cn}{minghuiliwang@tongji.edu.cn};
    \protect\href{mailto:lili@tongji.edu.cn}{lili@tongji.edu.cn};
    \protect\href{mailto:chenjie206@tongji.edu.cn}{chenjie206@tongji.edu.cn}).
    \IEEEcompsocthanksitem B.~Zhang is with the Research Center of Ubiquitous Sensor Networks, University of Chinese Academy of Sciences, Beijing 100049, China (e-mail: \protect\href{mailto:bxzhang@ucas.ac.cn}{bxzhang@ucas.ac.cn}).
    \IEEEcompsocthanksitem C.~Li is with the School of Engineering Science, Simon Fraser University, Burnaby, BC V5A 1S6, Canada (e-mail: \protect\href{mailto:li\_cheng@sfu.ca}{li\_cheng@sfu.ca}).
    }
}

\markboth{Journal of \LaTeX\ Class Files,~Vol.~14, No.~8, August~2020}%
{Shell \MakeLowercase{\textit{et al.}}: Bare Demo of IEEEtran.cls for IEEE Journals}

\maketitle

\begin{abstract}
Intelligent radio maps (IRMs) provide fine-grained spatial--spectral awareness for wireless network planning, spectrum sharing, and interference management. However, practical IRM construction in large-scale urban environments remains challenging due to sparse measurements, complex propagation effects, diverse 3D city morphologies, and limited computing resources. To address these issues, this paper proposes a closed-loop multi-stage framework for multi-scale 3D IRM construction in urban cities. In the initial stage, physics-aware environmental priors and sparse sensing measurements are integrated into a hybrid-conditioned diffusion transformer to generate a coarse IRM with uncertainty feedback. In iterative stages, an uncertainty-guided active sensing mechanism selects informative candidates based on an attention network, while newly acquired measurements are recursively incorporated into the inference step. A knowledge-distillation strategy is further introduced to reduce inference complexity and realize progressive IRM refinement with high computational efficiency. Moreover, we develop a comprehensive spectrum dataset covering diverse urban morphologies, height-aware blockages, mobile transmitters, and different receiver conditions. Extensive experiments show that the proposed framework outperforms representative baselines in inference accuracy, sensing-efficiency gain, and robustness, providing a practical real-time solution.
\end{abstract}


\section{Introduction}

\subsection{Background and Challenges}

The evolution of wireless networks toward massive connectivity, heterogeneous architectures, and increasingly dynamic spectrum usage calls for high-quality intelligent radio maps (IRMs), which reveal fine-grained spectral awareness across spatial and temporal domains~\cite{10948391,10025551}. By properly exploiting this spectrum knowledge, IRMs facilitate proactive network planning, support reliable spectrum sharing, realize interference mitigation, and empower a wide range of wireless applications, including emergency services, spectrum-based navigation, and Internet of Things (IoT) applications.
Specifically, a complete IRM construction process typically consists of two fundamental steps: \textit{(i)} spectrum sensing, which acquires spectral measurements in appropriate spatial and temporal domains; and \textit{(ii)} spectrum inference, which estimates a complete radio map from limited measurements. 

However, constructing high-quality IRMs in large-scale urban environments faces several challenges. 
First, spectral measurements are inherently sparsely distributed and expensive to collect, as sensing platforms, e.g., mobile Uncrewed Aerial Vehicles (UAVs) or manual devices, are constrained by accessibility to candidate locations and limited sensing energy. Consequently, constructing IRMs purely through dense spectrum sensing is both labor-intensive and infeasible for real-world deployment.
Second, in practical urban environments, radio propagation is strongly affected by height-dependent obstacles, including buildings, and terrain-related city morphologies. Moreover, sensing errors and interference dynamics also lead to strong spatial radio variations in a period of time. Thus, purely inference-based methods are unable to capture dynamically evolving IRMs. 
Third, practical deployment requires low-latency construction and continuous updates, requiring both sensing and inference steps to be computationally efficient. This is particularly important given the recent shift of IRM research from two-dimensional (2D) planar maps to large-scale three-dimensional (3D) representations, where multiple transmitter/receiver altitudes, building heights, and terrain elevations dramatically increase the scalability requirements of the IRM construction framework.

%%%这里我强调了spatiotemporal的定义是：一系列地图，不同的位置，不同的结果。持续采样，持续更新
%% 也跟大模型咨询了，认可了这样sequential更新来作为时序变化的观点。
%% \hl{please look at the explanation of spatiotemporal here}

To overcome the challenges of constructing spatiotemporal IRMs\footnote{In this paper, the term spatiotemporal IRM refers to a sequence of radio maps generated by different transmitter locations over time.} in large-scale urban environments, we propose a closed-loop multi-stage IRM construction framework that tightly couples sensing and inference steps. Specifically, in the initial stage, environmental priors, statistical techniques, and the state-of-the-art diffusion model are systematically integrated to generate coarse IRM estimates, which are consistent with physical measurements. In the subsequent iterative stages, a bi-directional feedback mechanism is established between sensing and inference, where the progressively refined IRM continuously guides informative sensing, while newly acquired measurements further improve inference accuracy. Meanwhile, a lightweight design is adopted to ensure computational efficiency, enabling practical deployment in urban regions.

\subsection{Related Work}
\label{subsec:works}
Existing IRM construction studies can be categorized into the following three major directions.
\subsubsection{Sensing-oriented IRM construction}

Sensing-oriented studies aim to acquire informative and reliable spectrum measurements under practical sensing budgets and accessibility constraints. Along this line, existing works mainly evolve along two closely related dimensions. 

\paragraph{Organization of sensing platforms}
Early studies of sensing-oriented IRM construction often relied on fixed sensing infrastructures, which provide controlled calibration and stable data quality, but suffer from high deployment cost and weak adaptability. More recent sensing paradigms move toward three complementary platform types. First, fixed-layout IoT nodes enable long-term monitoring with stable calibration and centrally configured sensing schedules~\cite{lyu2023adaptive,surekha2022spectrum,fu2025energy}. Second, ground-level crowdsensing leverages ubiquitous user devices and software-defined devices to collect large-scale measurements at relatively low cost, while introducing privacy, security, and heterogeneity challenges~\cite{smith2023sdr,celdran2024cyberspec,wang2023privacy}. Third, unmanned-agent-assisted sensing uses UAVs and other mobile agents to actively control sensing positions or trajectories, thereby improving coverage, controllability, and accessibility in large-scale 3D environments~\cite{zhu2023path,shrestha2023spectrum}.

\paragraph{Budget-constrained sensing strategies}
Beyond platform choice, another research trend is to determine informative sensing locations under limited energy and time budgets. Classic approaches include informative path planning, active surveying, and adaptive trajectory design, which attempt to maximize information gain per unit cost~\cite{9199788,9857777,9629332}. Recently, learning-based policies, especially reinforcement learning, have been introduced to support sequential sensing decisions under mobility constraints and long-horizon objectives~\cite{9354009,2020DeepSensing,9272371}. In parallel, multi-source fusion has become increasingly important because practical IRMs often aggregate measurements from heterogeneous devices with different sampling rates and noise statistics~\cite{smith2021novel}. 

Sensing-oriented methods improve measurement quality and spatial coverage under resource constraints, but they often treat downstream inference as a passive backend, making it difficult to exploit inference feedback for online IRM refinement.

\subsubsection{Inference-oriented IRM construction}
Inference-oriented methods aim to accurately construct fine-grained IRMs from sparse and heterogeneous measurements. Representative methodologies are categorized as follows.

\paragraph{Propagation-based methods}
These methods incorporate physical channel knowledge, such as path-loss laws, ray-tracing intuition, and shadowing-aware priors, into IRM inference. Their main advantage is strong physical interpretability and robustness under sparse observations \cite{xu2021radio,10584440}. However, they often rely on simplified channel assumptions or environment-specific calibration, making it difficult to accurately characterize complex urban morphology, multipath propagation, and dynamic spectrum variations.

\paragraph{Statistical-driven methods}
Statistical inference methods provide lightweight and mathematically grounded solutions. Commonly-used methods include Kriging interpolation, which explicitly exploit spatial correlation and naturally quantify predictive uncertainty \cite{2021Space}; hypothesis testing, which operates effectively under limited data and supports real-time performance \cite{9310285}; and tensor/matrix completion, which leverage low-rank structures across spatial, temporal, and spectral domains to improve inference accuracy as well as efficiency \cite{10679641,yin2024compressed}. Despite their solid theoretical foundations, the effectiveness may deteriorate due to 3D heterogeneous obstacles and time-varying radio dynamics.
\paragraph{Learning-enabled methods}
These methods learn nonlinear mappings from sparse sensing and environmental context to effectively estimate IRMs, which have become increasingly popular in recent years. Specifically, classical machine learning models such as random forests and K-nearest neighbors offer favorable generalization under limited data~\cite{9973373}. Deep convolutional architectures, such as RecNet~\cite{8422971} and RadioUNet~\cite{9354041}, reformulate radio mapping as an image-to-image transformation problem and enable fast feed-forward inference. More complicated networks further incorporate spatiotemporal dependencies and structural priors, including Supreme, RadioMapMotion, and RadioGAT~\cite{9644596,jia2025radiomapmotion,10682510}. In addition, generative models, such as GAN- and diffusion-based methods (e.g., Pix2Pix+, RecuGAN, 3D-RadioDiff, and RadioLAM~\cite{10227351,10637562,10963917,liu2025radiolam}), are widely used in large-scale 3D environments. Nevertheless, the performance of learning-enabled methods largely depends on the quality of training data.
\paragraph{Hybrid methods}
Hybrid inference methods combine physical models, statistical structure, and learning-based modules to balance accuracy, efficiency, and robustness. Representative examples include super-resolution pipelines with geographical interpolation, Kriging plus GAN refinement, tensor decomposition plus deep generative inference, physics-informed diffusion and multi-scale inpainting~\cite{deng2020radio,10335642,10634565,11278649,10791446}. These studies are particularly relevant to emerging 3D spatiotemporal IRMs, but suffers from high complexity and limited scalability when facing heterogeneous urban environments.
 
Although most inference-oriented methods can achieve accurate IRM estimation from fixed sparse measurements, they lack feedback-driven active sensing mechanisms.

\subsubsection{Jointly-designed IRM construction}

A growing yet still limited research trend jointly designs sensing and inference to construct high-quality IRMs. Instead of optimizing the two steps independently,  sensing constraints such as sensor placement, UAV accessibility and measurement budgets are explicitly coupled with the inference objective. Recent studies include time-variant inference with optimized distributed sensors~\cite{10902472}, robot-assisted indoor sensing with interpolation-guided GAN refinement~\cite{9031749}, 3D radio map construction under constrained UAV trajectories~\cite{10025551}, and first-predict-then-correct frameworks with measurement-guided inference~\cite{10838708}. However, most existing approaches remain in a single-stage setting without progressive refinement. Moreover, the interaction between sensing and inference is typically one-directional and lacks an explicit recursive feedback mechanism.

\subsection{Novelty and Contributions}

In real-world complex urban environments, the construction of high-quality IRMs is challenged by three key issues identified in the related work. Specifically, \textit{(i)} for the 3D paradigm shift, we introduce a cooperative sensing mechanism that integrates UAVs and ground-level manual devices, enabling geometry-aware data acquisition in urban morphologies; \textit{(ii)} to handle transmitter-induced spatiotemporal variations, we incorporate physics-informed priors together with a dedicated transformer, which not only preserves historical information through diffusion-driven memory but also enforces physically consistent IRM evolution; \textit{(iii)} for large-scale deployment, we further introduce a knowledge distillation strategy to compress the model while maintaining construction fidelity. 

To this end, we couple these components into a unified closed-loop multi-stage framework, where sensing and inference are tightly integrated through a bi-directional feedback mechanism, and the IRMs can be progressively updated and refined within multiple stages. The main contributions of this work are summarized as follows.

%%% 4个创新，字数精简：一是3维、大规模、时空动态irm的表达，二是具体的方法closed-loop multi-stage；三是自己构建数据集（pipeline）；四是实验验证结果。
%%%%0617创新点修改，注意相关字词的表述

\begin{itemize}

     \item We formulate a unified representation of 3D spatiotemporal IRMs, where complex urban morphology, mobile sensing agents, dynamic inference evolution, and physics-aware propagation priors are all explicitly modeled.
    
    \item We propose a novel closed-loop multi-stage IRM construction framework, where sensing and inference steps are jointly optimized through active feedback. In the initial stage, a dedicated Diffusion Transformer (DiT) is employed for inference to incorporate spatiotemporal dynamics and environmental priors, followed by progressive knowledge distillation to improve computational efficiency. In iterative stages, inference outputs guide subsequent sensing decisions, while newly acquired measurements continuously refine the inference model. 
    
    \item We develop a comprehensive IRM dataset for large-scale urban environments, which integrates realistic geographic maps, mobile transmitters, height-aware obstacle blockages, and 3D dynamic sensing. This dataset enables systematic training and evaluation of IRM construction methods under diverse real-world cities.
    
    \item We conduct extensive experiments to validate the proposed framework compared with various baselines, demonstrating significant improvements in IRM construction accuracy, robustness, and efficiency, thereby highlighting its engineering-oriented value.

\end{itemize}

Section~\ref{sec:system_model} describes the system model. Section~\ref{sec:method} details the proposed closed-loop multi-stage construction framework. Section~\ref{sec:experiments} presents the simulation process of the IRM dataset. Section~\ref{sec:performance_evaluation} provides a comprehensive evaluation of different construction methods. Finally, Section~\ref{sec:conclusion} concludes the paper.

\section{System Model}
\label{sec:system_model}
For clarity, the main notations used in the system model to describe the large-scale urban radio environment are summarized in Table~\ref{tab:notations}. Meanwhile, several key methodological parameters are also included for completeness.

\begin{table}[!t]
\caption{Main Notations Used Throughout the Paper}
\label{tab:notations}
\centering
\renewcommand{\arraystretch}{1.05}
\setlength{\tabcolsep}{4pt}
\begin{tabular}{p{0.13\columnwidth}p{0.52 \columnwidth}p{0.22\columnwidth}}
\toprule
\textbf{Symbol} & \textbf{Description} & \textbf{Range} \\
\midrule

$x$ &

Discrete planar coordinate along $x$-axis &

$\{1,\ldots,\mathbf X\}$

\\

$y$ &

Discrete planar coordinate along $y$-axis &

$\{1,\ldots,\mathbf Y\}$

\\

$z$ &
Discrete altitude coordinate along $z$-axis&
$\{1,\ldots,\mathbf Z\}$
\\

$f$ &
Frequency-band index &
$\{1,\ldots,\mathbf F\}$
\\

$\Omega$ &
Joint 4D spatial--spectral grid &
$\mathbf X\times\mathbf Y\times\mathbf Z\times\mathbf F$
\\

$\mathbf S$ &
Ground-truth IRM &
$\mathbb R^{\Omega}$
\\

$\hat{\mathbf S}$ &
Inferred IRM &
$\mathbb R^{\Omega}$
\\

$\mathbf M_b$ &
Building-occupancy mask &
$\{0,1\}^{\Omega}$
\\

$\mathbf M_g$ &
Geographical-context mask &
$\{0,1\}^{\Omega}$
\\

$\mathbf O$ &
Observation mask &
$\{0,1\}^{\Omega}$
\\

$\mathbf \Pi$ &
Sparse sensing-record tensor &
$\mathbb R^{\Omega}$
\\

$\mathbf E$ &
Sensing-noise tensor &
$\mathbb R^{\Omega}$
\\

$\mathbf V$ &
Uncertainty map &
$\mathbb R^{\Omega}$

\\

$k$ &
Index of spatial--spectral patches &
$\{1,\ldots,K\}$

\\

$\tau$ &
Index of diffusion step &
$\{1,\ldots,\tau^\star\}$
\\

$\ell$ &
Index of DiT blocks &
$\{1,\ldots,\ell^\star\}$
\\

$r$ &
Index closed-loop stages  &
$\{1,\ldots,r^\star\}$
\\

$i$ &
Index of sensing candidates &
$\{1,\ldots,I\}$
\\

\bottomrule
\end{tabular}
\end{table}

\subsection{Representation of IRMs}
\label{subsec:irm_representation}

We consider a height-aware and time-evolving IRM scenario, where the underlying urban morphology remains static,
while the transmitter location changes over time. Consequently, different height-aware spectrum conditions are generated from time-varying transmitter locations, resulting in a sequence of 3D IRMs. For notational convenience and engineering practicality, both the spatial domain and the spectrum domain are discretized. Specifically, the target area is partitioned into a 3D grid of size $\mathbf X \times \mathbf Y \times \mathbf Z$, while the available spectrum is divided into $\mathbf F$ frequency bands. Therefore, the joint four-dimensional (4D) grid can be denoted by $\Omega \triangleq \mathbf X\times \mathbf Y \times \mathbf Z \times \mathbf F$. Accordingly, each spatial--spectral entry is indexed by $(x,y,z,f)$. Under this discretization, the ground-truth IRM is represented by a 4D tensor:
\begin{equation}
\mathbf{S}=\big[\mathcal{S}_{x,y,z,f}\big]\in \mathbb{R}^{\mathbf X \times \mathbf Y \times \mathbf Z \times \mathbf F}=\mathbb{R}^{\Omega},
\label{eq:irm_tensor}
\end{equation}
where each entry $\mathcal{S}_{x,y,z,f}$ corresponds to a radio measurement, such as received signal strength indicator (RSSI), channel state information (CSI), or path loss, at a specific 3D location and frequency band. The complete collection of all entries constitutes the complete tensor $\mathbf S$, i.e., the ground-truth IRM. In practice, however, only a small fraction of entries can be sensed due to sensing cost, accessibility constraints, and computational overhead \cite{10948391,10025551}, which calls for an efficient IRM construction framework.

\subsection{Environmental prior and binary masks}
\label{subsec:physical_constraints}

The environmental prior is represented by a group of binary masks, denoted by $\mathbf{M}\in\{0,1\}^{\Omega}$.  In this group, there are two primary types: obstacle masks and sensing masks. 

On one hand, the obstacle masks include the building-occupancy mask $\mathbf M_b$, capturing the dominant height-dependent building blockage, and the geographical-context mask $\mathbf M_g$, describing non-building terrain structures that may influence sensing accessibility or radio propagation, such as parks, rivers, lakes, roads, tracks, railways, etc. To maintain notational consistency with the 4D IRM representation $\mathbf{S}$, both $\mathbf M_b$ and $\mathbf M_g$ are written over the spatial--spectral domain, although physical structures typically obstruct almost all conventional communication frequency bands. Accordingly, each entry in the obstacle masks is set to one if the obstacle occupies the corresponding spatial position, and zero otherwise, which remains constant over a given time horizon. On the other hand, the sensing masks contain the UAV sensing mask $\mathbf M_u$ and the manual-device sensing mask $\mathbf M_m$. Different from obstacle masks, these two binary masks record whether sensing has been performed at the corresponding 4D spatial--spectral entry. Therefore, $\mathbf M_u$ and $\mathbf M_m$ show the frequency awareness over a three-dimensional spatial domain and can be dynamically updated as new measurements are collected.

Moreover, the obstacle masks induce platform-dependent accessibility constraints for sensing masks. Specifically, to avoid possible collisions, UAV-based sensing is restricted from approaching buildings, but it can safely survey the airspace above buildings and ground-inaccessible terrains such as rivers and lakes \cite{11278649}. Accordingly, the reachable regions of UAVs are defined as $\mathcal{A}_{u}
\triangleq
\left\{
(x,y,z) : M_{b}(x,y,z,f) = 0, \ \forall f
\right\}$. As a complement, manual devices can enter building interiors, even across different floors, to acquire measurements. However, in most geographical contexts, manual devices can only sense the ground-level results. Thus, the corresponding reachable regions are defined by $\mathcal A_m
=
\left\{
(x,y,z):
M_b(x,y,z,f)=1
\right\}
\cup
\left\{
(x,y,0):
M_g(x,y,0,f)=0
\right\}$.  

Beyond the spatial accessibility constraint defined by $\mathcal{A}_u \cup \mathcal{A}_m$, the spectral dimension also introduces additional practical boundaries. In other words, although mobile UAVs and manual devices are typically capable of full-band sensing at various spatial positions \cite{10791446}, the total number of probed frequency bands is inherently constrained by $F$. 

\subsection{Sparse Sensing Measurements}
\label{subsec:sensing_meas}

Based on predefined sensing masks and obstacle masks, we use two additional tensors to represent sparse sensing measurements: the observation tensor $\mathbf O$ (where sparse sensing has been performed) and the sensing-record tensor $\mathbf \Pi$ (the corresponding measured values).

Specifically, the observation tensor is defined as the logical union of UAV-based and manual-device sensing masks:
\begin{equation}
\mathbf O
=
\mathbf M_u \lor \mathbf M_m
=
\min\{\mathbf M_u+\mathbf M_m,\mathbf 1\},
\label{eq:observation_mask}
\end{equation}
where $\lor$ denotes the element-wise logical OR operation. Therefore, $O_{x,y,z,f}=1$ indicates that the corresponding spatial--spectral entry has been sampled by at least one sensing platform\footnote{If the entry is simultaneously sampled by both UAVs and manual devices, the corresponding mask value remains one, thereby maintaining mathematical consistency with the binary sensing masks.}, while $O_{x,y,z,f}=0$ indicates that it remains unobserved. The number of sampled spatial--spectral entries is given by $N=
\left\|\mathbf O\right\|_0.$

The actual sensed radio values are stored in the sensing-record tensor $\mathbf\Pi\in\mathbb{R}^{\Omega}$, whose entries are defined as
\begin{equation}
\Pi_{x,y,z,f}
=
\begin{cases}
\mathcal{S}_{x,y,z,f}+\epsilon_{x,y,z,f},
& O_{x,y,z,f}=1,\\
0,
& O_{x,y,z,f}=0,
\end{cases}
\label{eq:sensing_record_tensor}
\end{equation}
where $\epsilon_{x,y,z,f}$ denotes the sensing noise. Equivalently, the relationship among three primary 4D tensors, i.e., the ground-truth IRM, all sparse sensing observations, and corresponding measured values, is shown as: 
\begin{equation}
\mathbf \Pi
=
\mathbf O
\odot
(\mathbf S
+
\mathbf E),
\label{eq:sensing_tensor_compact}
\end{equation}
where $\mathbf E$ denotes the sensing-noise tensor and $\odot$ represents the Hadamard product.

\subsection{Inference Objectives}
\label{subsec:infer_obj}

Based on the environmental prior masks $\mathbf M$, the observation tensor $\mathbf O$, and the sensing-record tensor $\mathbf \Pi$, the spectrum inference step has two tightly coupled objectives.

The primary objective is to recover the complete spatial--spectral tensor $\mathbf S$, i.e., construct the IRM over the entire grid $\Omega$ from sparse sensing measurements and environmental priors. Mathematically, the inferred IRM is represented as
\begin{equation}
\hat{\mathbf S}=
\mathcal F_{\Theta}\left(\mathbf \Pi,\mathbf O,\mathbf M\right),
\label{eq:reconstruction_model}
\end{equation}
where $\mathcal F_{\Theta}(\cdot)$ denotes the spectrum-inference model and $\Theta$ denotes its trainable parameters. To evaluate the quality of the inference methods, two criteria are considered simultaneously. First, the inferred values on the observed entries should remain  strictly consistent with the corresponding sensing records stored in $\mathbf \Pi$. Second, for the unobserved entries, the inference model is expected to minimize the discrepancy between inferred $\hat{\mathbf S}$ and the ground-truth $\mathbf S$. Accordingly, the inference objective is formulated as
\begin{equation}
\begin{aligned}
\arg\min_{\Theta}
\Bigl(
\lambda_{\mathrm{obs}}
\|\mathbf O\odot(\hat{\mathbf S}-\mathbf \Pi)\|_F^2
+
\|(\mathbf 1-\mathbf O)\odot(\hat{\mathbf S}-\mathbf S)\|_F^2
\Bigr).
\end{aligned}
\label{eq:inference_objective}
\end{equation}
where the coefficient $\lambda_{\mathrm{obs}}$ balances the relative importance of these two objectives.

Besides constructing the IRM itself at this stage, another important objective is to generate an uncertainty map that quantifies the reliability of inferred radio values and provides feedback to choose new sensing candidates in the subsequent stage. Analogous to the IRM tensor $\mathbf S$, the uncertainty map is represented as a 4D tensor $\mathbf V \in \mathbb{R}^{\Omega}$ over the spatial--spectral domain, defined as
\begin{equation}
\mathbf V
=
[V_{x,y,z,f}]
=
\mathcal U_{\Phi}
\!\left(
\mathbf \Pi,
\mathbf O,
\mathbf M
\right),
\label{eq:uncertainty_generation}
\end{equation}
where $\mathcal U_{\Phi}(\cdot)$ denotes the uncertainty-estimation model and $\Phi$ represents its trainable parameters. Intuitively, for each spatial-spectral entry, the larger $V_{x,y,z,f}$ may indicate insufficient observations or higher inference ambiguity.

%%%%%明天根据这个新的定义，开始写methodology，inner和outer 循环没问题，方法的部分进一步精简阐述。

\subsection{Radio Propagation Model}
\label{subsec:prop_model}

The obstacle masks not only determine sensing accessibility, but also affect the physical radio propagation process. Specifically, the obstacle masks $\mathbf M_b$ and $\mathbf M_g$ jointly influence both large-scale and small-scale channel characteristics, thereby providing another valuable input prior. Moreover, the specific propagation model derived from obstacle masks provides a coarse physics-aware estimate, which regularizes the inference process and reduces physically implausible predictions, particularly in diffusion-based methods \cite{bastek2025physics}.

From the perspective of large-scale fading, obstacle structures affect the Line-of-Sight (LoS)/Non-Line-of-Sight (NLoS) condition between transmitters and receivers, leading to different path-loss and shadowing behaviors. In particular, dense buildings usually introduce additional attenuation, whereas geographical contexts such as parks, rivers, and tracks produce morphology-dependent blockage effects. Therefore, the average received power at location $p=(x,y,z)$ and frequency band $f$ can be approximately expressed as
\begin{equation}
P_r(p,f)
=
P_t(f)
-
\mathrm{PL}(d_p,f)
-
L^{\mathrm{obs}}_{p,f}
\label{eq:large_scale_prior}
\end{equation}
where $P_t(f)$ denotes the transmit power, $d_p$ denotes the Euclidean distance between the transmitter and location $p$, and $\mathrm{PL}(d_p,f)$ follows standardized LoS/NLoS path-loss formulations, such as the 3rd Generation Partnership Project (3GPP) UMa channel model \cite{3gpp38901}. The residual attenuation term $L^{\mathrm{obs}}_{p,f}$ captures additional losses induced by building blockage, diffraction, and terrain-related fading based on empirically validated propagation models \cite{iturp1812}. 

For small-scale fading, obstacle masks further affect multipath propagation and channel visibility. Generally, urban regions with dense building distributions show higher channel uncertainty and larger fading fluctuations, while open regions are more likely to exhibit stable LoS-dominant propagation.

\begin{figure*}[htbp]
    \centering  \includegraphics[width=2.05\columnwidth]{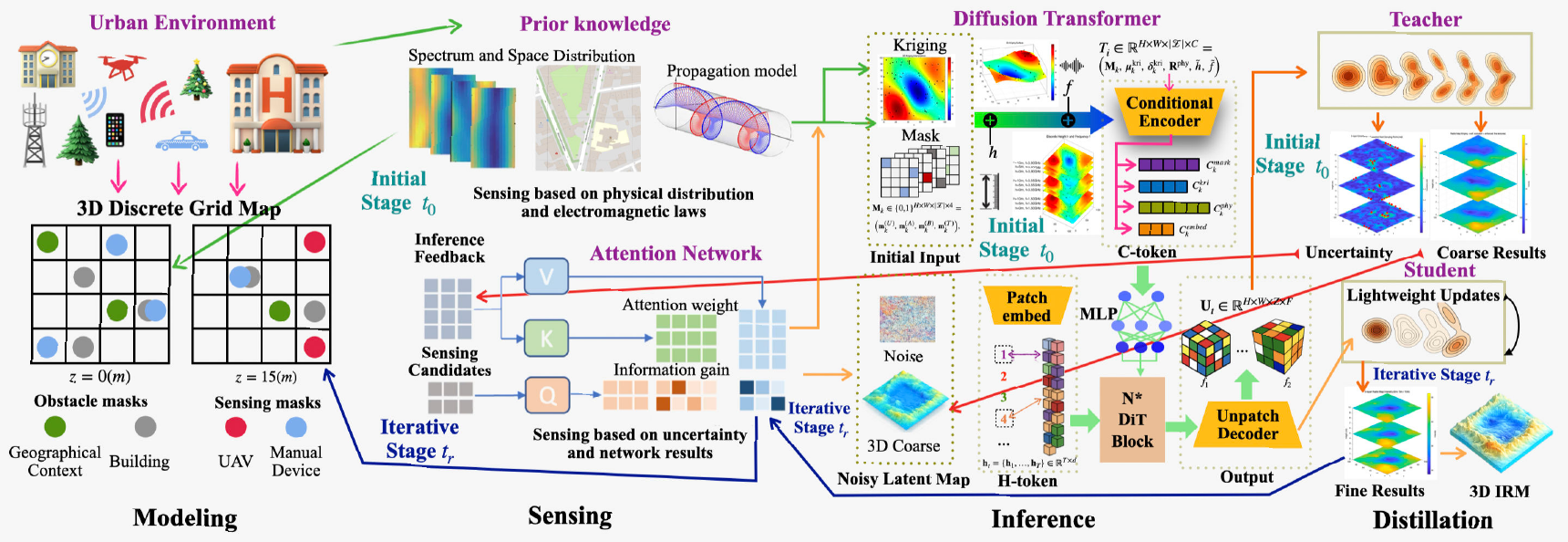}
    \caption{Closed-loop multi-stage IRM construction framework.}
    \label{1-closed_loop_framework}
\end{figure*}

\section{Methodology}
\label{sec:method}

To satisfy practical requirements, we develop a closed-loop multi-stage IRM construction framework based on the system model, which can generate 3D IRMs and continuously update them over time. The overall framework can be understood from two complementary perspectives. \textit{(i)} Outer-loop Progressive Stage Refinement: The proposed IRM construction framework consists of one initial stage followed by multiple iterative stages. The initial stage exploits physical priors including environmental knowledge, obstacle distributions, and initial sparse measurements to generate a coarse IRM estimate. Subsequently, iterative stages progressively refine inference models and update IRMs based on additional sensing measurements. Moreover, to ensure practical applicability as the data volume increases, a knowledge-distillation mechanism is introduced during iterative stages, allowing lightweight student models to replace computationally expensive teacher models while preserving inference accuracy. \textit{(ii)} Inner-loop Feedback Mechanism: Each stage contains two tightly coupled steps, namely spectrum sensing and inference. 
Specifically, the first-step sensing acquires informative radio observations under physical accessibility and environmental constraints, thereby providing reliable ground truth for training inference models. In return, based on the updated sensing measurements, the second-step inference generates fine-grid IRMs together with uncertainty feedback, which is subsequently fed back to guide the next-stage sensing decision. 

The intuitive diagram of the closed-loop
multi-stage IRM construction framework is shown in Fig.~\ref{1-closed_loop_framework}.

\subsection{Initial Stage Sensing: Physics-Aware Sparse Acquisition}
\label{subsec:first_sensing}

In the initial stage, the first sensing step aims to acquire sparse measurements at time instant $t_0$, thereby establishing the observation basis for the subsequent IRM inference step. Following the spatial--spectral discretization introduced in Section~\ref{subsec:irm_representation}, the building-occupancy mask $\mathbf M_b$ and geographical-context mask $\mathbf M_g$ are treated as static environmental priors throughout the considered IRM construction process. In contrast, the UAV sensing mask $\mathbf M_u^{(t)}$ and manual-device sensing mask $\mathbf M_m^{(t)}$ evolve over time as new measurements are collected. Accordingly, the sensing locations in the initial stage are evenly distributed over the space while satisfying physical accessibility constraints given by obstacle occupancy masks. 

Meanwhile, the initial observation mask $\mathbf O^{(t_0)}$ contains in total $N_{t_0}=\left\|\mathbf O^{(t_0)}\right\|_0$ sensing results.  After the corresponding radio measurements are collected at certain spatial–spectral entries, based on \eqref{eq:sensing_tensor_compact}, the sensing-record tensor is updated as
\begin{equation}
\mathbf \Pi^{(t_0)}
=
\mathbf O^{(t_0)}
\odot
\left(
\mathbf S^{(t_0)}
+
\mathbf E^{(t_0)}
\right).
\label{eq:initial_sensing_tensor}
\end{equation}

Beyond the directly collected measurements, a lightweight prior-construction method can be employed to generate an initial sketched IRM. In this work, ordinary Kriging \footnote{Other classical spectrum inference methods, such as matrix completion, $k$-nearest neighbors, and hypothesis-testing, can also be employed as alternatives to obtain the sketched IRM.} is adopted to exploit the spatial correlation among sparse observations and produce a coarse IRM estimate, expressed as
\begin{equation}
\breve{\mathbf S}^{(t_0)}
=\mathcal G\left(\mathbf \Pi^{(t_0)},\mathbf O^{(t_0)},\mathbf M_b,\mathbf M_g\right),
\label{eq:sketched_map_initial}
\end{equation}
where $\mathcal G(\cdot)$ denotes the Kriging-based operator. The resulting sketched map $\breve{\mathbf S}^{(t_0)}$ is not regarded as the final IRM, but as an additional structured prior that can facilitate and regularize the training process of the learning-based inference model $\mathcal F_{\Theta}(\cdot)$. Additionally, the coarse physics-aware estimate based on an empirical propagation model and obstacle masks serves as a complementary prior to the Kriging-based sketched IRM.

Consequently, the sensing step in the initial stage obtains $N_{t_0}$ measurements and generates three categories of outputs: environmental priors, sensing-record tensors $\mathbf \Pi^{(t_0)}$, and coarse structured priors $\breve{\mathbf S}^{(t_0)}$. These outputs jointly provide the input basis for the subsequent inference step.

\subsection{Initial Stage Inference: DiT-Based Networks}

Given the aforementioned sensing outputs, the inference step at the initial stage aims to construct a coarse-grained IRM and provide timely uncertainty-aware feedback for the next sensing step. Unlike conventional spectrum inference methods, the proposed framework needs to simultaneously capture spatial, spectral, and temporal dependencies induced by heterogeneous 3D sensing, electromagnetic propagation laws, and complex urban morphology.

Therefore, to effectively capture these dependencies, a Diffusion Transformer (DiT) network is adopted. Specifically, the transformer backbone captures spatial and temporal variations, conditional tokens encode sensing observations and structured priors, and diffusion-based latent refinement progressively improves inference quality while preserving physical consistency. Accordingly, inspired by \cite{peebles2023scalable}, the proposed inference method is introduced through three key modules: conditional-token construction, hidden-token evolution, and uncertainty-aware map generation.

\subsubsection{Construction of Conditional Tokens}

The purpose of the conditional tokens (\emph{C}-tokens) is to encode sensing observations and structured priors obtained from the initial sensing stage, thereby providing explicit guidance for the subsequent diffusion process. To organize the heterogeneous inputs obtained at $t_0$ into a unified representation, we first concatenate them into a time-variant conditional tensor:
\begin{equation}
\mathbf T_{\mathrm c}^{(t_0)}
=
\operatorname{Cat}
\Big[
\mathbf O^{(t_0)},
\mathbf \Pi^{(t_0)},
\breve{\mathbf S}^{(t_0)},
\mathbf M_b,
\mathbf M_g,
\mathbf S_{\mathrm{phy}}
\Big],
\label{eq:conditional_tensor}
\end{equation}
where $\mathbf S_{\mathrm{phy}}$ represents a coarse physics-aware estimate generated from the empirical propagation model defined in \eqref{eq:large_scale_prior}, which is jointly determined by transmitter configurations, path-loss characteristics, and obstacle-induced attenuation based on $\mathbf M_b$ and $\mathbf M_g$.

Then, the conditional tensor needs to be tokenized for modulation. Specifically, $\mathbf T_{\mathrm c}^{(t_0)}$ is partitioned into $K$ non-overlapping spatial--spectral patches. For the $k$-th patch, a multi-layer perceptron (MLP) is employed to project the local feature into a $d_t$-dimensional token embedding, recorded as
\begin{equation}
\mathbf c_k^{(t_0)}
=
\operatorname{MLP}_{\mathrm c}
\!\left(
\operatorname{Vec}
\big(
\mathbf T_{\mathrm c}^{(t_0)}
|_{\mathcal B_k}
\big)
\right)
+
\mathbf e_k^{\mathrm{pos}},
\label{eq:conditional_token_generation}
\end{equation}
where $\operatorname{Vec}(\cdot)$ denotes vectorization, and $\mathbf e_k^{\mathrm{pos}}$ denotes the positional embedding. Accordingly, the conditional-token sequence is written as
\begin{equation}
\mathbf C^{(t_0)}
=
\left[
\mathbf c_1^{(t_0)},
\ldots,
\mathbf c_K^{(t_0)}
\right]
\in
\mathbb R^{K\times d_t},
\label{eq:conditional_token_sequence}
\end{equation}
where $K$ denotes the number of spatial--spectral patches and $d_t$ represents the embedding dimension of each conditional token. The resulting token sequence is subsequently used to modulate the diffusion transformer.

\subsubsection{Evolution of Hidden Tokens}

Unlike the \emph{C}-tokens that encode sensing observations and structured priors, the hidden tokens (\emph{H}-tokens) are designed to represent the evolving latent IRM throughout the diffusion process. To avoid ambiguity, different from the stage index $r$, we use $\tau$ to denote the diffusion step, while $\ell$ denotes the index of DiT blocks in the backbone.

First, the Kriging-based sketched IRM and the sparse measurements obtained from the sensing step at the initial stage are concatenated and encoded into a compact latent representation:
\begin{equation}
\mathbf Z^{(t_0)}
=
\mathcal E_{\mathrm{lat}}
\Big(
\operatorname{Cat}
\big[
\breve{\mathbf S}^{(t_0)}, \mathbf \Pi^{(t_0)}
\big]
\Big),
\label{eq:latent_template}
\end{equation}
where $\mathcal E_{\mathrm{lat}}(\cdot)$ denotes a lightweight latent encoder. Following the standard diffusion process, Gaussian perturbations are then gradually injected into the latent representation:
\begin{equation}
\mathbf x_{\tau}^{(t_0)}
=
\sqrt{\bar{\alpha}_{\tau}}
\mathbf Z^{(t_0)}
+
\sqrt{1-\bar{\alpha}_{\tau}}
\boldsymbol{\varepsilon},
\qquad
\boldsymbol{\varepsilon}
\sim
\mathcal N(\mathbf 0,\mathbf I),
\label{eq:first_diffusion_latent}
\end{equation}
where $\bar{\alpha}_{\tau}$ denotes the cumulative noise-scheduling coefficient.

Subsequently, similar to the conditional tensor, the noisy latent representation is also partitioned into $K$ non‑overlapping spatial–spectral patches, where an MLP-based embedding layer is used to map the local feature into the hidden-token space:
\begin{equation}
\mathbf h_{k,0}^{(t_0)}
=
\operatorname{MLP}_{\mathrm h}
\left(
\operatorname{Vec}
\left(
\mathbf x_{\tau}^{(t_0)}
\big|_{\mathcal B_k}
\right)
\right)
+
\mathbf e_k^{\mathrm{pos}}.
\label{eq:hidden_token_generation}
\end{equation}

In contrast to \emph{C}-tokens, which remain fixed and only depend on the input priors at time $t_0$, \emph{H}-tokens evolve across transformer layers as they are progressively refined by the DiT blocks and feed forward operations. Accordingly, the $\ell$-th hidden-token sequence is formed as
\begin{equation}
\mathbf H_\ell^{(t_0)}
=
\left[
\mathbf h_{1,\ell}^{(t_0)},
\mathbf h_{2,\ell}^{(t_0)},
\ldots,
\mathbf h_{K,\ell}^{(t_0)}
\right]
\in\mathbb R^{K\times d_h},
\label{eq:hidden_token_sequence}
\end{equation}
where $d_h$ is the hidden embedding dimension, $\mathbf h_{k,\ell}^{(t_0)}$ denotes the hidden token corresponding to the $k$-th spatial-spectral patch when applying the $\ell$-th DiT block.

To inject conditional information into the DiT backbone, the sequence of  \emph{C}-tokens is first aggregated via average pooling:
\begin{equation}
\bar{\mathbf c}^{(t_0)}
=
\frac{1}{K}\sum_{k=1}^{K}\mathbf c_k^{(t_0)},
\end{equation}
and then combined with the diffusion-step embedding $\mathbf e_\tau$, feeding into a modulation network:
\begin{equation}
\Delta_\ell^{(t_0)}
=
\operatorname{Mod}_\ell
\big(
\bar{\mathbf c}^{(t_0)} + \mathbf e_\tau
\big),
\end{equation}
where $\operatorname{Mod}_\ell(\cdot)$ generates the adaptive scale, shift, and residual-gating parameters used to modulate the multi-head self-attention (MSA) and feed-forward (FFN) branches of the $\ell$-th DiT block. Specifically, let $\mathbf H_{\ell-1}^{(t_0)}$ denote the input hidden-token sequence for the $\ell$-th DiT block. The self-attention branch is formulated as
\begin{equation}
\begin{aligned}
\widetilde{\mathbf H}_{\ell-1}^{(t_0)}
&=
\operatorname{LN}
\big(
\mathbf H_{\ell-1}^{(t_0)}
\big),\\
\mathbf U_{\ell}^{(t_0)}
&=
\mathbf H_{\ell-1}^{(t_0)}
+
\boldsymbol{\alpha}_{\ell}^{\mathrm{msa}}
\odot
\operatorname{MSA}
\Big(
(1+\boldsymbol{\gamma}_{\ell}^{\mathrm{msa}})
\odot
\widetilde{\mathbf H}_{\ell-1}^{(t_0)}
+
\boldsymbol{\beta}_{\ell}^{\mathrm{msa}}
\Big).
\end{aligned}
\label{eq:dit_msa_update_refined}
\end{equation}
where $\operatorname{LN}(\cdot)$ denotes layer normalization, and
$\boldsymbol{\alpha}_{\ell}^{\mathrm{msa}}$,
$\boldsymbol{\gamma}_{\ell}^{\mathrm{msa}}$,
$\boldsymbol{\beta}_{\ell}^{\mathrm{msa}}$
are the corresponding scale, shift, and residual modulation parameters for the MSA branch. Analogously, the feed-forward branch is then updated as
\begin{equation}
\begin{aligned}
\widehat{\mathbf U}_{\ell}^{(t_0)}
&=
\operatorname{LN}
\big(
\mathbf U_{\ell}^{(t_0)}
\big),\\
\mathbf H_{\ell}^{(t_0)}
&=
\mathbf U_{\ell}^{(t_0)}
+
\boldsymbol{\alpha}_{\ell}^{\mathrm{ffn}}
\odot
\operatorname{FFN}
\Big(
(1+\boldsymbol{\gamma}_{\ell}^{\mathrm{ffn}})
\odot
\widehat{\mathbf U}_{\ell}^{(t_0)}
+
\boldsymbol{\beta}_{\ell}^{\mathrm{ffn}}
\Big).
\end{aligned}
\label{eq:dit_ffn_update_refined}
\end{equation}
where the FFN-related modulation parameters are defined in a similar manner as the MSA branch.

Through the repeated interaction between MSA and FFN branches under conditional modulation, DiT blocks continuously update. In other words, the latent IRM representation from \emph{H}-tokens is progressively refined toward a physically coherent inference with high accuracy.

%%%马上从这里开始修改
\subsubsection{Uncertainty-aware Map Generation}

At diffusion step $\tau$, the final hidden-token representation $\mathbf H_{\ell^\star}^{(t_0)}$ is shared by two prediction heads for denoising and uncertainty estimation. 

For denoising, the final hidden representation $\mathbf H_{\ell^\star}^{(t_0)}$ is mapped to a noise estimate $\hat{\boldsymbol{\varepsilon}}_{\tau}^{(t_0)}$, which approximates the injected noise $\boldsymbol{\varepsilon}$. The denoised latent is then obtained via the standard reverse diffusion step:
\begin{equation}
\hat{\mathbf Z}^{(t_0)}
=
\frac{
\mathbf x_{\tau}^{(t_0)}
-
\sqrt{1-\bar{\alpha}_{\tau}}\,
\hat{\boldsymbol{\varepsilon}}_{\tau}^{(t_0)}
}{
\sqrt{\bar{\alpha}_{\tau}}
}.
\label{eq:denoised_latent}
\end{equation}
Subsequently, the coarse IRM at initial stage $t_0$ is then inferred through a lightweight decoder:
\begin{equation}
\hat{\mathbf S}^{(t_0)}
=
\mathcal D_{\mathrm{lat}}\!\left(\hat{\mathbf Z}^{(t_0)}\right),
\label{eq:first_coarse_map}
\end{equation}
where $\mathcal D_{\mathrm{lat}}(\cdot)$ consists of a shallow convolutional refinement module to restore spatial--spectral consistency.

For uncertainty estimation, we adopt a stability-based formulation over consecutive denoising steps:
\begin{equation}
\boldsymbol{\Sigma}^{(t_0)}
=
\left(
\hat{\mathbf S}_{\tau}^{(t_0)}
-
\hat{\mathbf S}_{\tau-1}^{(t_0)}
\right)^{\odot 2},
\label{eq:first_variance_tensor}
\end{equation}
where $\hat{\mathbf S}_{\tau}^{(t_0)}$ and $\hat{\mathbf S}_{\tau-1}^{(t_0)}$ denote two adjacent intermediate reconstructions during the reverse diffusion process. Larger values indicate higher prediction instability. To incorporate the measurement consistency, the uncertainty map is defined as
\begin{equation}
\mathbf V^{(t_0)}
=
\boldsymbol{\Sigma}^{(t_0)}
+
\lambda_{\mathrm r}
\mathbf O^{(t_0)}
\odot
\left(
\hat{\mathbf S}^{(t_0)}-\mathbf \Pi^{(t_0)}
\right)^{\odot 2}.
\label{eq:first_uncertainty_map}
\end{equation}
Generally, higher values of $V_{x,y,z,f}^{(t_0)}\in \mathbf V^{(t_0)}$ indicate either elevated inference instability or significant discrepancy between inferred and observed value. 

Overall, the learning-based IRM inference at initial stage can be unified as a parametric mapping $\mathcal F_{\Theta}(\cdot)$, where $\Theta$ includes token encoders, DiT backbone, and decoding heads.

\subsection{Iterative Stages Sensing: Active Acquisition}
\label{sec:attn_sensing}

In the iterative sensing stages, we introduce a unified temporal index $r \in \{1,2,...,r^\star\}$ to denote the current sensing iteration. At each time instant $t_r$, the sensing policy is actively executed based on the historical information available up to the previous stage $t_{r-1}$, including the inferred IRM, observation mask, and uncertainty estimates. This formulation naturally induces a closed-loop interaction between inference and sensing steps, where the inference output at stage $t_{r-1}$ directly shapes the sensing strategy at stage $t_{r}$.

To illustrate this process, we consider the first iterative stage at time instant $t_1$ as an example, following the initial stage $t_0$. The available environmental information consists of three key components: the fixed obstacle masks $\mathbf{M}_b$ and $\mathbf{M}_g$, the uncertainty map $\mathbf{V}^{(t_0)}$, and the coarse inferred IRM $\hat{\mathbf{S}}^{(t_0)}$, which serve as the basis for subsequent active sensing decisions. At this stage, the sensing step is formulated as a selection problem over a candidate set, indexed by $i \in \{1,2,\ldots,I\}$. The goal is to identify $N^{(t_1)}$ informative candidates within the set. Motivated by the scaled dot-product attention mechanism in Transformer architectures~\cite{vaswani2017attention},  we design an attention-based network to enable active and intelligent selection. Specifically, the sensing candidate coordinates, environmental context, and uncertainty feedback are embedded as the query (Q), key (K), and value (V) using multilayer perceptron (MLP), respectively. Mathematically, it can be recorded as $\mathbf q_i^{(t_1)} = \mathrm{MLP}_q(\mathbf r_i^{(t_1)}); \mathbf k_i^{(t_1)} = \mathrm{MLP}_k(\mathbf b_i^{(t_1)}); \mathbf v_i^{(t_1)} = \mathrm{MLP}_v(\mathbf u_i^{(t_1)}). $ 

For sensing candidates, $\mathbf r_i^{(t_1)} = [x_i, y_i, z_i, f_i] \in \Omega$ denotes the feasible spatial--spectral coordinate of candidate $i$, and $\mathbf b_i^{(t_1)}$ represents the corresponding environmental context, which encodes both static obstacle constraints and dynamic inference priors, defined as
\begin{equation}
\mathbf b_i^{(t_1)} =
\operatorname{Cat}
\left[
\mathbf M_{b},
\mathbf M_{g},
\hat{\mathbf S}^{(t_0)}
\right].
\end{equation}
In addition, $\mathbf u_i^{(t_1)} = V_i^{(t_0)}$ quantifies the uncertainty associated with the corresponding spatial--spectral location, thereby characterizing its expected informativeness for active sensing.

Accordingly, the attention weight between candidate $i$ and all candidates is then computed as
\begin{equation}
\eta_{ij}^{(t_1)}
=
\frac{
\exp\left((\mathbf q_i^{(t_1)})^{\top}\mathbf k_j^{(t_1)} / \sqrt{d_a}\right)
}{
\sum_{j=1}^{I}
\exp\left((\mathbf q_i^{(t_1)})^{\top}\mathbf k_j^{(t_1)} / \sqrt{d_a}\right)
}.
\end{equation}
where $d_a=\dim(\mathbf q_i)=\dim(\mathbf k_j)$ denotes attention embedding dimension. Based on the attention distribution, uncertainty information is aggregated in a context-aware manner:
\begin{equation}
\mathbf o_i^{(t_1)} = \sum_{j=1}^{I} \eta_{ij}^{(t_1)}\mathbf v_j^{(t_1)},
\end{equation}
which enables each candidate to incorporate global environmental correlations rather than relying solely on local uncertainty estimates. Finally, the acquisition score of each candidate is computed as
\begin{equation}
s_i^{(t_1)} =
\mathbf w_s^{\top}
\left[
\mathbf q_i^{(t_1)} ; \mathbf o_i^{(t_1)}
\right]
+ b_s,
\label{final_attention_network}
\end{equation}
where $\mathbf w_s$ and $b_s$ are learnable parameters.

Based on Eq.~\eqref{final_attention_network}, the sensing step in iterative stage $t_1$ selects the top-$N_{t_1}$ candidates with the highest acquisition scores. In other words, the new observation mask is recorded by $\mathbf O^{(t_1)}=\mathbf M_u^{(t_1)} \lor \mathbf M_m^{(t_1)}$. Accordingly, the sensing-record tensor $\mathbf \Pi^{(t_1)}$ is updated, providing a broader observation basis for the subsequent inference step in iterative stage $t_1$.

Moreover, the proposed attention-based sensing mechanism is recursively applied to all subsequent iterative stages $t_r$ until the ending stage. Although the corresponding query, key, and value embeddings vary across stages due to the continuously updated observation masks, refined IRMs, and dynamic uncertainty estimates, the candidate-selection network remains unchanged. Thus, the sensing policy can be repeatedly exploited.

\subsection{Iterative Stages Inference: Knowledge-Distilled Refinement}

To reduce the computational burden introduced by repeatedly executing the full-scale DiT model, a knowledge-distillation mechanism is used for the inference steps throughout the iterative stages. Specifically, the high-capacity DiT network employed at the initial stage is regarded as a teacher model, denoted by $\mathcal F_{\Theta_T}$. Subsequently, a lightweight student model $\mathcal F_{\Theta_S}$ is trained to mimic the teacher behavior while requiring substantially lower computational complexity. Moreover, the student model simultaneously learns both the IRM construction and the uncertainty feedback generated by the teacher, formulating the knowledge distillation objective as
\begin{equation}
\mathcal L_{\mathrm{KD}}
=
\lambda_s
\left\|
\hat{\mathbf S}_{S}
-
\hat{\mathbf S}_{T}
\right\|_F^2
+
\lambda_v
\left\|
\mathbf V_{S}
-
\mathbf V_{T}
\right\|_F^2,
\end{equation}
where $(\hat{\mathbf S}_{T},\mathbf V_{T})
=\mathcal F_{\Theta_T}(\cdot)$ and $(\hat{\mathbf S}_{S},\mathbf V_{S})=\mathcal F_{\Theta_S}(\cdot)$ denote the outputs of the teacher and student networks under identical inputs, respectively, and $\lambda_s$ and $\lambda_v$ are balancing coefficients.

To further improve inference efficiency, the student adopts a compressed DiT architecture with fewer DiT blocks, reduced hidden embedding dimensions, and fewer diffusion steps. After convergence, the distilled student model is recursively used throughout all subsequent iterative stages to efficiently infer the IRM and uncertainty map. Taking the iterative stage at time $t_r$ as an example, the knowledge-distilled student model takes updated sensing step outcome, fixed obstacle masks, and previously inferred IRM at $t_{r-1}$ as initialization priors. Thus, refined IRM is generated as
\begin{equation}
\hat{\mathbf S}^{(t_r)}
=
\mathcal F_{\Theta_S}
\Big(
\hat{\mathbf S}^{(t_{r-1})},
\mathbf \Pi^{(t_r)},
\mathbf O^{(t_r)},
\mathbf M_b,
\mathbf M_g
\Big).
\end{equation}
Meanwhile, an updated uncertainty map $\mathbf V^{(t_r)}$ is simultaneously produced and fed back to the attention-based network, guiding the sensing step in the subsequent stage at time $t_{r+1}$. Consequently, the sensing and inference steps form a recursive closed-loop process, where newly collected  measurements continuously improve the inferred IRM, while the refined IRM further enhances future sensing decisions.

In conclusion, by executing the teacher model only at the initial stage and repeatedly reusing the distilled student model afterwards, the proposed framework substantially reduces computational overhead while preserving most of the inference capability, which facilitates practical deployment of IRMs in large-scale urban regions.

\section{Dataset Generation and Experimental Setup}
\label{sec:experiments}

This section presents a novel self-constructed dataset for 3D spatiotemporal IRMs, which reflects realistic urban radio environments by jointly modeling transmitter mobility, height-aware obstacle distributions, and temporal spectrum variations. The generated dataset subsequently serves as ground truth for evaluating the proposed closed-loop multi-stage IRM construction framework against other baseline methods. The key simulation parameters are summarized in Table~\ref{tab:simulation_parameters}.

\begin{table}[htbp]
\centering
\caption{Key simulation parameters}
\label{tab:simulation_parameters}
\begin{tabular}{@{}l l@{}}
\toprule
\textbf{Parameter} & \textbf{Value}\\
\midrule

\textbf{General settings} &\\

Number of cities &
50 urban regions
\\

Map source &
OpenStreetMap
\\

Discretized grid &
$500\times500$
\\

Per grid resolutions &
$6\mathrm{m}\times6\mathrm{m}$
\\

Number of TX &
3
\\

Geographical context &

Buildings, Tracks, Parks, Rivers, etc.

\\

Ground-truth &
WinProp-Based simulator
\\

\midrule

\textbf{Spatial-domain settings} &\\

Spatial resolution &
$6\,\mathrm{m}/\mathrm{grid}$
\\

TX movement   & 10 trajectories per city

\\

TX height &
30 m
\\

Receiver heights &
varied in $\{0,15,30\}$ m
\\

Obstacle heights &
within $[0,100]$ m
\\

\midrule

\textbf{Spectrum-domain settings} &\\

Carrier frequency &
3.75 GHz
\\

Radio metric &
RSSI (dBm)
\\

TX transmitting power &
46 dBm
\\

Sector antennas &
3 sectors, 18 dBi gain
\\

\midrule

\textbf{Time-evolving sensing settings} &\\

Total sensing candidates &
2500
\\

Patch-level sensing sparsity &
$\{1.6\%, 2.4\%, 3.2\%, 4.0\%, 4.8\%\}$

\\

\bottomrule
\end{tabular}
\end{table}

\subsection{Physical Map Preparation}
\label{subsec:basic_assumptions}

%% 总起句；
%%% 下面三段分别对应：原始图，遮挡高度图，以及tx移动图
Since a height-aware and time-evolving IRM scenario is considered, the underlying physical map should explicitly capture the temporal evolution of the relative positions between radio transmitters and sensing receivers (UAVs and users), as well as three-dimensional obstructions caused by buildings and diverse geographical contexts. The real-world physical maps are first obtained from OpenStreetMap. To ensure diversity, 50 urban regions from different countries worldwide are selected, each covering an approximate planar area of $3\,\mathrm{km} \times 3\,\mathrm{km}$. 

To incorporate height-aware morphological characteristics into the original planar OpenStreetMap, the following height-assignment criteria are introduced. \textit{(i)} For buildings, if an explicit OpenStreetMap height tag is available, the corresponding height is used; otherwise, if the floor (levels) tag exists, it is converted into height using a $3~\mathrm{m}$ per-floor assumption; otherwise, the height is estimated from the building footprint area with bounded perturbation. \textit{(ii)} The height of different geographical contexts varies: roads, rivers, and lakes are assigned zero obstruction height as non-blocking entities; public parks and green spaces are modeled as low-density urban structures with an equivalent height of $5~\mathrm{m}$, while overpasses and high tracks are uniformly represented as $10~\mathrm{m}$ obstructions within their footprints for simplicity. For cities with significant elevation variations, e.g., the hilly regions around Victoria Peak in Hong Kong, the height maps are further update using contour information, shown in \hl{Appendix B}.

\begin{figure}[!t]
\centering
\setlength{\tabcolsep}{0pt}
\renewcommand{\arraystretch}{0.52}
\scriptsize
\begin{tabular}{@{}cccc@{}}
Shanghai & Berlin & Nairobi & Chicago

\\
\noalign{\vskip -0.00001mm}

\includegraphics[width=0.235\columnwidth]{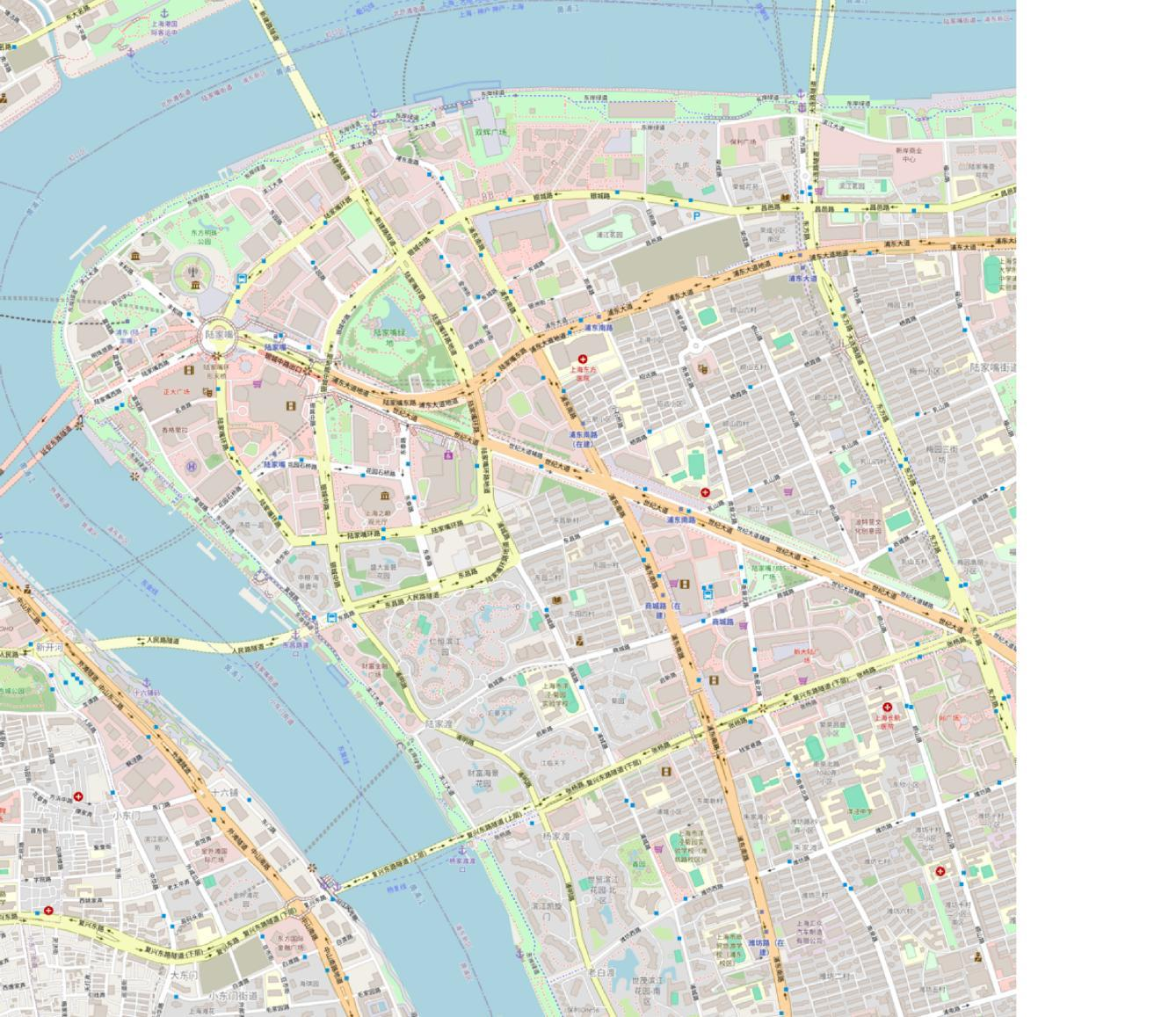} &
\includegraphics[width=0.235\columnwidth]{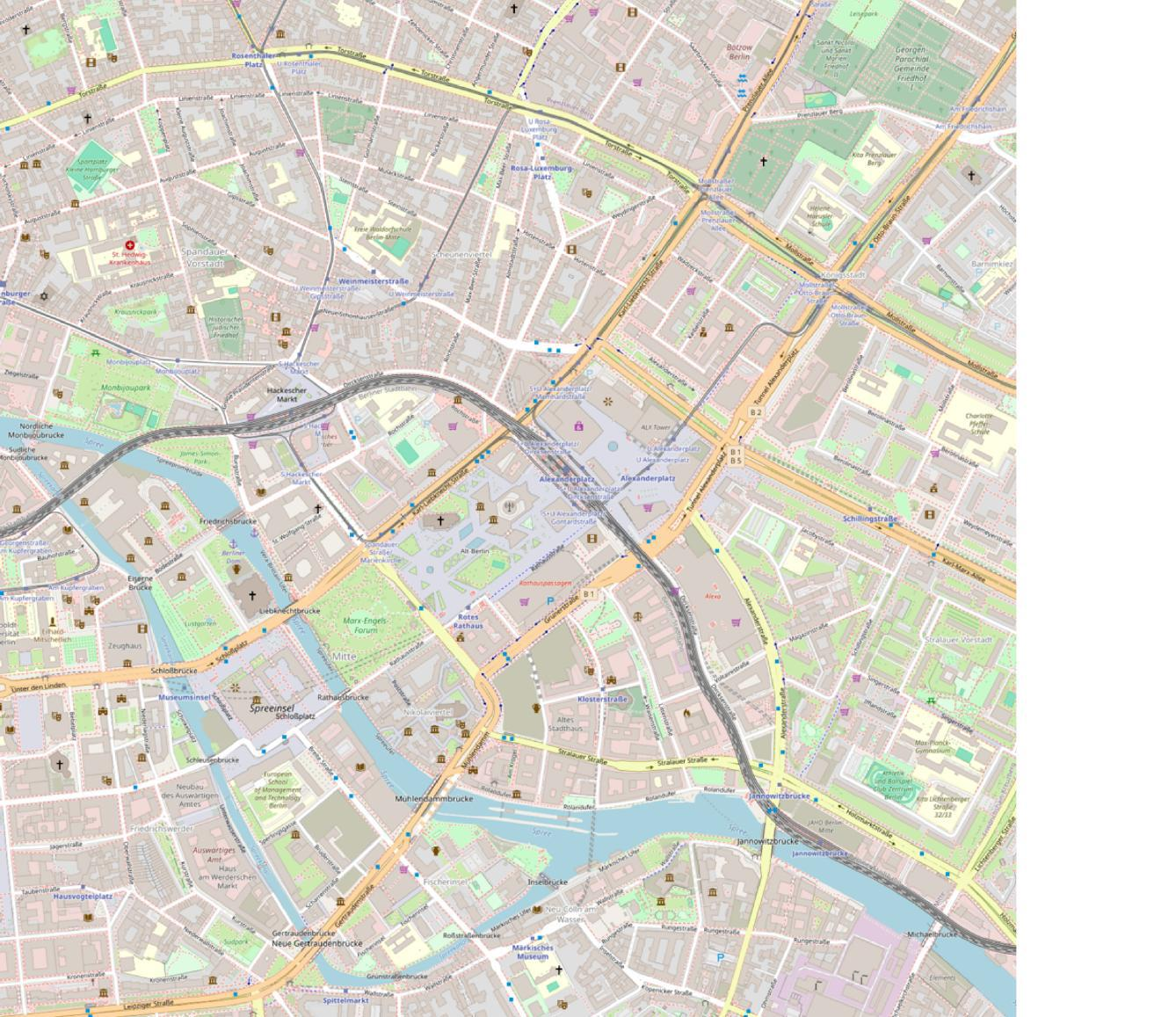} &
\includegraphics[width=0.235\columnwidth]{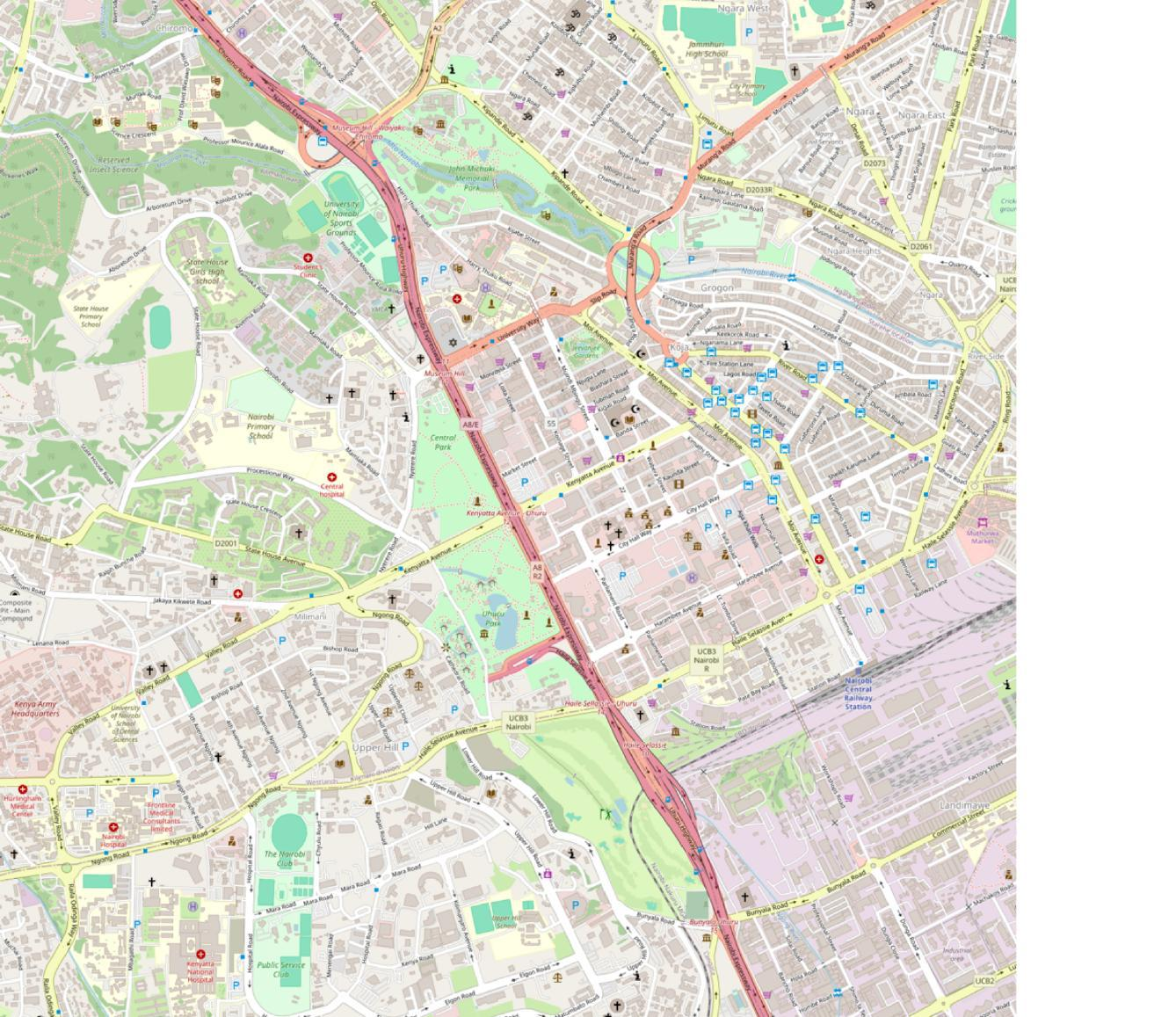} &
\includegraphics[width=0.235\columnwidth]{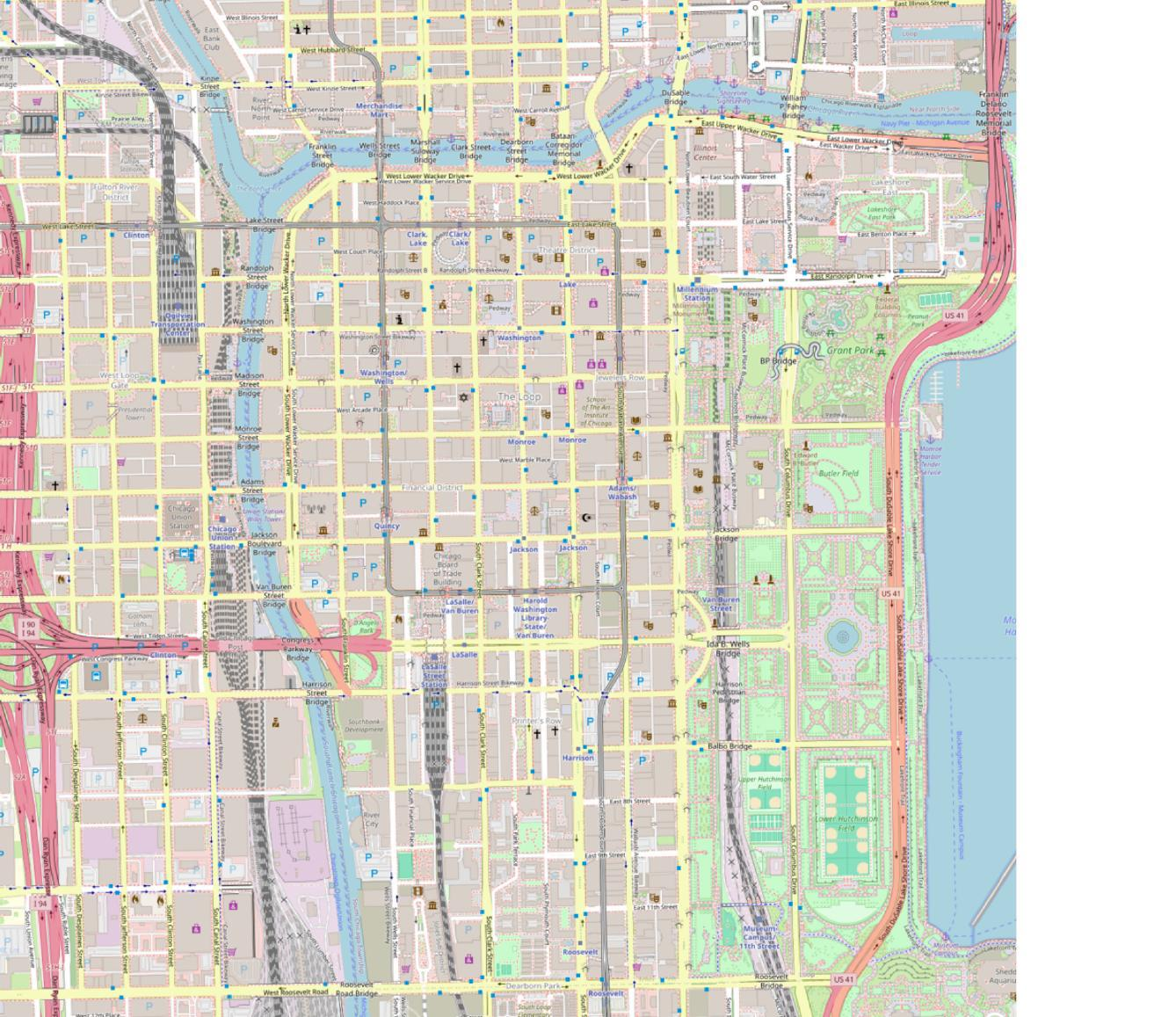} \\

\multicolumn{4}{c}{\scriptsize \textbf{(a)} Initial OpenStreetMap.}\\[-0.1mm]

\includegraphics[width=0.235\columnwidth]{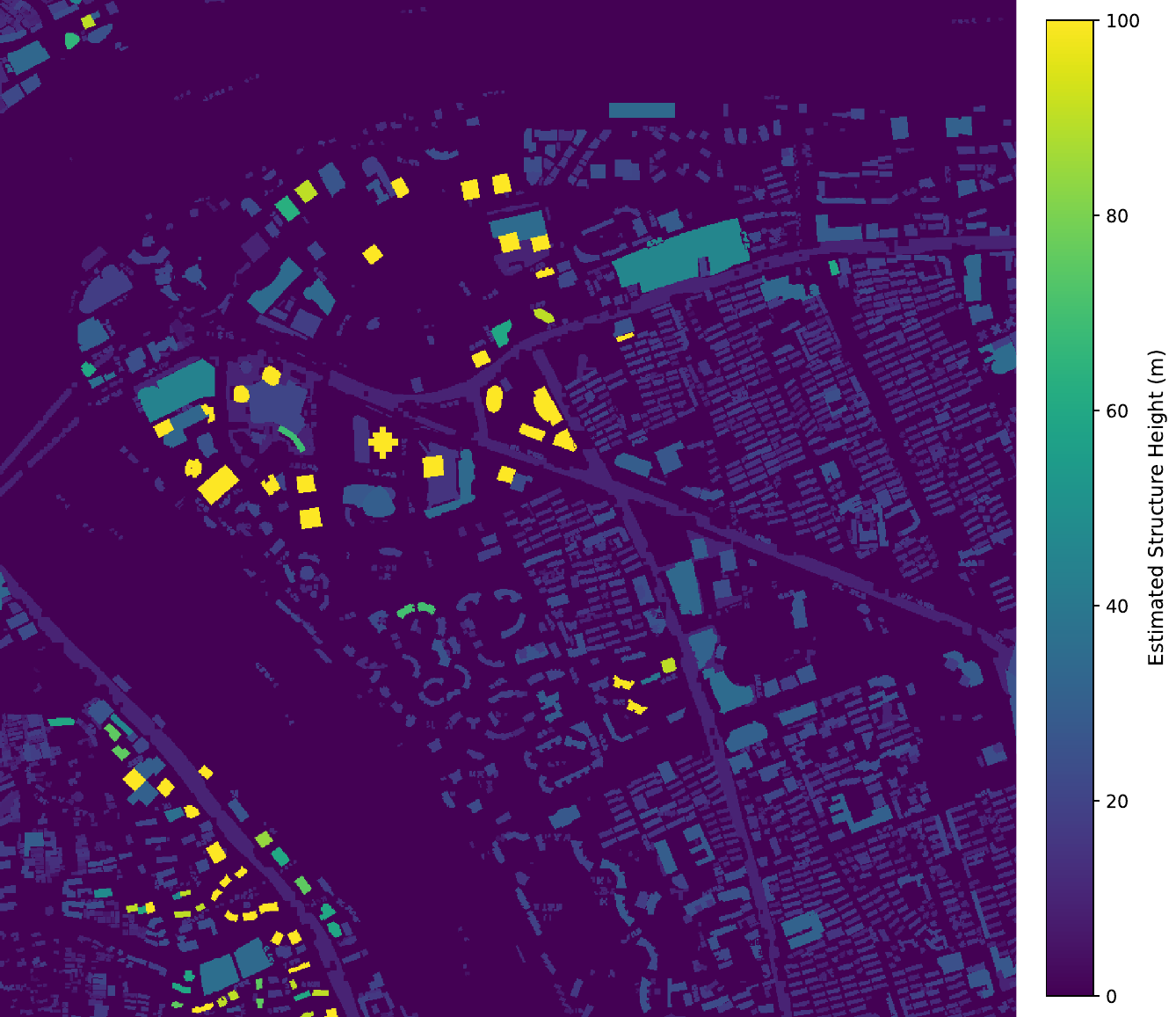} &
\includegraphics[width=0.235\columnwidth]{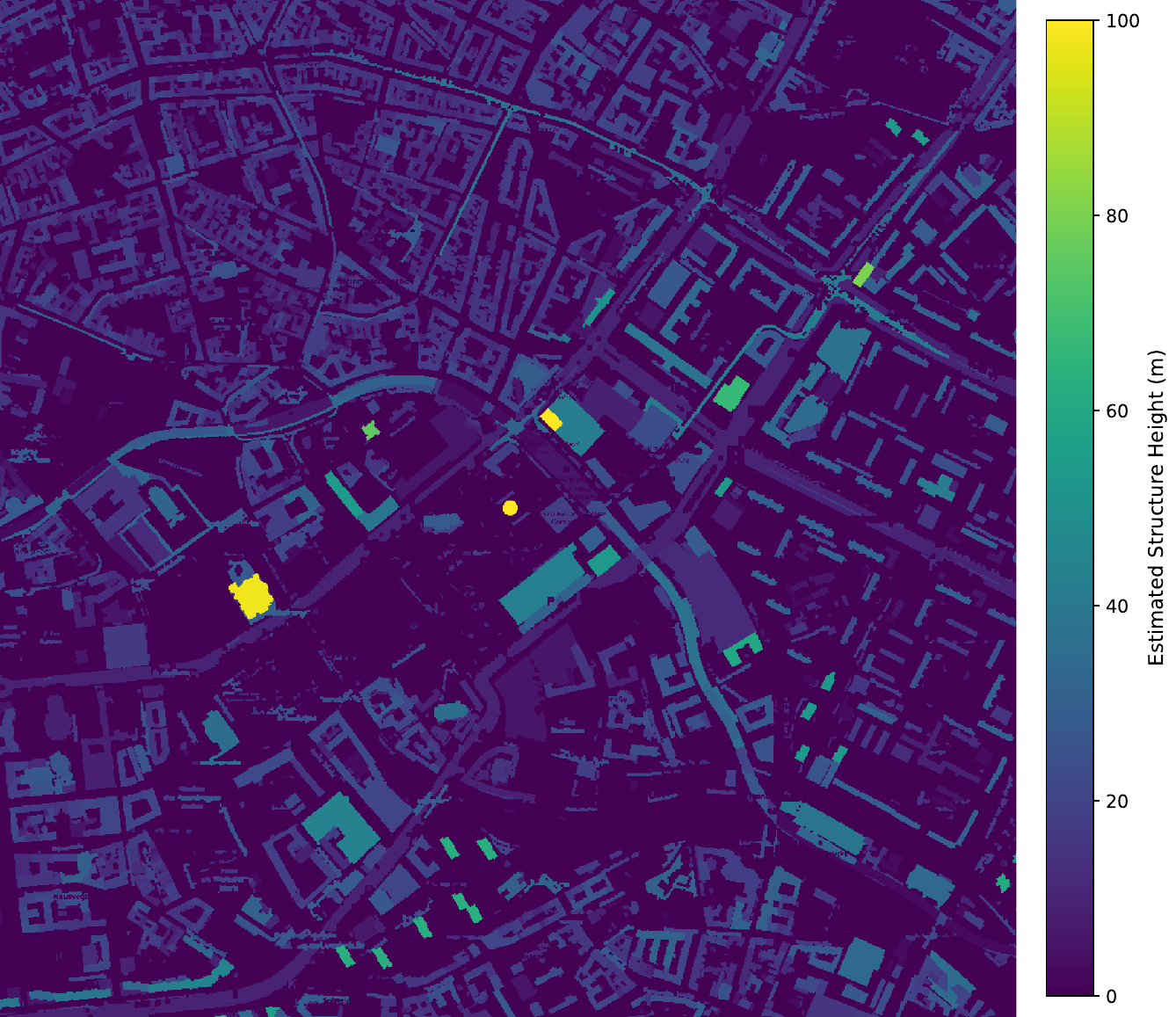} &
\includegraphics[width=0.235\columnwidth]{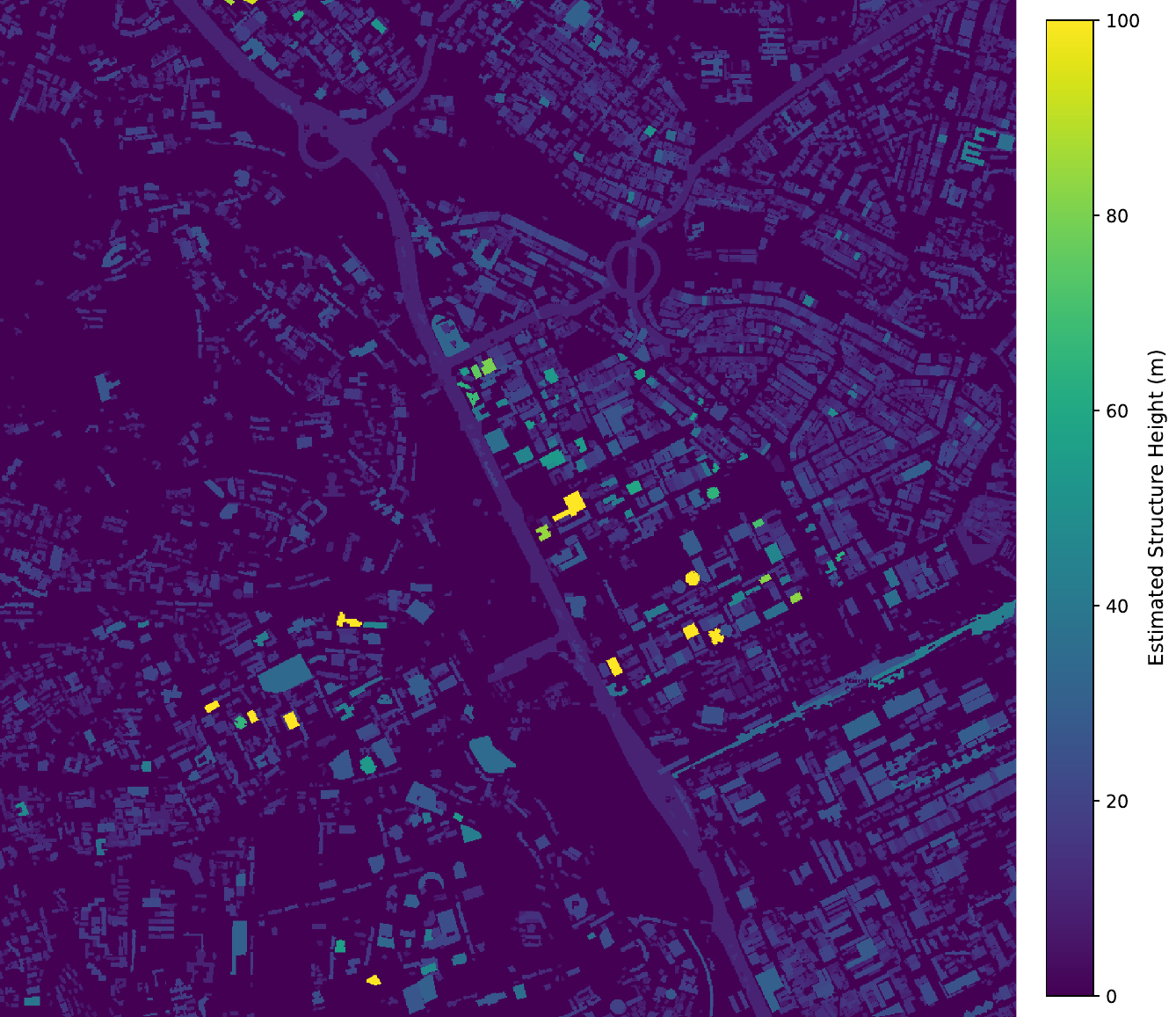} &
\includegraphics[width=0.235\columnwidth]{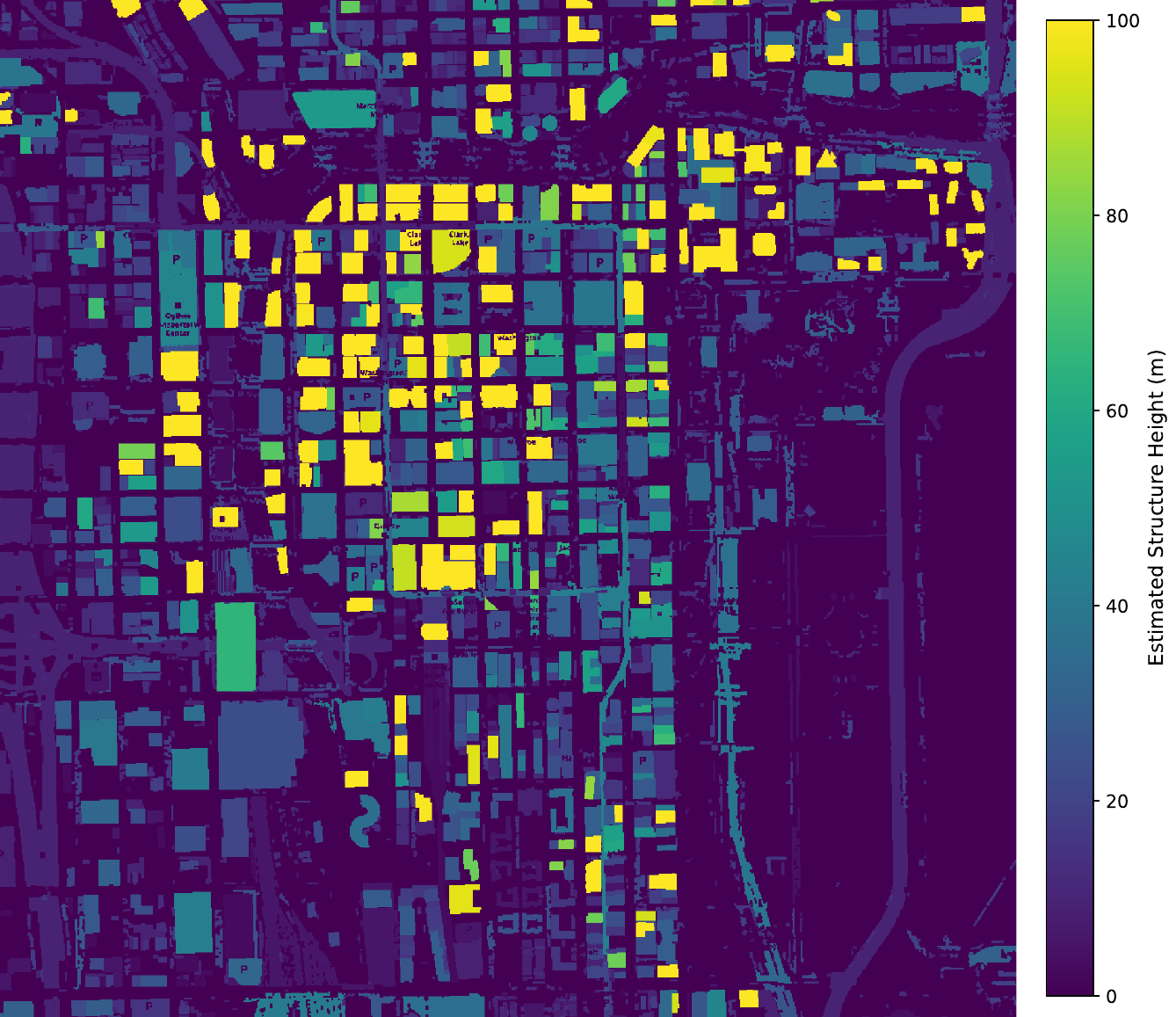} \\

\multicolumn{4}{c}{\scriptsize \textbf{(b)} Height map generated based on the height-estimation criteria.}\\[-0.001mm]

\includegraphics[width=0.235\columnwidth]{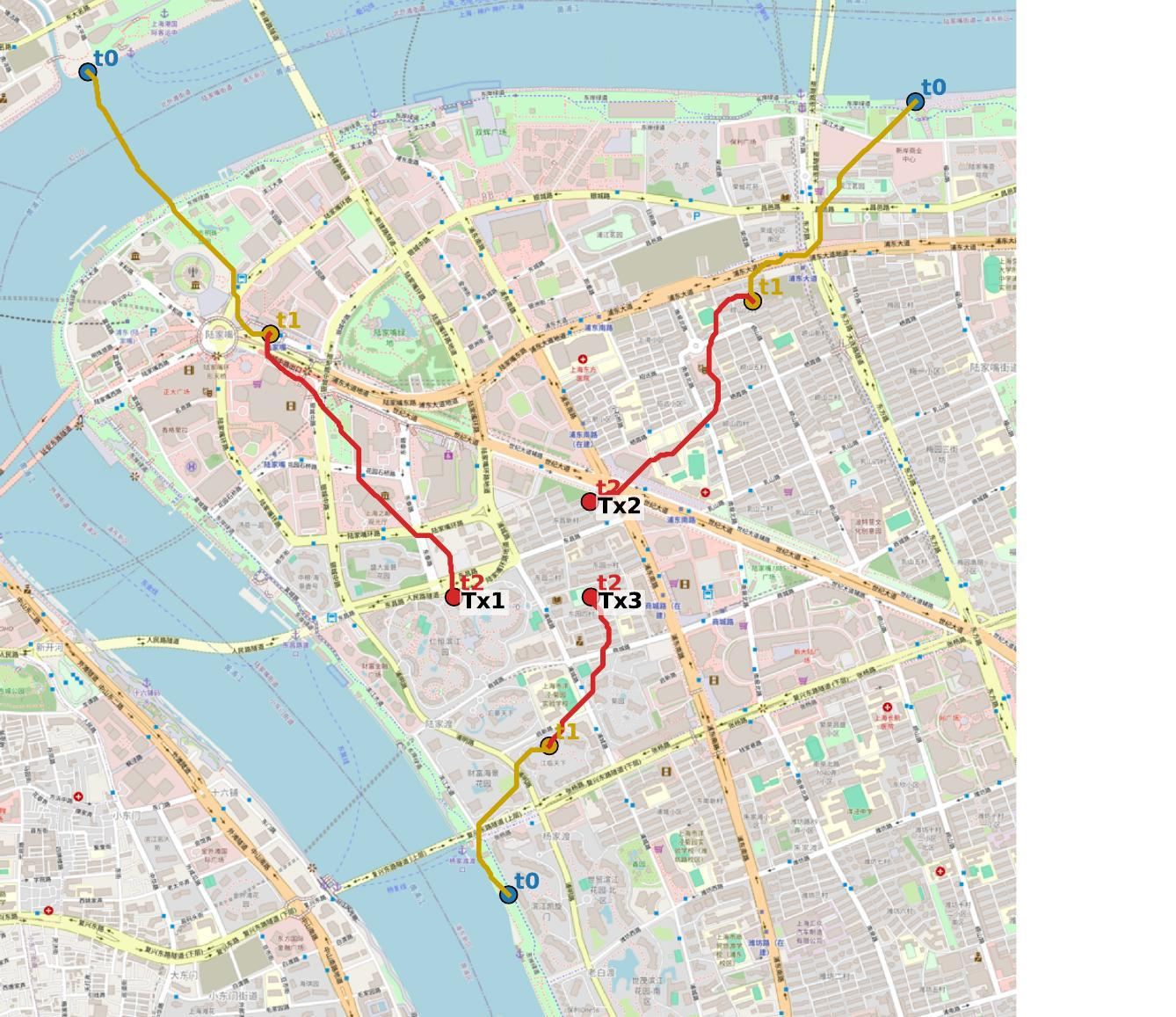} &
\includegraphics[width=0.235\columnwidth]{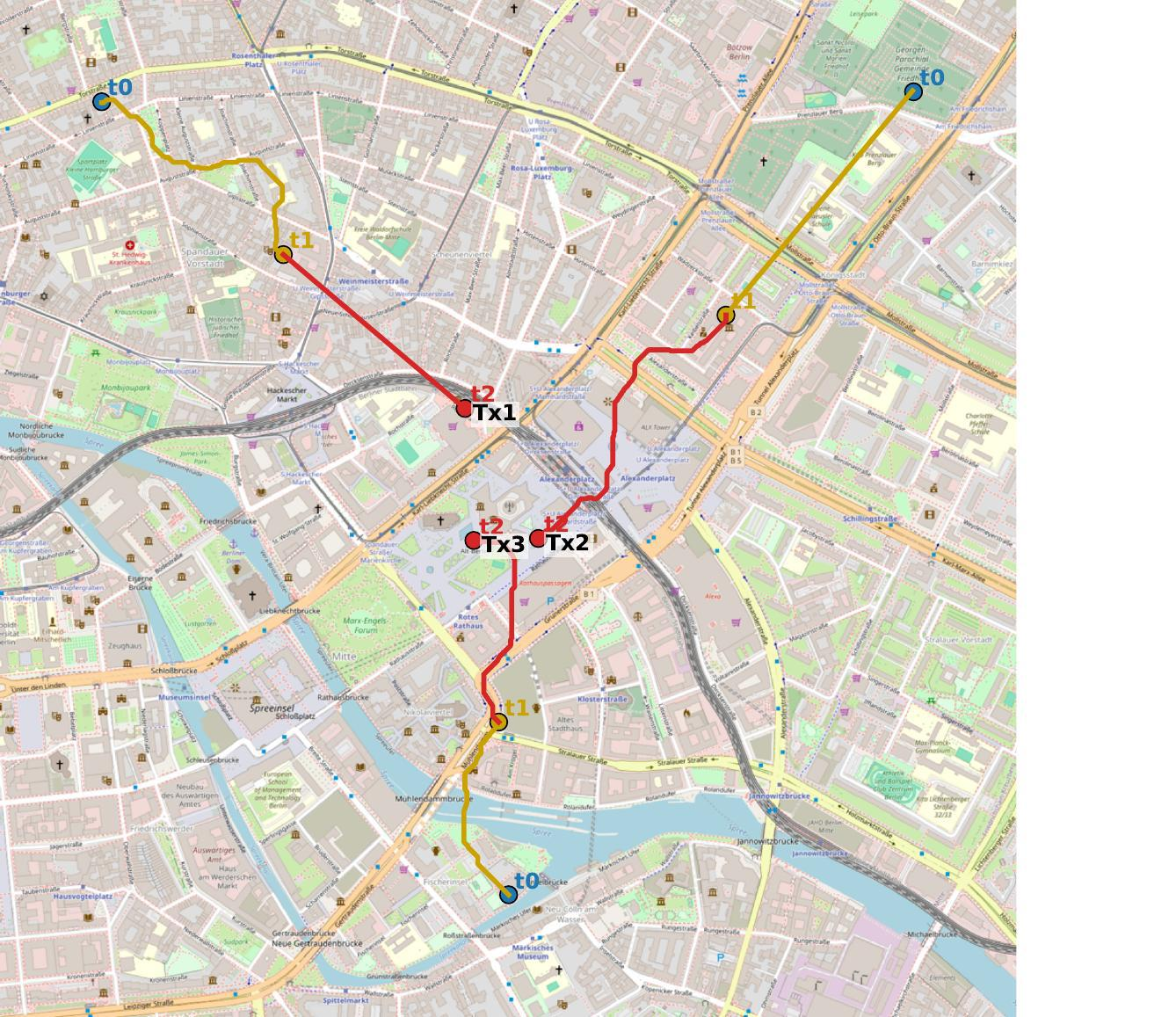} &
\includegraphics[width=0.235\columnwidth]{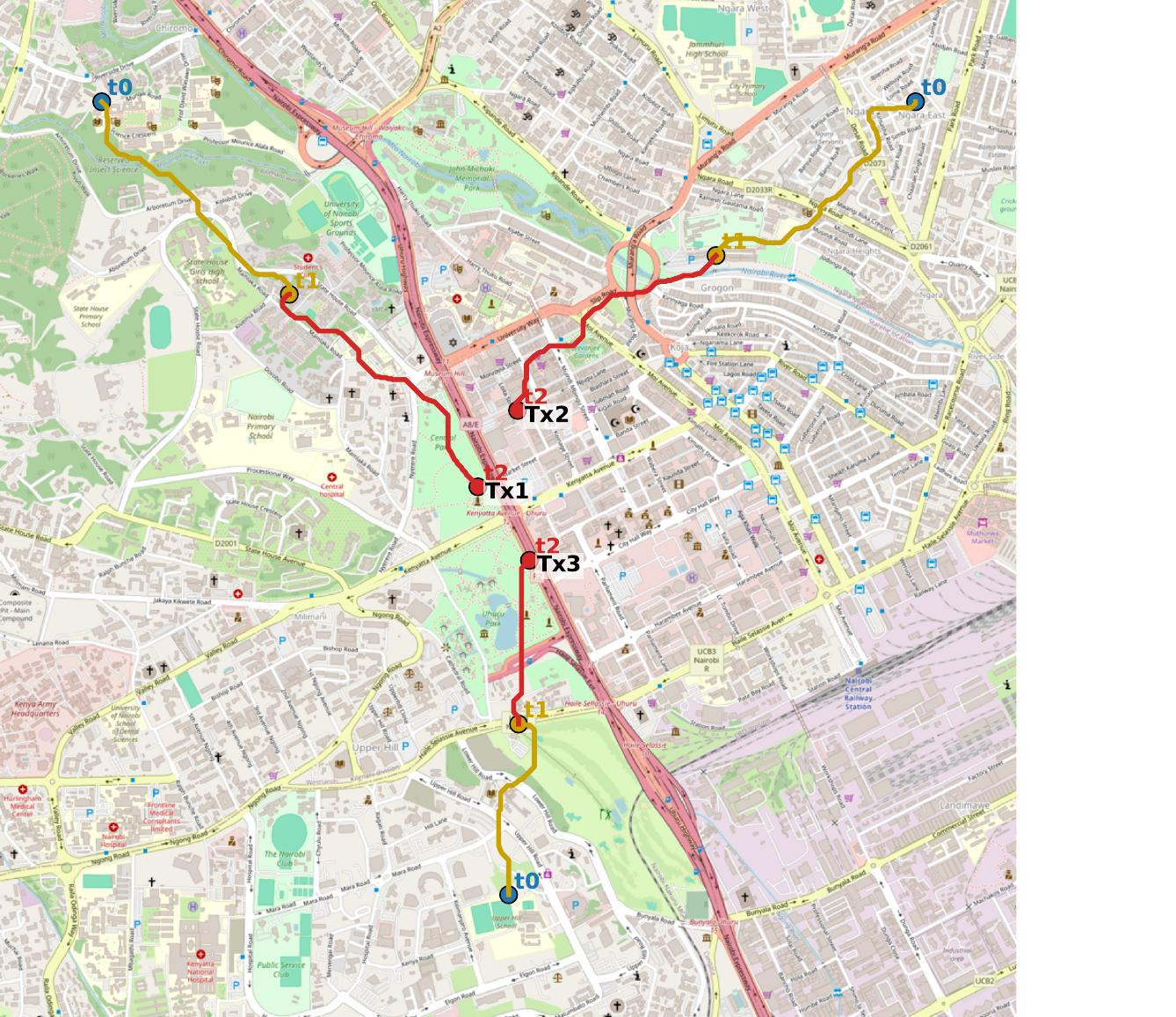} &
\includegraphics[width=0.235\columnwidth]{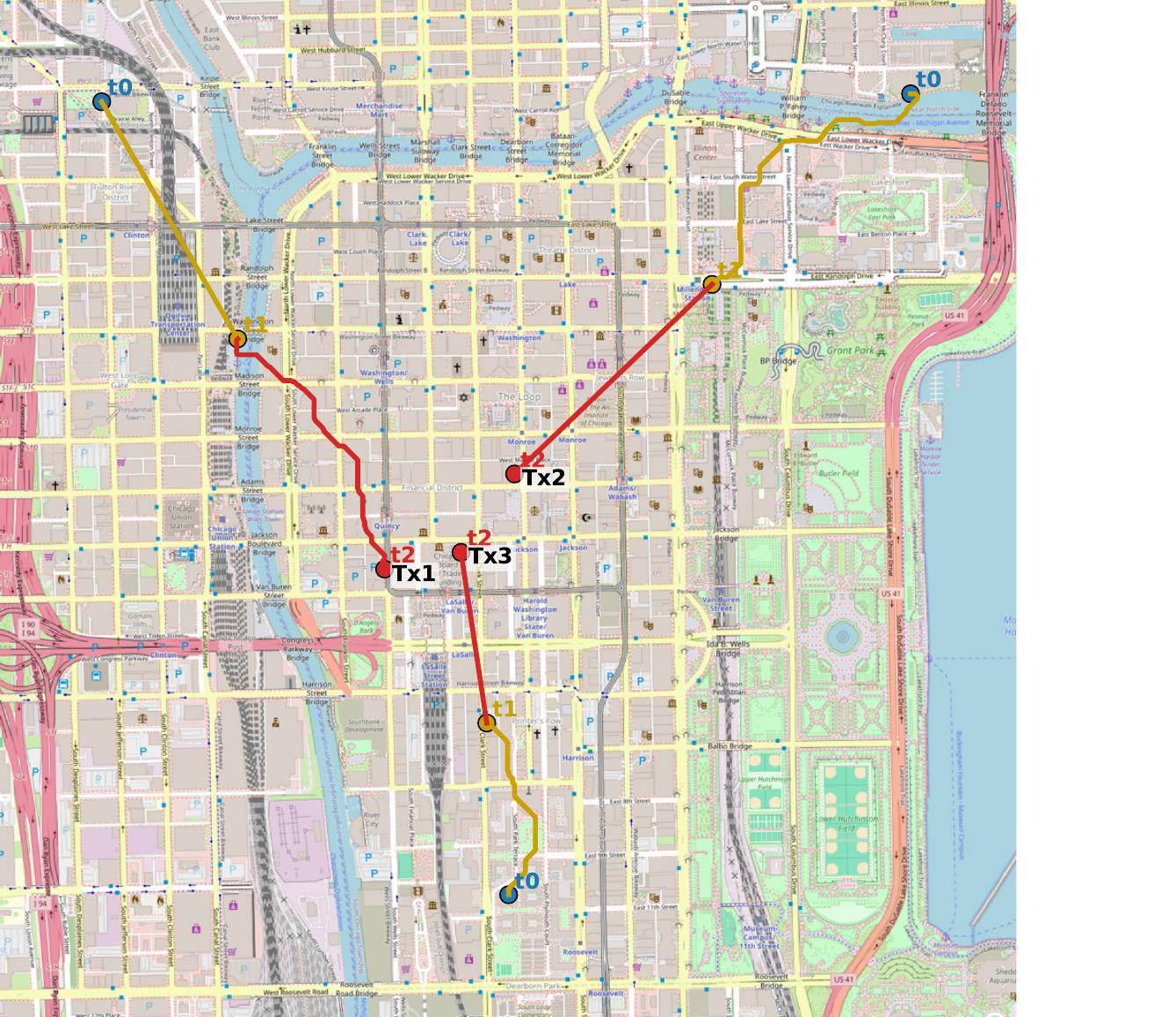} \\

\multicolumn{4}{c}{\scriptsize \textbf{(c)} TX movement trajectory and corresponding sampling locations.}
\end{tabular}

\caption{Examples of the height-aware data-generation pipeline for Shanghai, Berlin, Nairobi, and Chicago. Panels (a)--(c) show the OpenStreetMap, estimated height maps, and road-constrained TX movement layouts, respectively.}
\label{fig:dataset_examples}
\end{figure}

In each urban region, the number of transmitters (TX) considered in prior studies varies depending on the target application and simulation complexity. Specifically, RadioUNet, 3D-RadioDiff, R2Net and relevant works adopt a single transmitter~\cite{9354041,10963917,11505057}, whereas EPD/EPC inpainting and RadioGAT employ three transmitters~\cite{10424680,10682510}, and some cellular-network simulations further consider four transmitters~\cite{10127968}. Therefore, three simultaneously active transmitters are adopted for simulating the proposed dataset as a practical tradeoff between realism and computational cost. Moreover, the temporal evolution is introduced through the mobility of TXs. Specifically, the heights of the three TXs are fixed at $30~\mathrm{m}$ for all cities, while each TX follows a unique planar trajectory aligned with road constraints. To construct a sequence of IRMs, at each time step $t_0$, $t_1$, and $t_2$, the three simultaneously active TXs form one TX state, and each TX state generates one corresponding IRM, as shown in Fig.~\ref{fig:dataset_examples}(c).

%%%介绍图片展示结果
Due to space limitations, we present four representative cases, i.e., Shanghai, Berlin, Nairobi, and Chicago, because they represent distinct morphology types with different building densities and geographical contexts. The original OpenStreetMap and corresponding height map are shown in Fig.~\ref{fig:dataset_examples}(a) and Fig.~\ref{fig:dataset_examples}(b), respectively. Moreover, an outside-to-center trajectory pattern is shown in Fig.~\ref{fig:dataset_examples}(c). Detailed visualizations of 50 cities and their corresponding features are provided in \hl{Appendix A}.

\subsection{Ground-truth IRM Generation}
\label{subsec:dataset_generation}

Based on the constructed height-aware physical maps and time-evolving TX trajectories, 3D spatiotemporal IRMs are generated and treated as the ground-truth dataset for the following network training and evaluation. Specifically, inspired by the BRAT-LAB dataset \cite{bratlab}, we develop a customized simulation engine to produce realistic ground-truth IRMs under diverse urban morphologies, transmitter locations, and sensing heights, using the commercial electromagnetic software WinProp \cite{altairwinprop}.

\begin{figure}[!t]
\centering
\setlength{\tabcolsep}{0pt}
\renewcommand{\arraystretch}{0.60}
\scriptsize
\begin{tabular}{@{}cccc@{}}

\includegraphics[width=0.235\columnwidth]{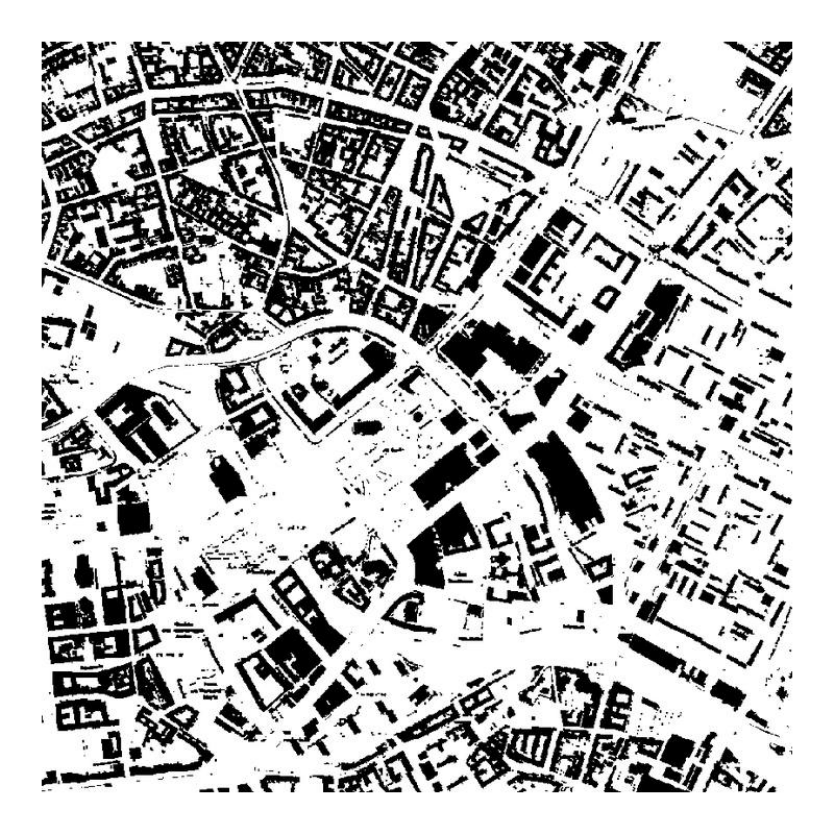} &
\includegraphics[width=0.235\columnwidth]{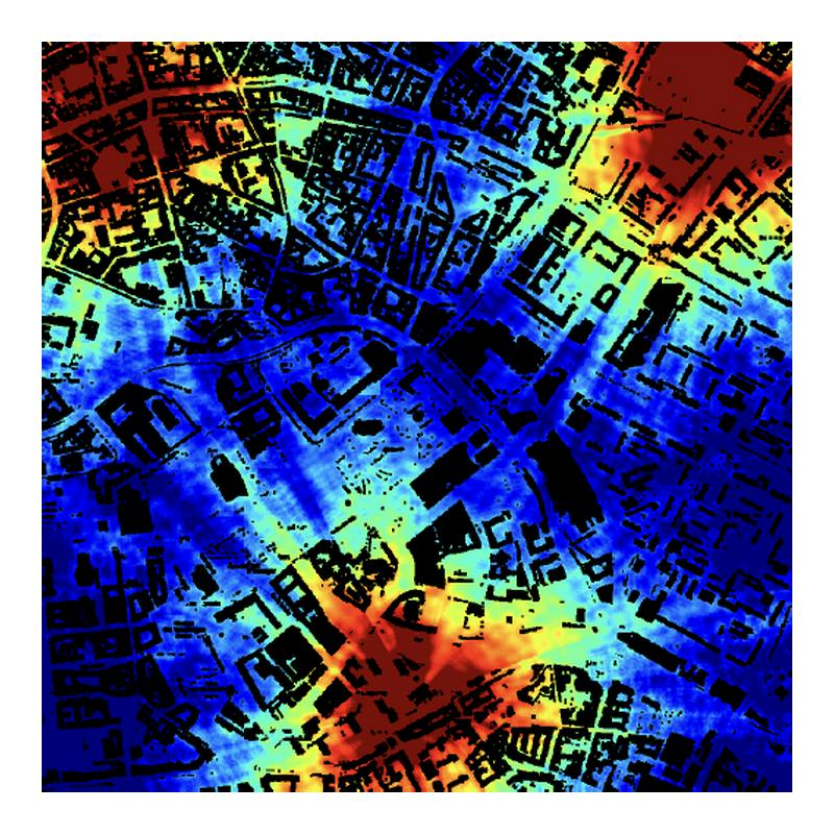} &
\includegraphics[width=0.235\columnwidth]{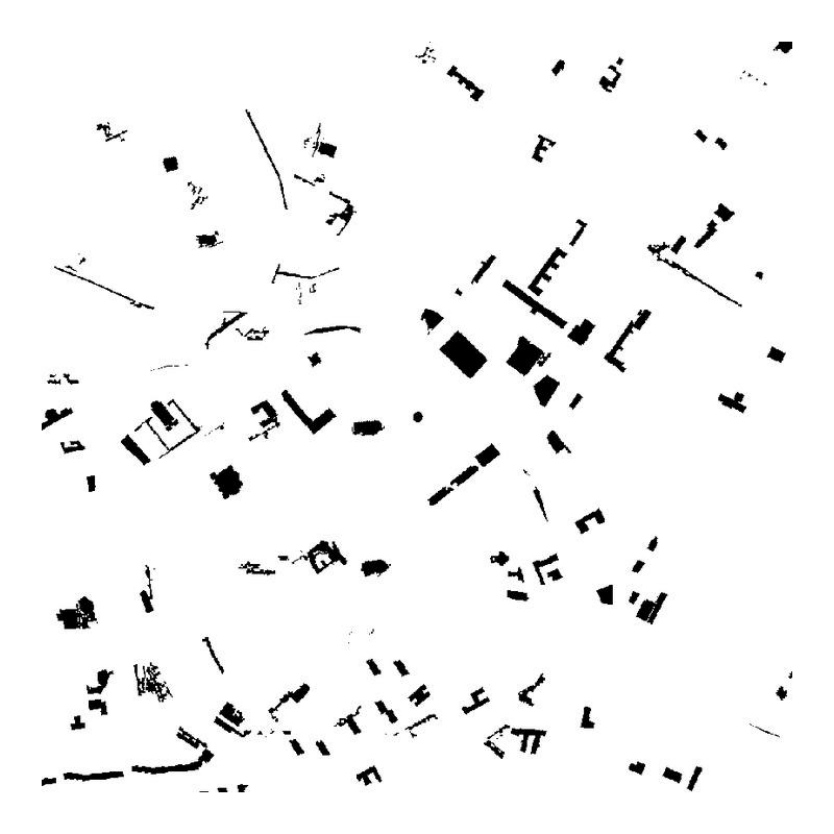} &
\includegraphics[width=0.235\columnwidth]{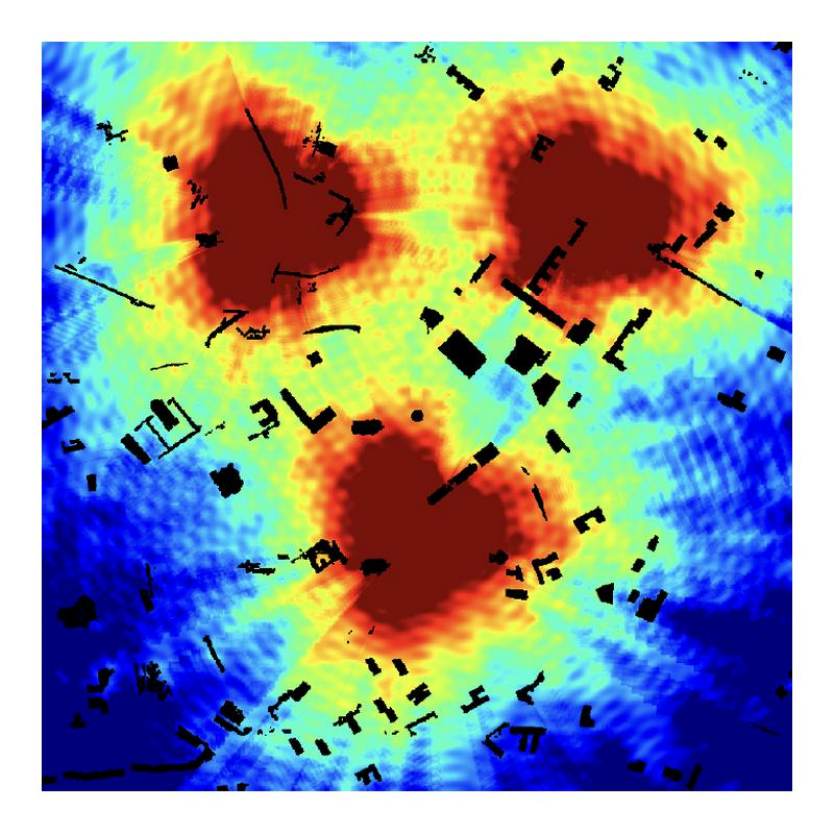} \\

(a) Obstacle mask &
(b) Ground-truth IRM &
(c) Obstacle mask &
(d) Ground-truth IRM

\end{tabular}

\caption{Simulated obstacle masks and ground-truth IRMs for Berlin. Panels (a) and (b) depict peripheral transmitters at $t_0$ with a receiver height of $0$ m, whereas panels (c) and (d) depict central transmitters at $t_1$ with a receiver height of $30$ m.}
\label{fig:berlin_groundtruth_examples}
\end{figure}

In each simulated ground-truth IRM, from the spatial perspective, the horizontal radio field spans the entire urban region, while the vertical dimension is discretized into three representative layers, i.e., $z\in \{0,15,30\}$~m, which correspond to ground-level, medium-altitude, and elevated spectrum features, respectively. Meanwhile, the temporal evolution is induced by
transmitter mobility, where different transmitter locations generate
a sequence of IRMs under the same urban morphology. In other words, changing the TX state from $t_0$ to $t_2$ results in a time-evolving map $\mathbf{S}^{(t)}$ with static obstacle masks $\mathbf M_b$ and $\mathbf M_g$, while the associated sensing masks $\mathbf M_u$ and $\mathbf M_m$ are continuously updated. Consequently, each IRM simultaneously captures the spatial obstacle variation, height-dependent propagation, and temporal evolution of the underlying radio environment. 
The detailed parameters of large-scale attenuation, small-scale fading, and antenna patterns are provided in \hl{Appendix B}. For demonstration, two representative spatiotemporal IRMs from Berlin Mitte are selected, as shown in Fig.~\ref{fig:berlin_groundtruth_examples}. The effective obstacle masks and corresponding ground-truth IRMs reveal that both transmitter mobility and receiver height significantly affect radio propagation patterns.

\subsection{Training Configuration}
\label{subsec:training_configuration}

The complete ground-truth dataset consists of 50 urban regions, each containing 10 independent TX moving trajectories. Moreover, in each trajectory, we select three temporal locations to generate IRMs at three sensing heights. Consequently, the complete dataset contains $50 \times 10 \times 3 \times 3 = 4500$ ground-truth IRMs. To avoid information leakage caused by geographically similar samples, the dataset is partitioned at the city level rather than the individual-map level. Specifically, to evaluate the proposed closed-loop multi-stage IRM construction framework, 40 cities (3600 IRMs) are used for training, 5 cities (450 IRMs) for validation, and the remaining 5 cities (450 IRMs) for testing, respectively.

For the closed-loop sensing, the initial stage is guided by physical priors and obstacle-avoidance constraints, where sensing locations are uniformly distributed across the accessible region. Subsequently, the sensing steps are iteratively selected, guided by the proposed attention-based network. Specifically, the network employs a single scaled dot-product attention layer followed by a linear scoring head, whose embedding dimension is set to 64, and each branch consists of a two-layer MLP with ReLU activation. Moreover, although the radio-propagation simulation and ground-truth IRMs are constructed on a $500\times500$ spatial grid, directly using this fine-grained discretization for training the attention-based network would introduce excessive sensing candidates and cause convergence difficulties. Therefore, we adopt a patch-level sensing representation, where each $10\times10$ spatial region is treated as one sensing patch. Any accessible grid within the patch is marked as 1 by sensing masks, while the averaged IRM value over this patch is recorded as the unified measurement value in $\mathbf \Pi$. In this way, the whole map contains at most $(500/10)^2=2500$ patches as sensing candidates\footnote{The actual number of sensing candidates is smaller, since some patches are infeasible to reach due to obstacle constraints.}. The initial number of sensing patches based on environmental priors is set to 40 ($1.6\%$ sensing sparsity). Subsequently, for each iterative stage, the sensing step progressively acquires 20 additional patches at each iteration until reaching 120 ($4.8\%$ sensing sparsity). In other words, the experimental sensing sparsity levels are set as $\mathcal R_{\rho}=\{1.6\%, 2.4\%, 3.2\%, 4.0\%, 4.8\%\}$.

For the closed-loop inference, for the initial stage, the sketch map $\breve{\mathbf S}^{(t_0)}$ is constructed using Ordinary Kriging with an exponential variogram model, while the DiT backbone of the teacher model consists of $\ell^\star=12$ transformer blocks, the diffusion process adopts $\tau^\star=100$ denoising steps, the hidden feature dimension is set to $d_h=512$, and training process is optimized using AdamW with a learning rate of $10^{-4}$ and cosine decay. Then for iterative stages, the student inference model is distilled, employing only 4 DiT blocks and 20 denoising steps with 256 hidden dimensions. The fewer network parameters are also optimized using AdamW with a learning rate of $5\times10^{-5}$.

All data processing, numerical simulation, and network training are conducted on a workstation with an Intel Core i7-11800H processor, 16 GB RAM, and an NVIDIA GeForce RTX 4060 with 8 GB GDDR6 memory.

\section{Evaluations}
\label{sec:performance_evaluation}

\subsection{Evaluation Metrics}

\label{subsec:evaluation_metrics}

To comprehensively evaluate different IRM construction methods, we consider four complementary metrics: inference accuracy, robustness across heterogeneous conditions, sensing-efficiency gain, and construction time. 

\subsubsection{Inference Accuracy}
When $\rho \in \mathcal{R}_\rho$ denotes the sensing sparsity, $\mathbf{O}^{(\rho)}$ is defined as the corresponding observation mask. Accordingly, the inference error between ground-truth and constructed IRMs is computed by:

\begin{equation}
\mathcal{M}_{\mathrm{acc}}(c,\xi,z,\rho)
=
\frac{
\left\|
\left(\mathbf{1}-\mathbf{O}^{(\rho)}\right)
\odot
\left(\hat{\mathbf{S}}_{c,\xi,z}^{(\rho)} - \mathbf{S}_{c,\xi,z}\right)
\right\|_{F}
}{
\left\|
\left(\mathbf{1}-\mathbf{O}^{(\rho)}\right)
\odot
\mathbf{S}_{c,\xi,z}
\right\|_{F} + \epsilon
}.
\label{ele:acc}
\end{equation}
where $c$, $\xi$, and $z$ represent the city index, transmitter location and receiver height, respectively. $\epsilon$ is a small constant used to prevent numerical instability, while $\|\cdot\|_{F}$ denotes the Frobenius norm. A lower value indicates better inference accuracy.

\subsubsection{Sensing-Efficiency Gain}

To quantify how effectively the inference accuracy improves with increased sensing budget, we define a sensing-efficiency gain metric that measures the marginal benefit of additional sensing samples. Unlike single-step improvement, the accumulated gain over multiple sparsity intervals is used for evaluation. Specifically, we define $\{\rho_n\}_{n=1}^{N}$ as the ordered elements of sensing
sparsity set $\mathcal{R}_\rho$, where $N = \|\mathcal{R}_\rho\|_0$. Conditioned on a given city $c \in \mathcal{C}$, the sensing-efficiency gain metric is defined as:

\begin{equation}
\mathcal{M}_{\mathrm{eff}}(c,\xi,z)
=
\frac{1}{N-1}
\sum_{n=2}^{N}
\frac{
\Delta \mathcal{M}_{\mathrm{acc}}(c,\xi,z,\rho_n)
}{
\Delta\rho + \epsilon
}.
\label{ele:gain}
\end{equation}
$\Delta \mathcal{M}_{\mathrm{acc}}(c,\xi,z,\rho_n)
=
\mathcal{M}_{\mathrm{acc}}(c,\xi,z,\rho_{n-1})
-
\mathcal{M}_{\mathrm{acc}}(c,\xi,z,\rho_{n})$ denotes the finite improvement of inference accuracy between two different sensing sparsity levels. A higher value indicates a stronger capability to exploit
additional measurements in the closed-loop IRM construction framework.

\subsubsection{Robustness across Heterogeneous Conditions}

To evaluate robustness, we measure the stability of inference accuracy under heterogeneous deployment environments. Specifically, we fix the transmitter location, receiver height, and sensing sparsity, and evaluate the aggregated variance of $\mathcal{M}_{\mathrm{acc}}$ across different cities. When the city index $c \in \{1,2,\ldots,\mathcal{C}\}$, the robustness metric is computed as:
\begin{equation}
\mathcal{M}_{\mathrm{rob}}(\xi,z,\rho)
=
\frac{1}{|\mathcal{C}|}
\sum_{c=1}^{|\mathcal{C}|}
\left(
\mathcal{M}_{\mathrm{acc}}(c,\xi,z,\rho)
-
\bar{\mathcal{M}}_{\mathrm{acc}}(\xi,z,\rho)
\right)^2.
\label{ele:rob}
\end{equation}
where
$
\bar{\mathcal{M}}_{\mathrm{acc}}(\xi,z,\rho)
=
\frac{1}{|\mathcal{C}|}
\sum_{c=1}^{|\mathcal{C}|}
\mathcal{M}_{\mathrm{acc}}(c,\xi,z,\rho)
$
denotes the mean IRM inference error across all cities. A lower value of $\mathcal{M}_{\mathrm{rob}}$ indicates more stable performance under heterogeneous urban environments.

\subsubsection{Construction Time}

We evaluate computational efficiency using the average IRM construction time. Since most IRM inference models require additional computation as the sensing sparsity increases, the construction time is averaged over all sparsity levels in $\mathcal{R}_{\rho}$. Let $\varsigma_{m,n}$ denote the IRM construction time of the $m$-th test sample under the $n$-th sensing sparsity level. This metric is defined as

\begin{equation}
\mathcal{M}_{\mathrm{time}}
=
\frac{1}
{M_{\mathrm{test}} \cdot N}
\sum_{m=1}^{M_{\mathrm{test}}}
\sum_{n=1}^{N}
\varsigma_{m,n}.
\label{metric-time}
\end{equation}
where $M_{\mathrm{test}}$ denotes the number of test rounds. A lower value of $\mathcal{M}_{\mathrm{time}}$ indicates higher construction efficiency and better real-time applicability.

\subsection{Compared Baselines}
\label{subsec:compared_algorithms}

We compare the proposed framework with four representative baselines. Specifically, the classic statistically methods are Kriging and K-nearest neighbors (KNN), where Kriging exploits spatial spectrum correlation through variogram modeling, while KNN estimates spectrum values using spatial coordinates, transmitter distance, receiver height, and blockage-related features. The learning-based baselines include 3D-RadioDiff~\cite{10963917}, which contains an altitude-conditioned diffusion model, and RadioLAM~\cite{liu2025radiolam}, which incorporates physical priors into a large-model-inspired framework for fine-grained 3D IRM estimation. Moreover, these baselines inspire some modules of the proposed framework, thereby providing strong references for comparative evaluation.

\newcommand{\resultrownote}[4]{\multicolumn{9}{c}{\scriptsize #1, TX location #2, $z=#3~\mathrm{m}$, $\rho$=#4\%}\\[-0.10mm]}
\newcommand{\resultrownoteletter}[5]{\multicolumn{9}{c}{\scriptsize \textbf{(#1)}~#2, TX location #3, $z=#4~\mathrm{m}$, $\rho$=#5\%}\\[-0.10mm]}
\begin{figure*}[!htp]
\centering
\setlength{\tabcolsep}{0.20pt}
\renewcommand{\arraystretch}{0.50}
\begin{tabular}{@{}ccccccccc@{}}
\scriptsize TX location & \scriptsize Ground Truth & \scriptsize Sensing Measurements & \scriptsize Kriging & \scriptsize KNN & \scriptsize RadioDiff & \scriptsize  RadioLAM & \scriptsize Proposed & \scriptsize Uncertainty \\[-0.2mm]
\includegraphics[width=0.101\textwidth]{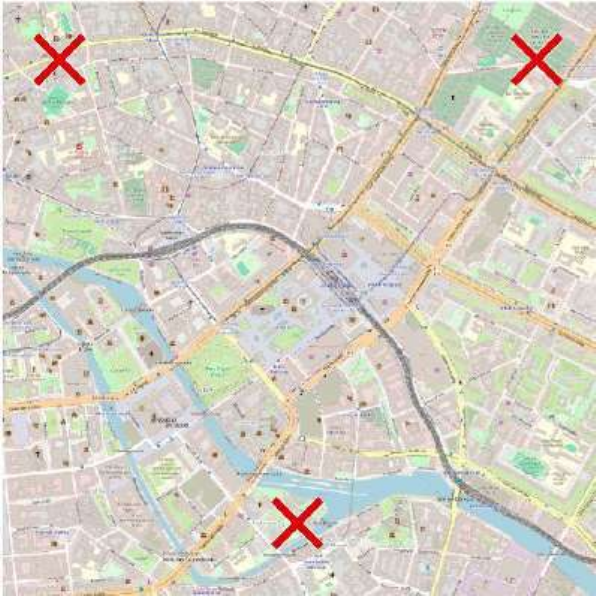} &
\includegraphics[width=0.101\textwidth]{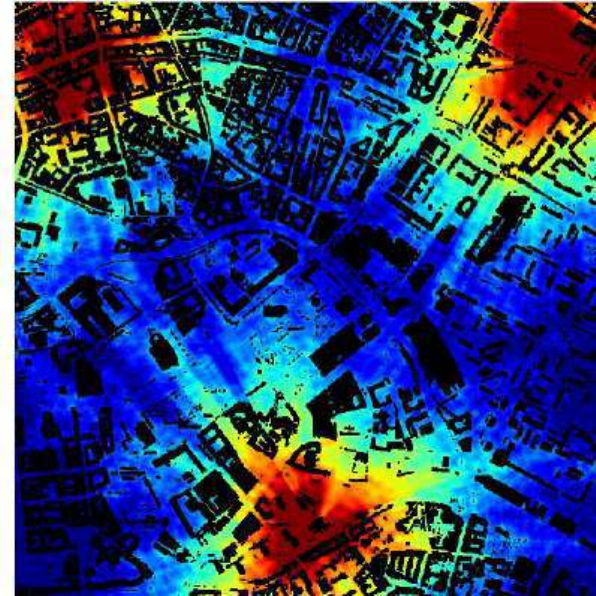} &
\includegraphics[width=0.101\textwidth]{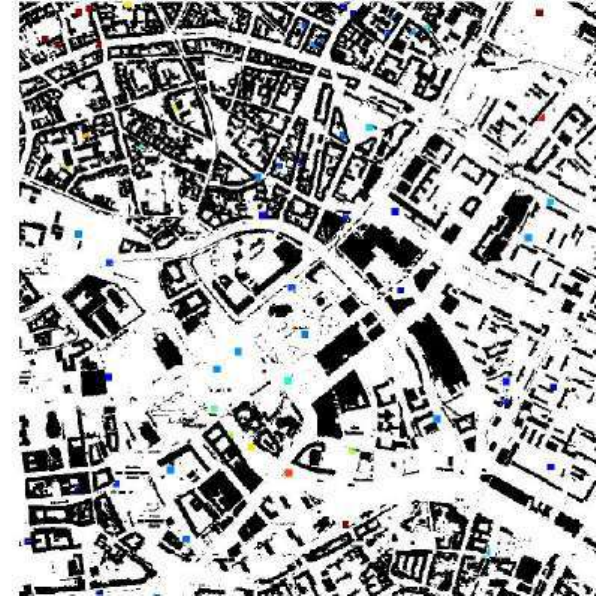} &
\includegraphics[width=0.101\textwidth]{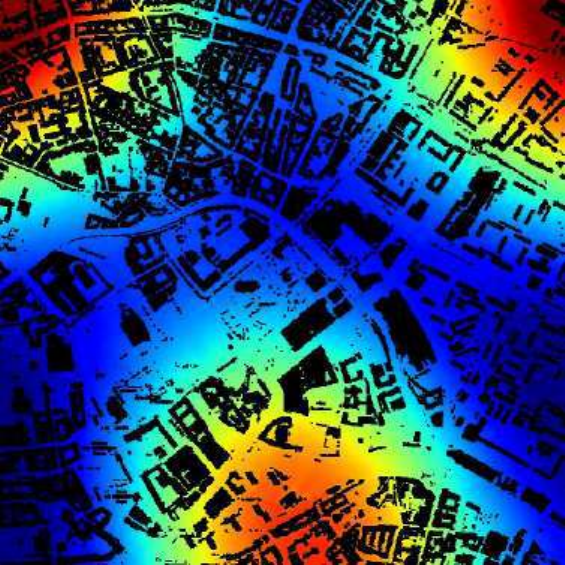} &
\includegraphics[width=0.101\textwidth]{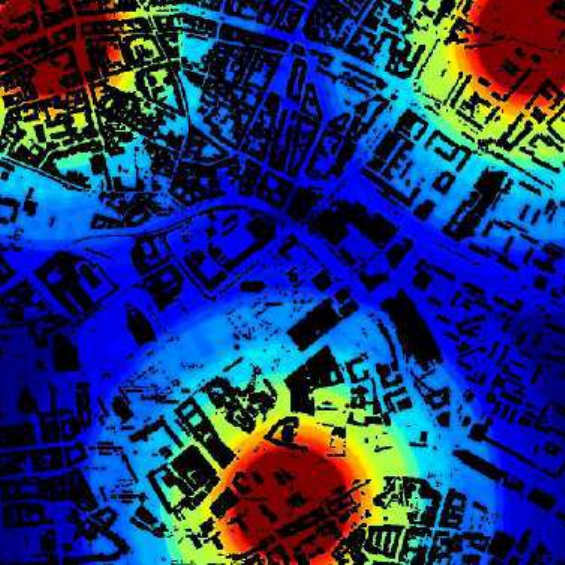} &
\includegraphics[width=0.101\textwidth]{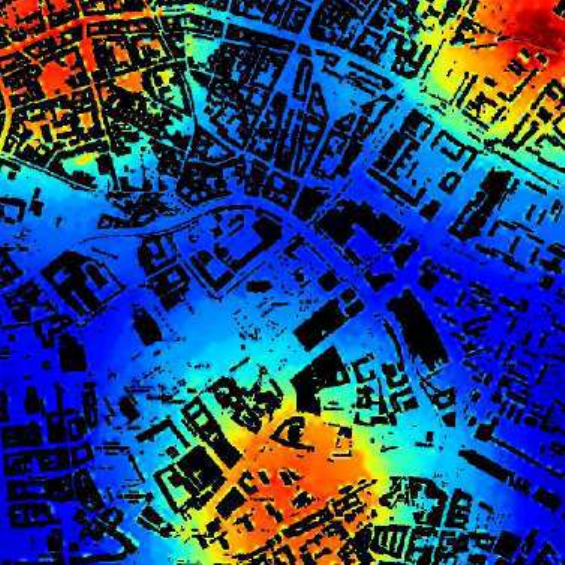} &
\includegraphics[width=0.101\textwidth]{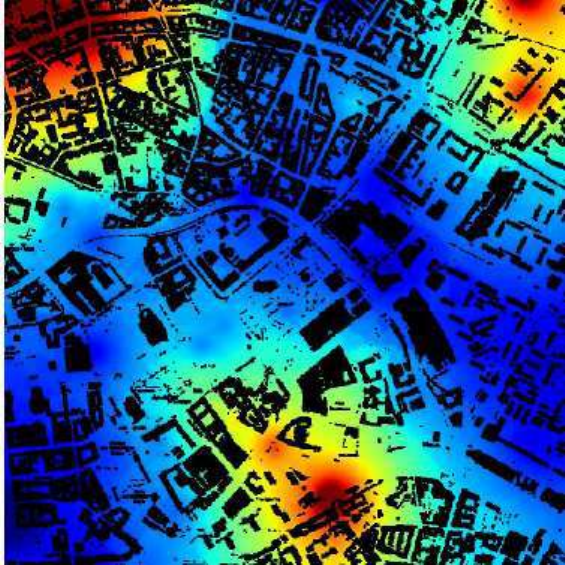} &
\includegraphics[width=0.101\textwidth]{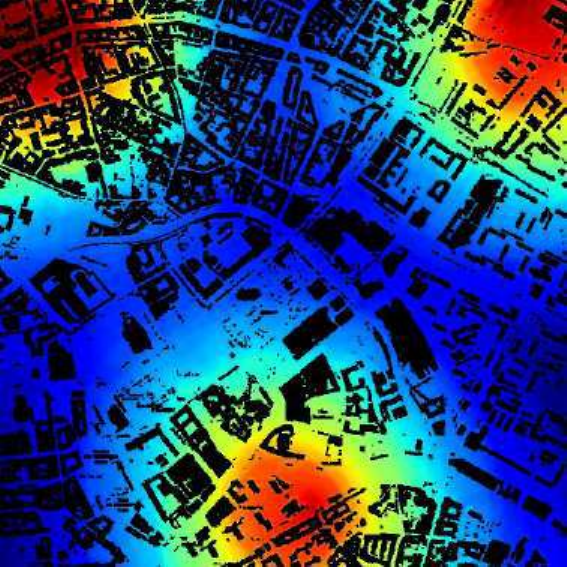} &
\includegraphics[width=0.101\textwidth]{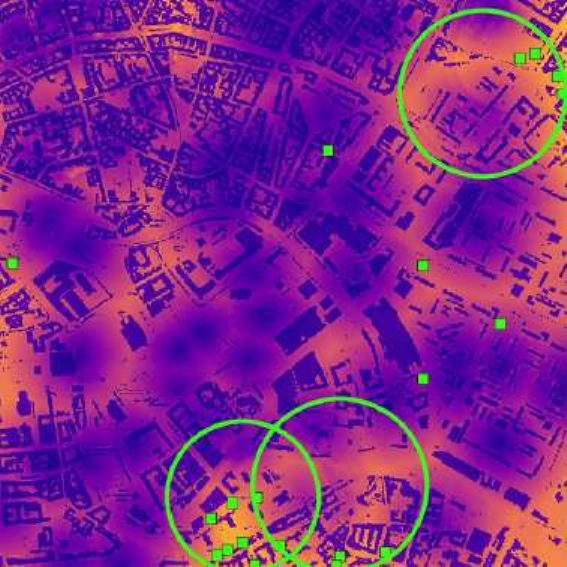} \\[0.10mm]
\resultrownoteletter{a}{Berlin Mitte}{1}{0}{2.4}
\includegraphics[width=0.101\textwidth]{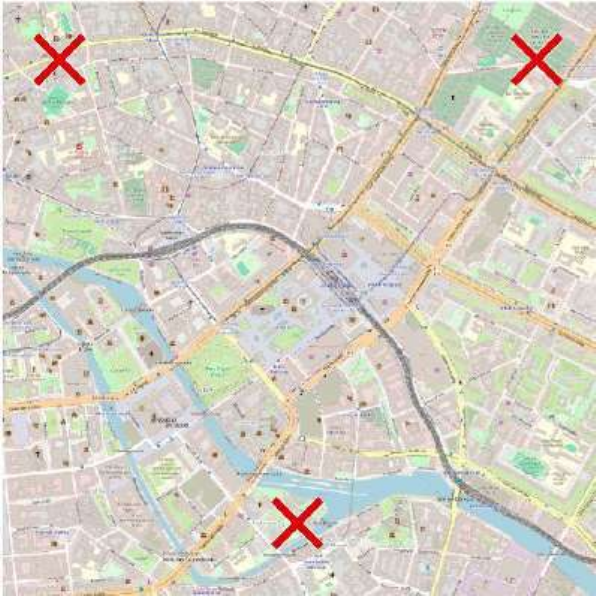} &
\includegraphics[width=0.101\textwidth]{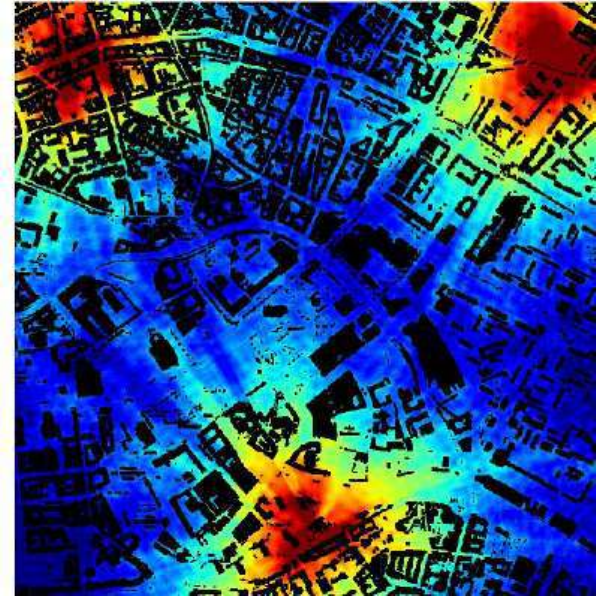} &
\includegraphics[width=0.101\textwidth]{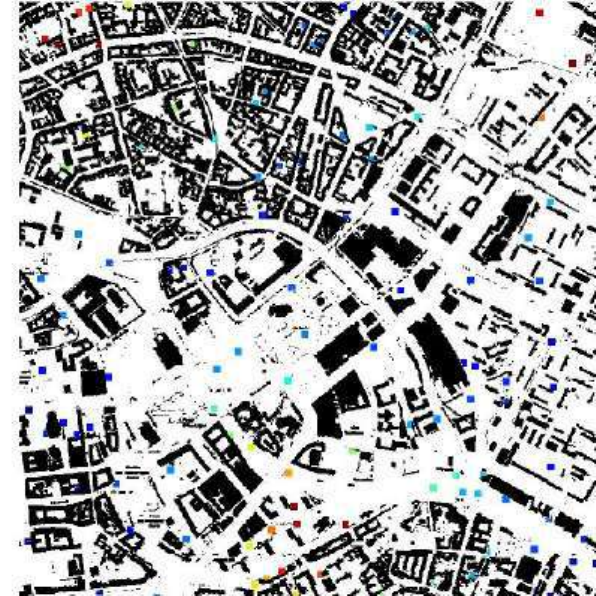} &
\includegraphics[width=0.101\textwidth]{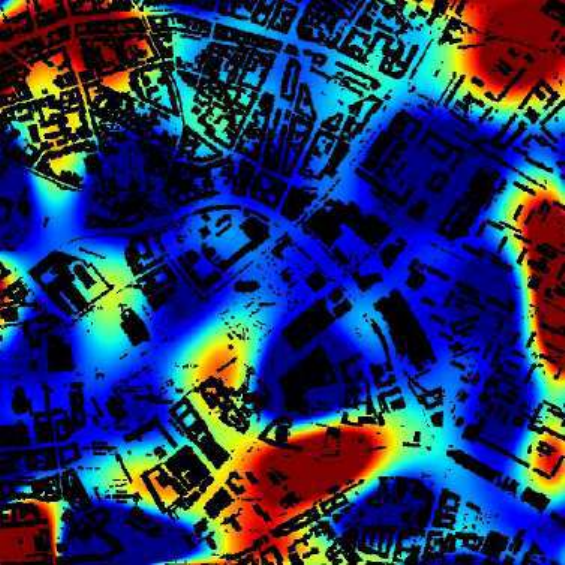} &
\includegraphics[width=0.101\textwidth]{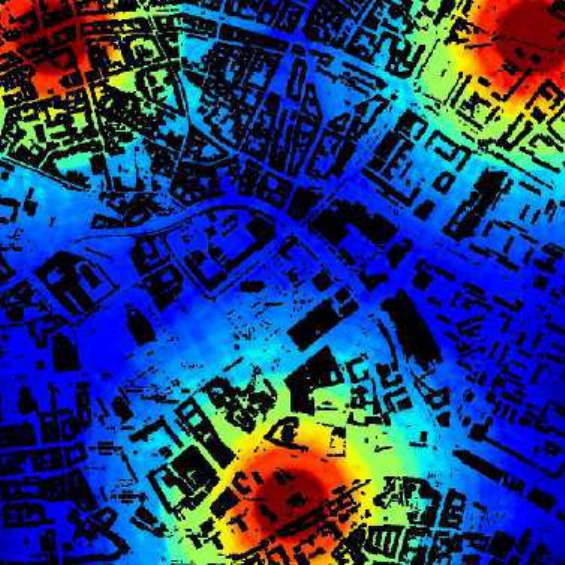} &
\includegraphics[width=0.101\textwidth]{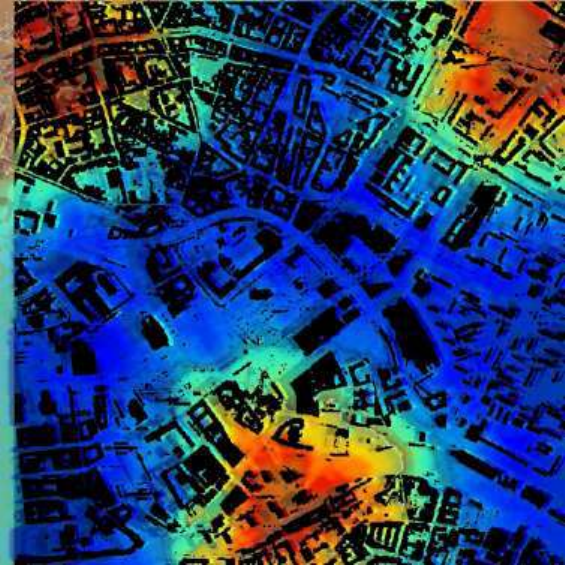} &
\includegraphics[width=0.101\textwidth]{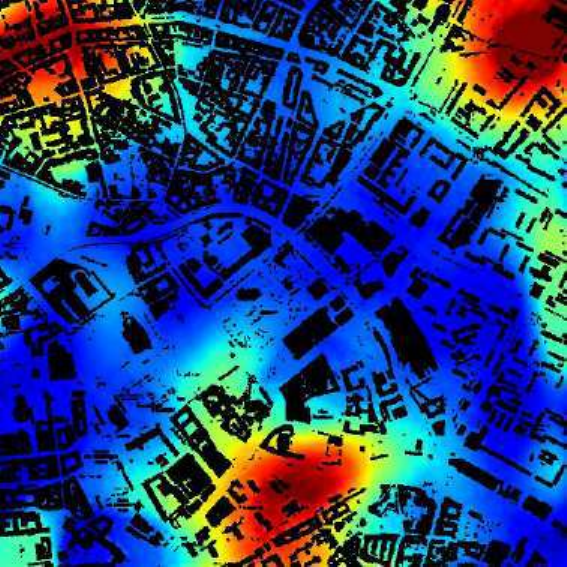} &
\includegraphics[width=0.101\textwidth]{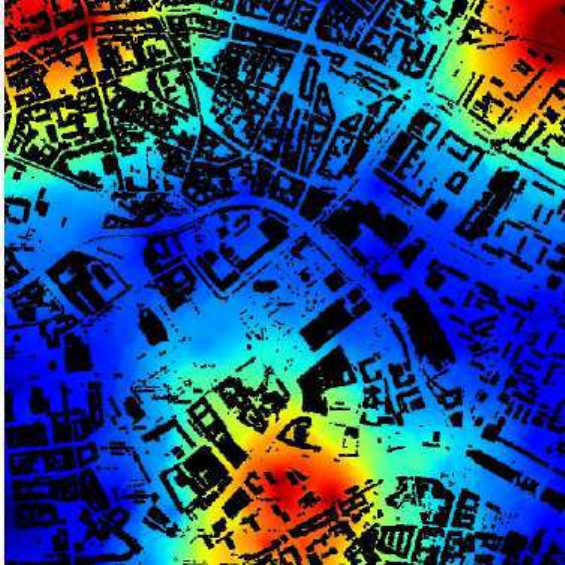} &
\includegraphics[width=0.101\textwidth]{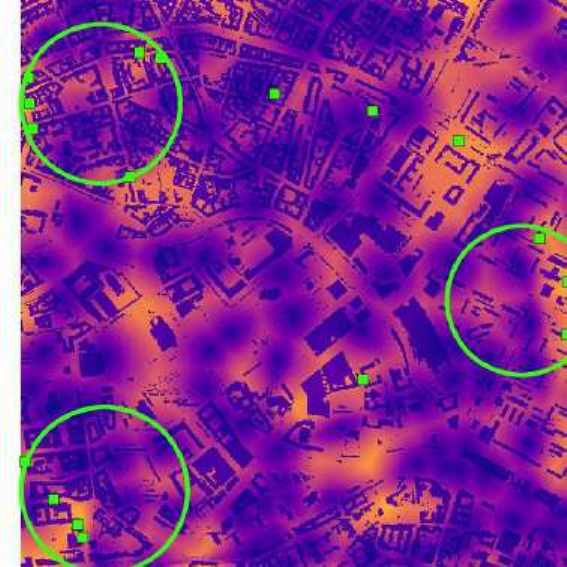} \\[0.10mm]
\resultrownoteletter{b}{Berlin Mitte}{1}{0}{4.8}
\includegraphics[width=0.101\textwidth]{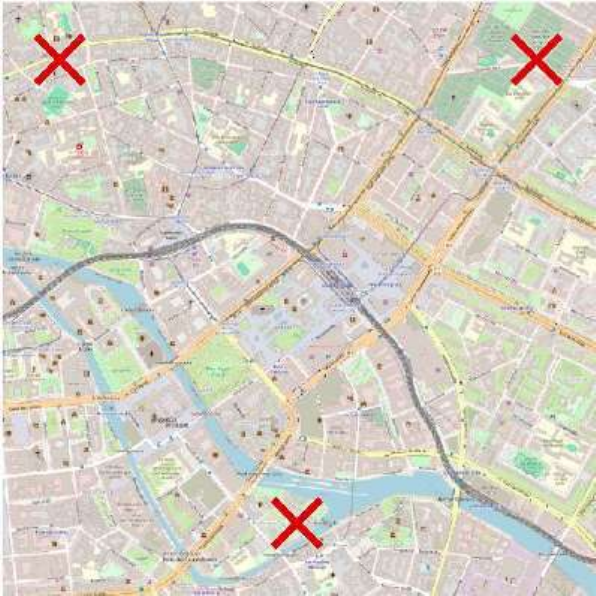} &
\includegraphics[width=0.101\textwidth]{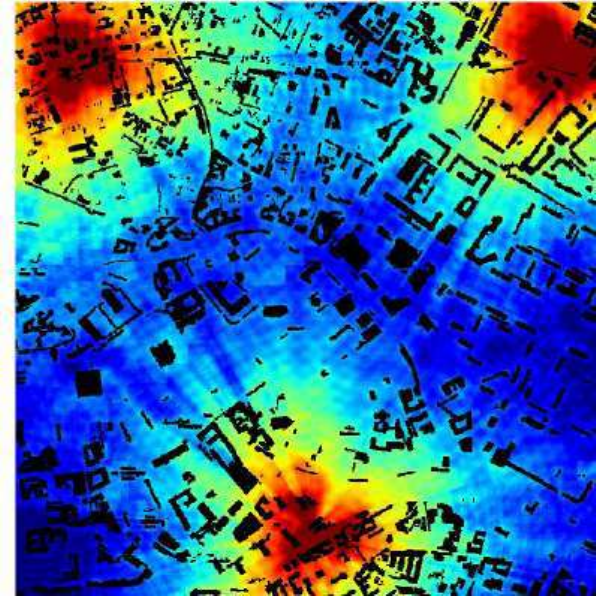} &
\includegraphics[width=0.101\textwidth]{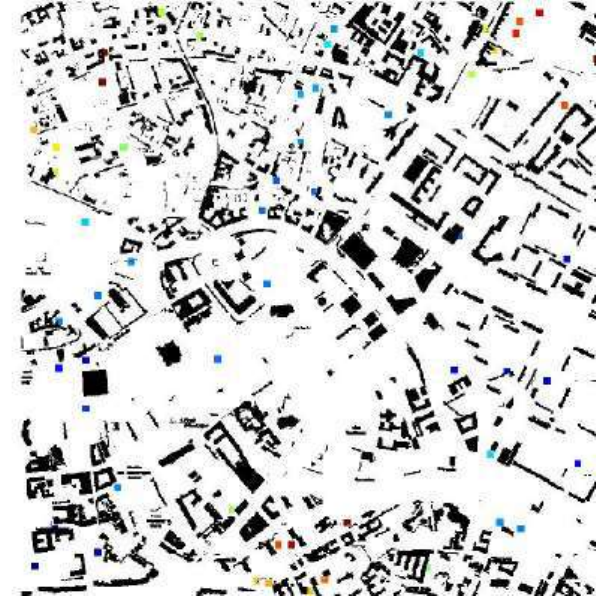} &
\includegraphics[width=0.101\textwidth]{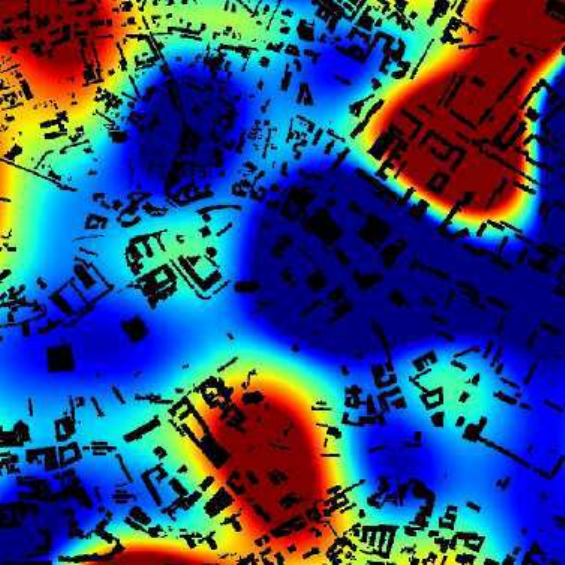} &
\includegraphics[width=0.101\textwidth]{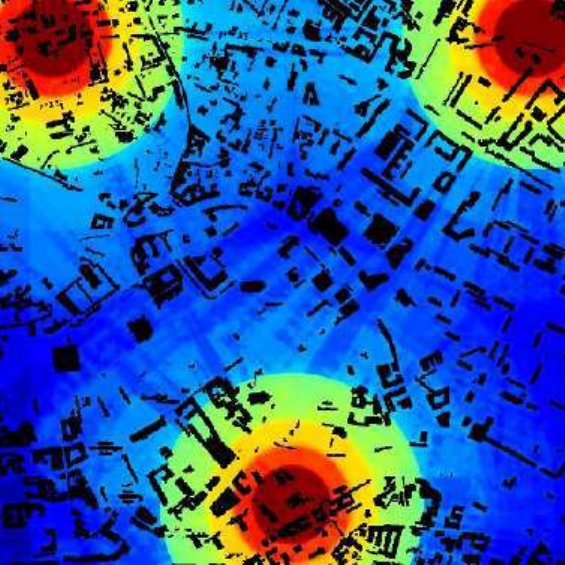} &
\includegraphics[width=0.101\textwidth]{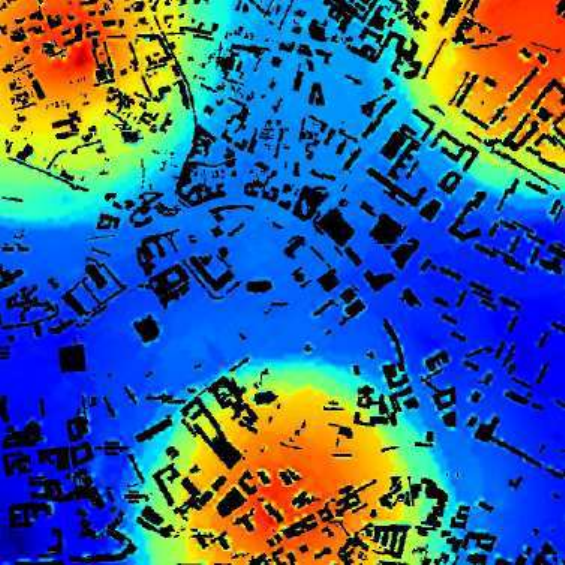} &
\includegraphics[width=0.101\textwidth]{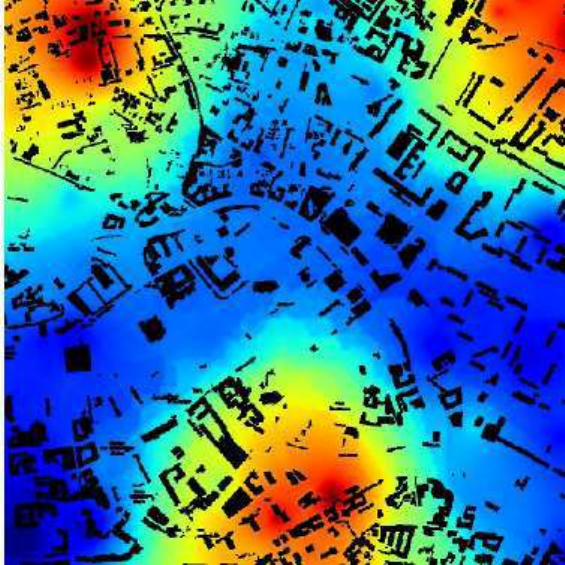} &
\includegraphics[width=0.101\textwidth]{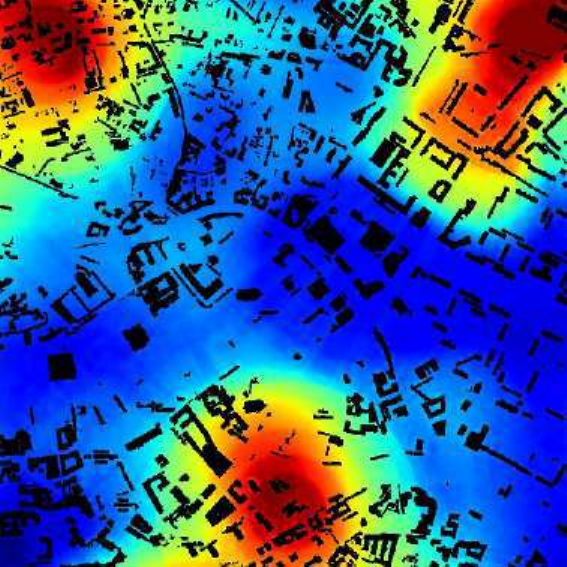} &
\includegraphics[width=0.101\textwidth]{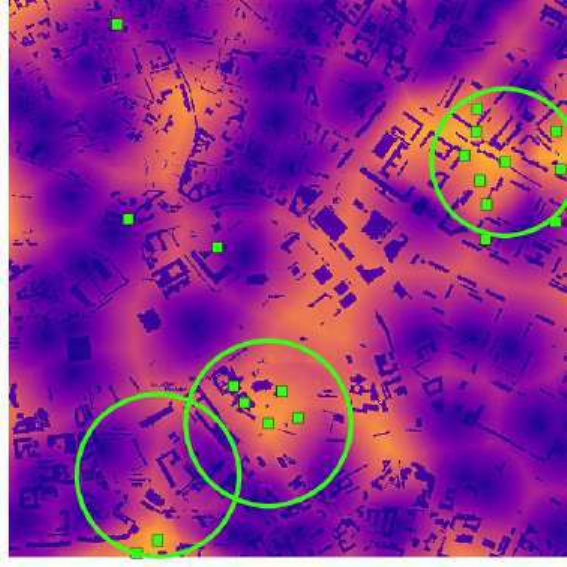} \\[0.10mm]
\resultrownoteletter{c}{Berlin Mitte}{1}{15}{2.4}
\includegraphics[width=0.101\textwidth]{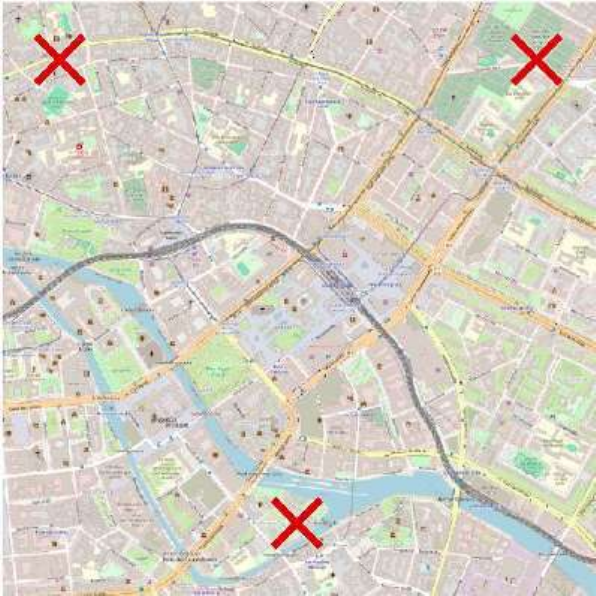} &
\includegraphics[width=0.101\textwidth]{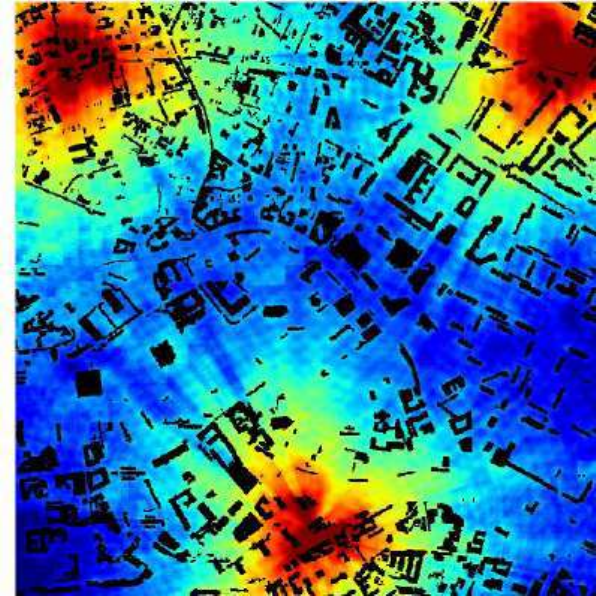} &
\includegraphics[width=0.101\textwidth]{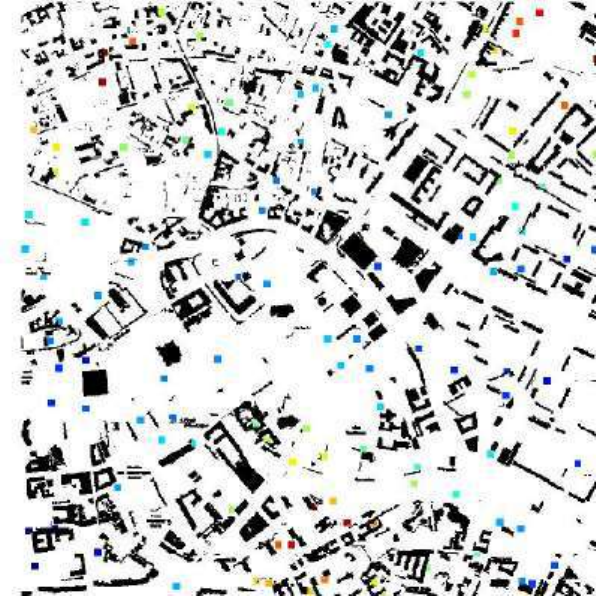} &
\includegraphics[width=0.101\textwidth]{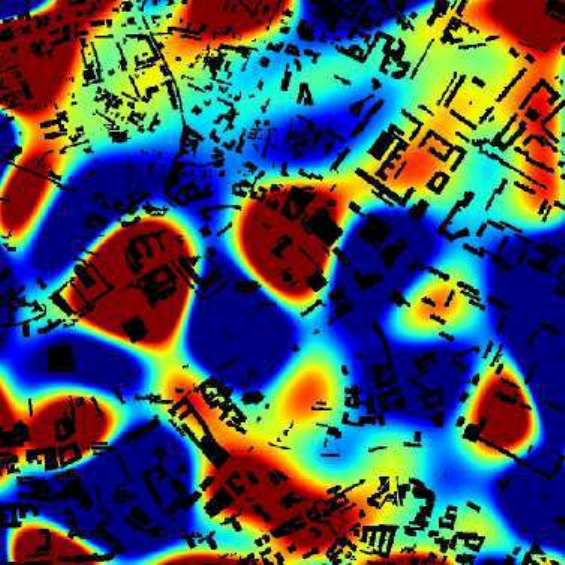} &
\includegraphics[width=0.101\textwidth]{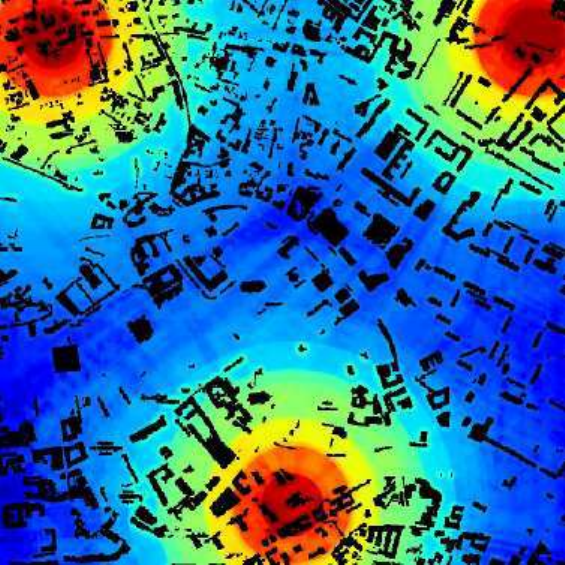} &
\includegraphics[width=0.101\textwidth]{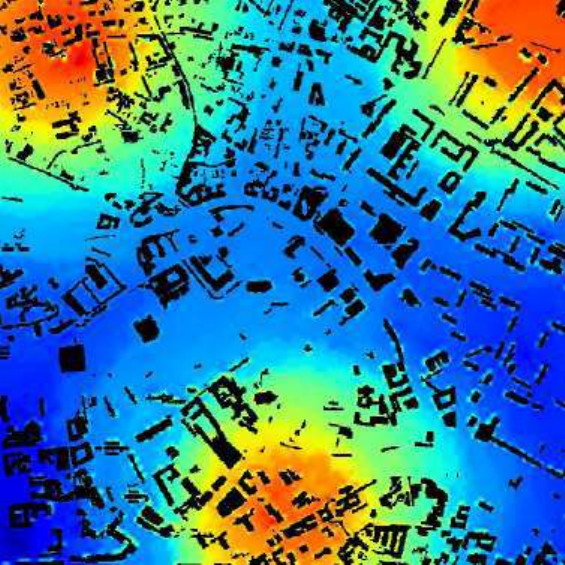} &
\includegraphics[width=0.101\textwidth]{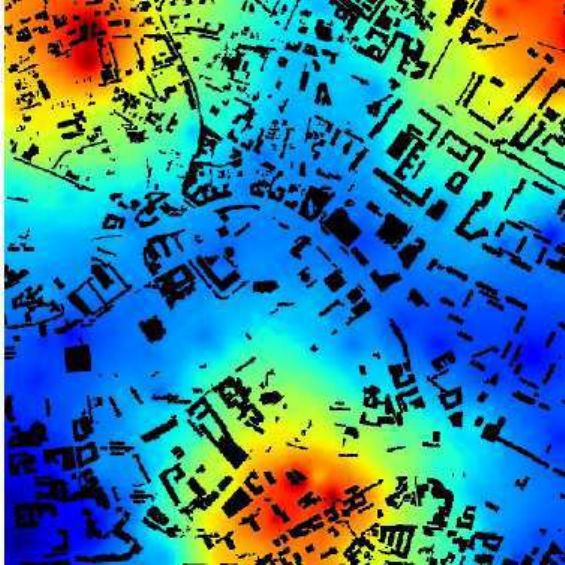} &
\includegraphics[width=0.101\textwidth]{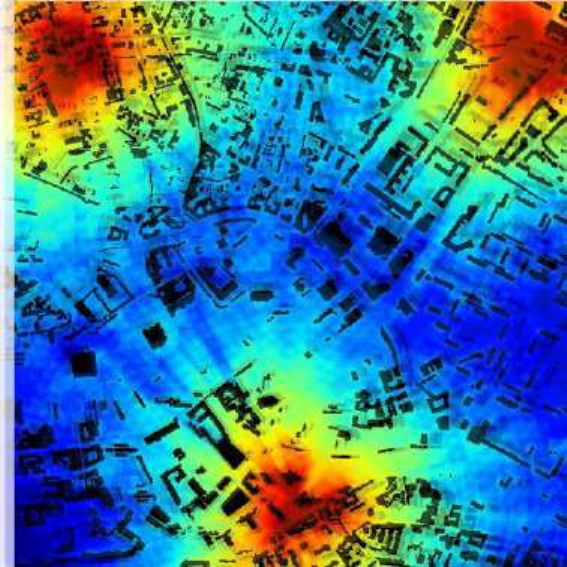} &
\includegraphics[width=0.101\textwidth]{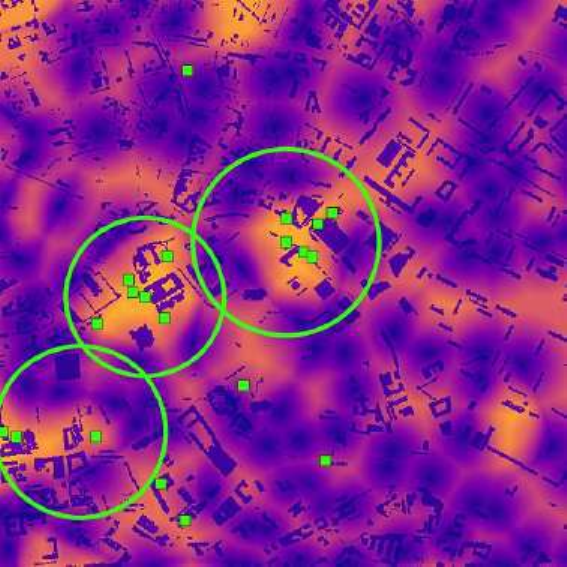} \\[0.10mm]
\resultrownoteletter{d}{Berlin Mitte}{1}{15}{4.8}
\includegraphics[width=0.101\textwidth]{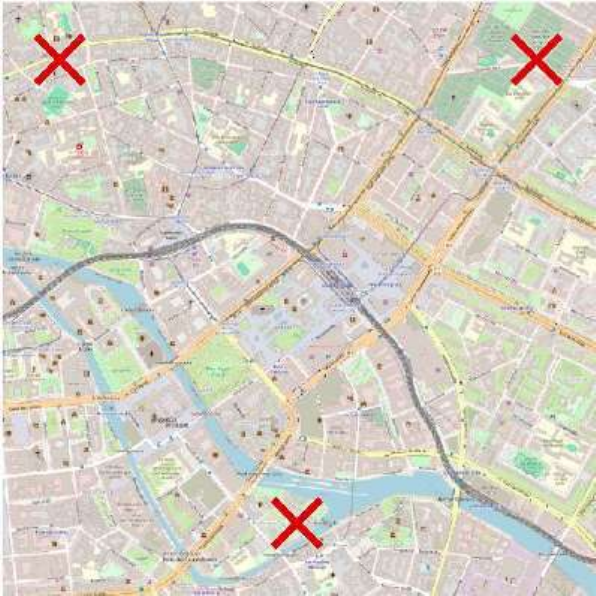} &
\includegraphics[width=0.101\textwidth]{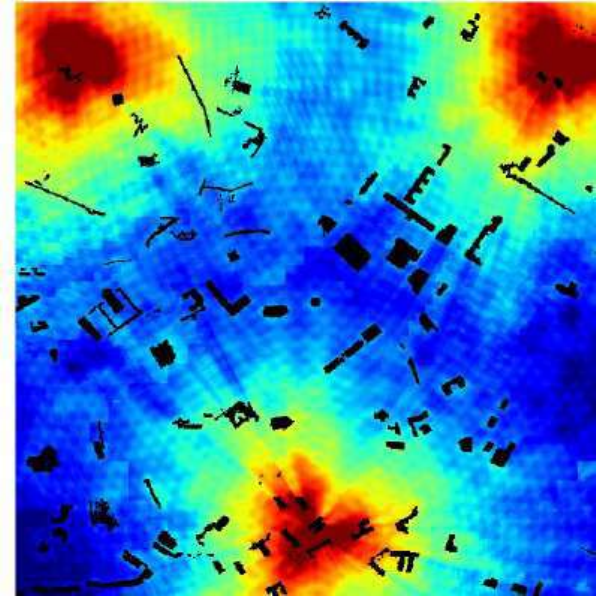} &
\includegraphics[width=0.101\textwidth]{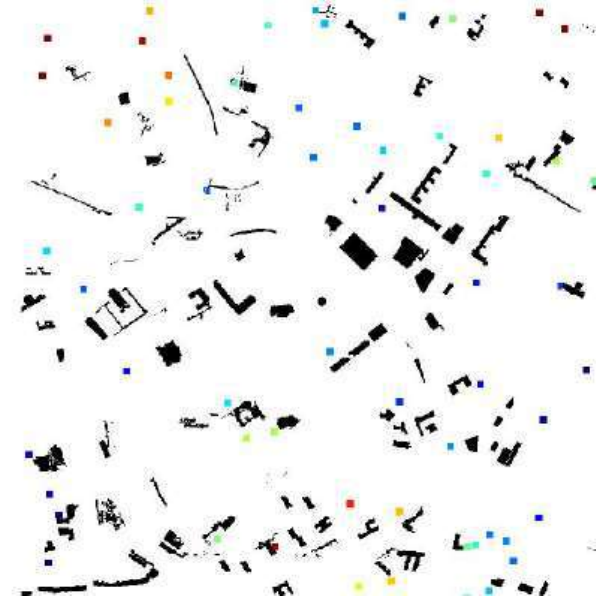} &
\includegraphics[width=0.101\textwidth]{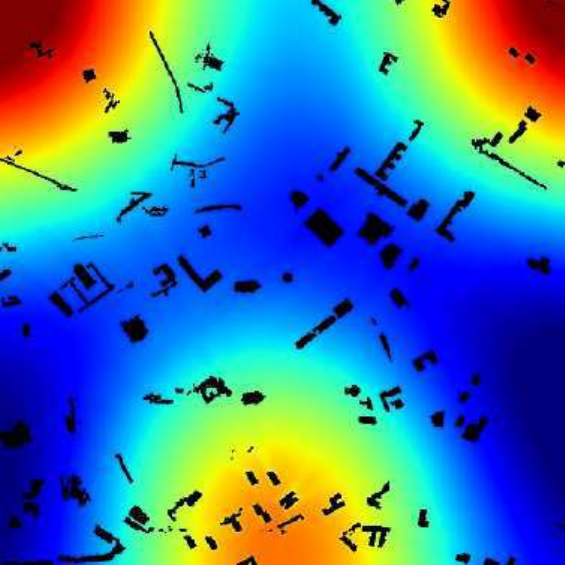} &
\includegraphics[width=0.101\textwidth]{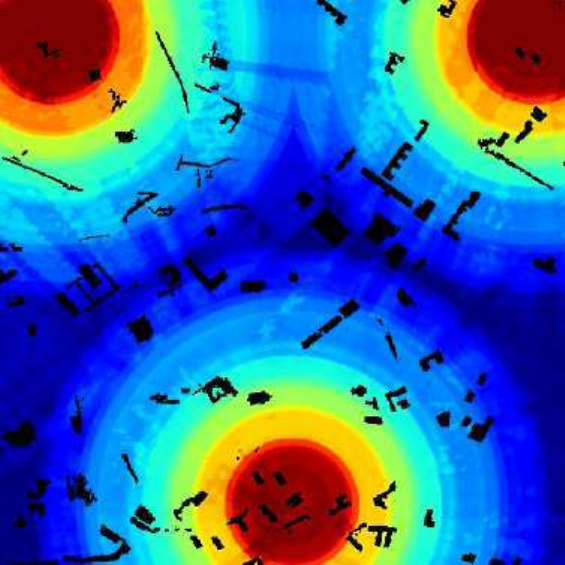} &
\includegraphics[width=0.101\textwidth]{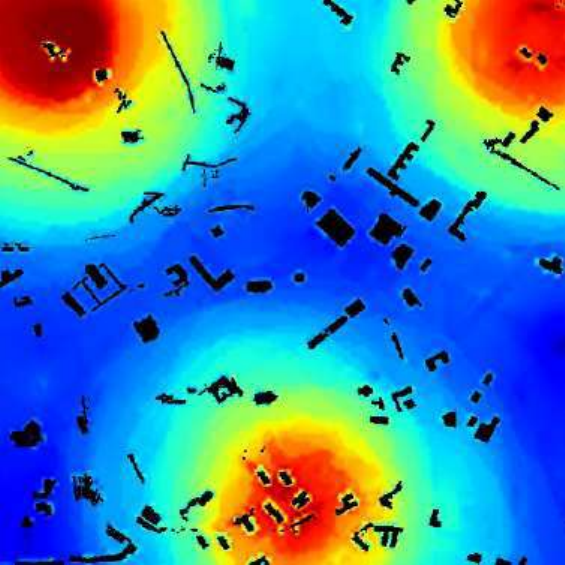} &
\includegraphics[width=0.101\textwidth]{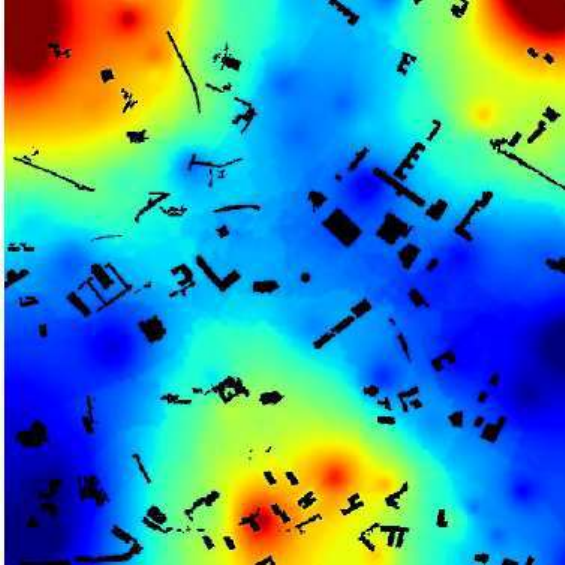} &
\includegraphics[width=0.101\textwidth]{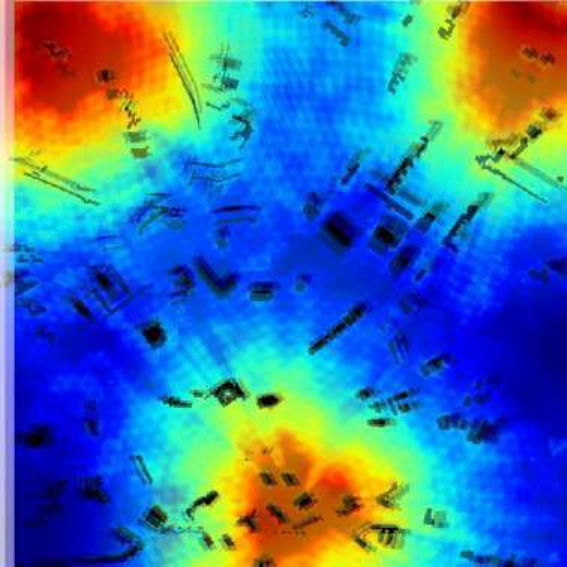} &
\includegraphics[width=0.101\textwidth]{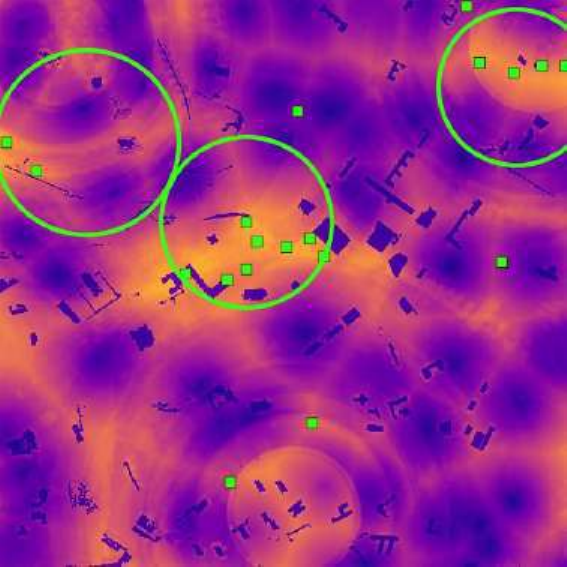} \\[0.10mm]
\resultrownoteletter{e}{Berlin Mitte}{1}{30}{2.4}
\includegraphics[width=0.101\textwidth]{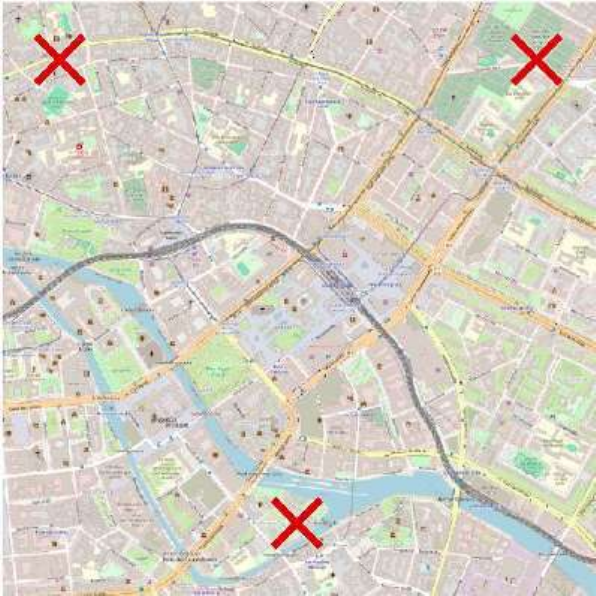} &
\includegraphics[width=0.101\textwidth]{figures/reshape_experiment_analysis/berlin_tx1_r06_h30m_t0_ratio_30pct_gt.pdf} &
\includegraphics[width=0.101\textwidth]{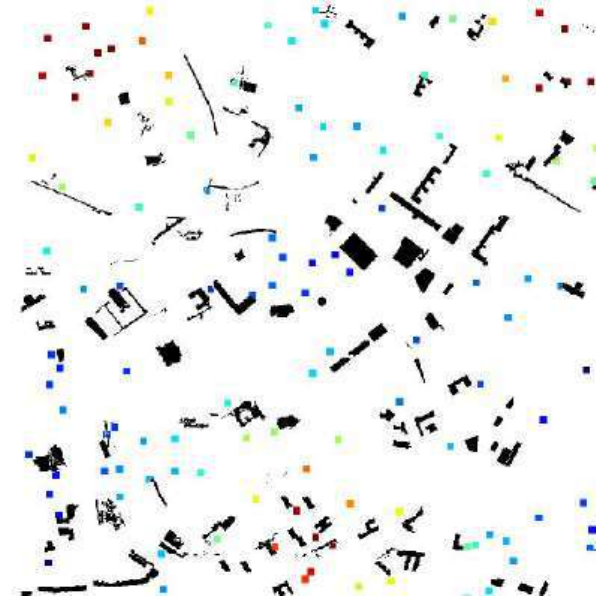} &
\includegraphics[width=0.101\textwidth]{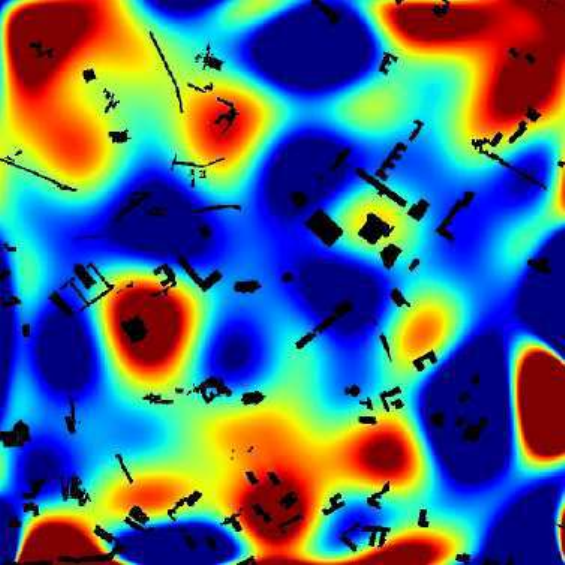} &
\includegraphics[width=0.101\textwidth]{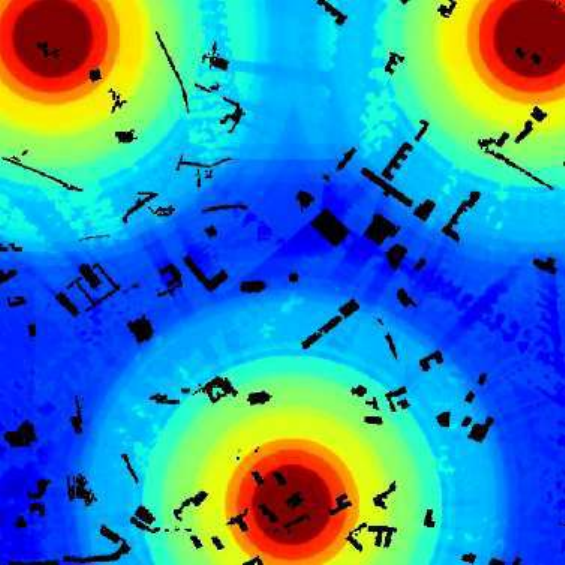} &
\includegraphics[width=0.101\textwidth]{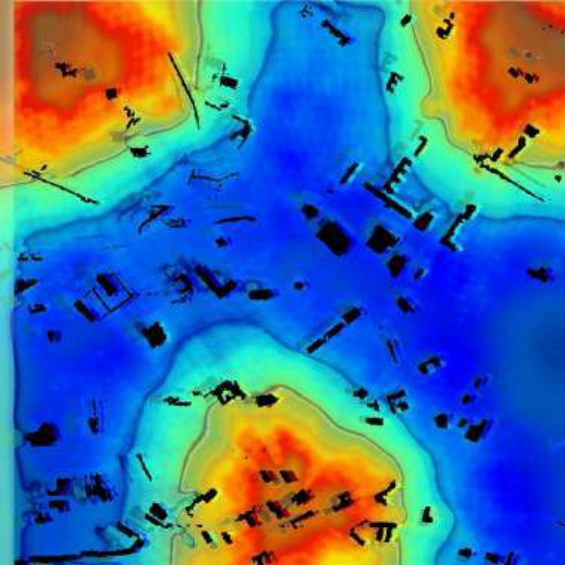} &
\includegraphics[width=0.101\textwidth]{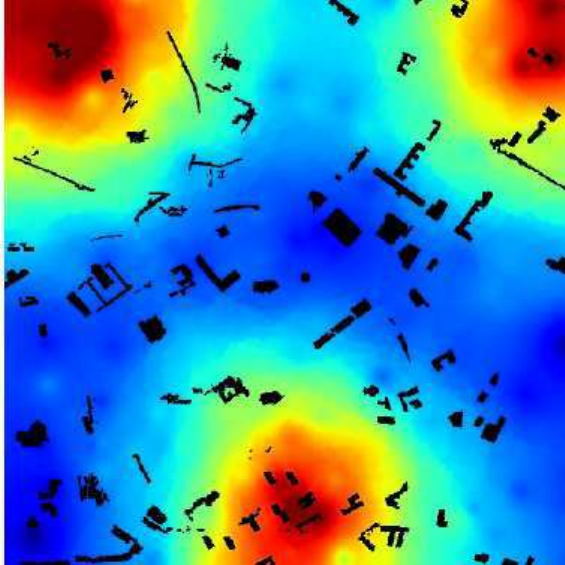} &
\includegraphics[width=0.101\textwidth]{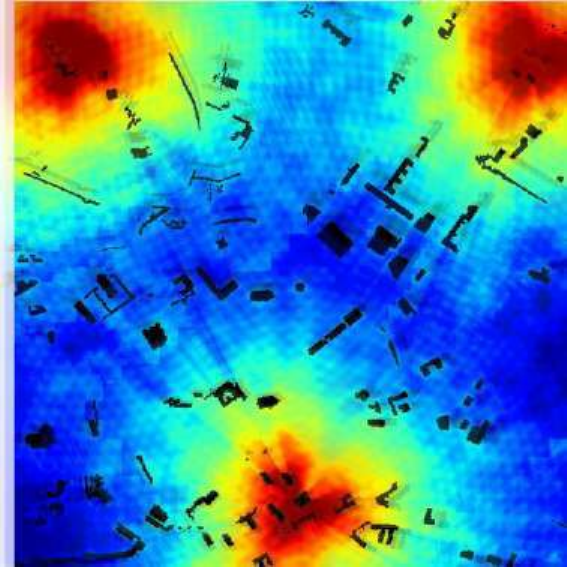} &
\includegraphics[width=0.101\textwidth]{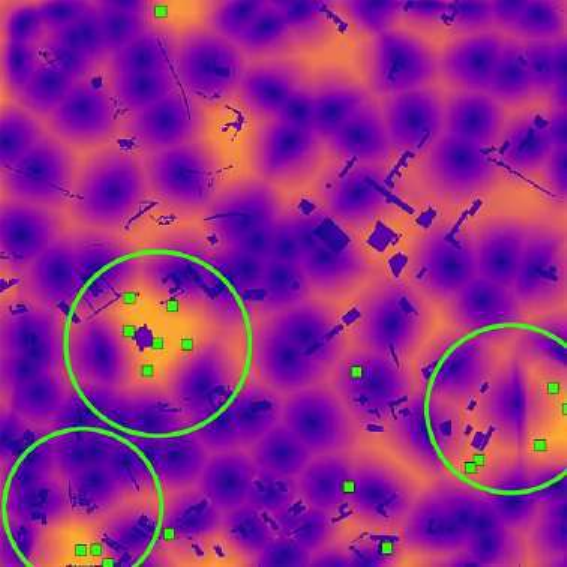} \\[0.10mm]
\resultrownoteletter{f}{Berlin Mitte}{1}{30}{4.8}
\end{tabular}
\caption{Closed-loop reconstruction examples for Berlin Mitte under transmitter location $\xi=1$. Panels (a)-(f) correspond to six combinations of receiver height $z$ and sensing ratio $\rho$}
\label{fig:berlin_tx1_result_grid}
\end{figure*}

\subsection{Performance Analysis}
\label{subsec:comparative_results}

%%%% 这里我有一个小巧思，就是高度，位置以及采样sparsity这一共有三个指标，如果三个指标都会变化，则非常难以对比，所以我剩下三个对比表格中，每个表格固定一个变量，然后去分析其他两个，这样充分且直观.
%% 三个变量是：TX位置（不同的电磁环境），RX高度（建筑物遮挡不同），感知的稀疏度（多少点被采样）

Due to space limitations, we only present Berlin Mitte as a representative urban region for intuitive visual comparison among different IRM construction methods. Additional results for other cities with different urban morphologies are provided in \hl{Appendix B}.

First, we evaluate the inference accuracy based on the fixed transmitter location, while the receiver height and sensing sparsity vary. The corresponding results are illustrated in Fig.~\ref{fig:berlin_tx1_result_grid}. Among various comparison baselines, several similar trends can be clearly observed: \textit{(i)} as the receiver height increases, the blockage effects caused by buildings and geographical contexts gradually weaken, resulting in smoother radio fields; \textit{(ii)} the uncertainty maps provide an intuitive interpretation of the proposed feedback mechanism, i.e., the highly uncertain regions (green circles in the last column) will generate more sensing candidates to acquire sparse measurements (colorful squares in column 3) for the subsequent stage; \textit{(iii)} all inference methods improve the quality of IRMs when the sensing sparsity increases significantly.
However, different methods exhibit varying inference accuracy as the receiver height and sensing sparsity level change. Specifically, Kriging is sensitive to local outliers, while KNN tends to overlook blockage-induced propagation effects. Although RadioDiff and RadioLAM generate visually appealing results with reasonable obstacle awareness, the proposed framework achieves more accurate fine-grained inference, particularly in accurately characterizing the spectrum intensity near the transmitter and shadowing effects around large buildings. 
To further provide quantitative evidence, repeated experiments are conducted and the corresponding inference accuracy results are computed based on \eqref{ele:acc}. Table~\ref{tab1:accuracy_results} presents the details, which further validate the visual comparisons and demonstrate the effectiveness of the proposed closed-loop framework.

\begin{table}[!t]
\centering
\caption{Quantitative comparison of inference accuracy ($\mathcal{M}_{\mathrm{acc}} \downarrow, \quad \xi=1$).}
\label{tab1:accuracy_results}
\resizebox{\columnwidth}{!}{
\begin{tabular}{lccccc}
\toprule
\textbf{Scenario}
& \textbf{Kriging}
& \textbf{KNN}
& \textbf{RadioDiff}
& \textbf{RadioLAM}
& \textbf{Proposed}
\\
\midrule

$z=0$ m, $\rho=2.4\%$
& 8.51 & 7.12 & 4.78 & 4.94 & \textbf{2.77}
\\

$z=0$ m, $\rho=4.8\%$
& 4.73 & 4.45 & 2.18 & 2.30 & \textbf{1.13}
\\

$z=15$ m, $\rho=2.4\%$
& 6.26 & 6.91 & 4.32 & 4.21 & \textbf{2.58}
\\

$z=15$ m, $\rho=4.8\%$
& 4.34 & 4.18 & 1.77 & 1.96 & \textbf{0.97}
\\

$z=30$ m, $\rho=2.4\%$
& 6.12 & 6.08 & 4.62 & 4.78 & \textbf{2.51}
\\

$z=30$ m, $\rho=4.8\%$
& 3.54 & 3.22 & 1.49 & 1.56 & \textbf{1.11}
\\

\bottomrule
\end{tabular}}
\end{table}

\begin{figure*}[!t]
\centering
\setlength{\tabcolsep}{0.20pt}
\renewcommand{\arraystretch}{0.50}
\begin{tabular}{@{}ccccccccc@{}}
\scriptsize TX location & \scriptsize Ground Truth & \scriptsize Sensing Measurements & \scriptsize Kriging & \scriptsize KNN & \scriptsize RadioDiff & \scriptsize  RadioLAM & \scriptsize Proposed & \scriptsize Uncertainty \\[-0.2mm]
\includegraphics[width=0.101\textwidth]{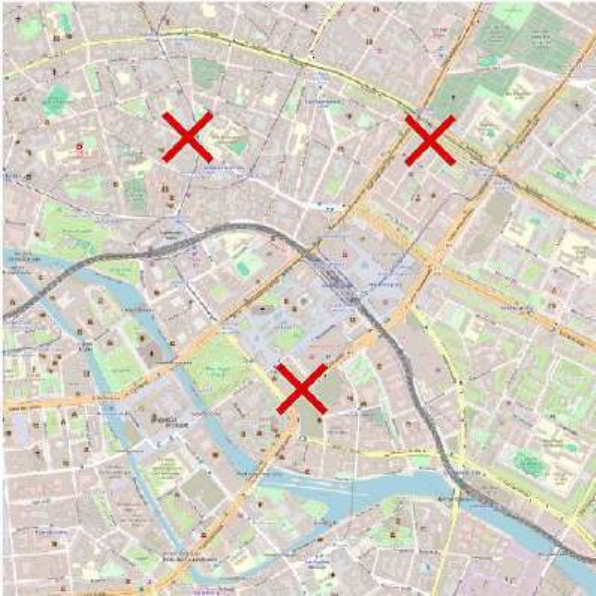} &
\includegraphics[width=0.101\textwidth]{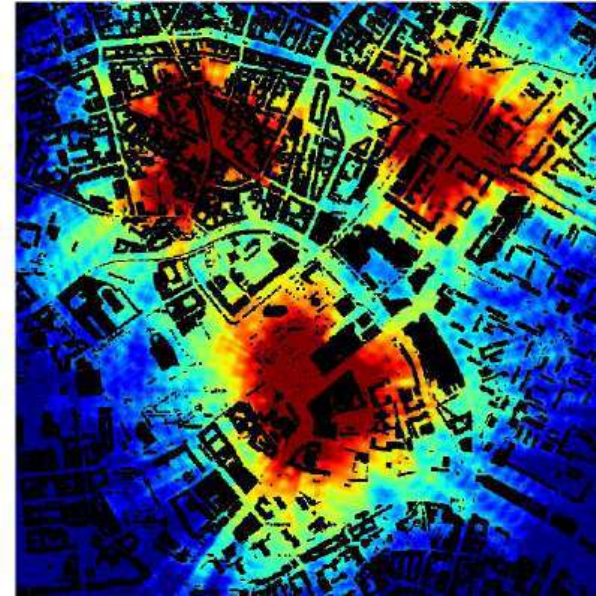} &
\includegraphics[width=0.101\textwidth]{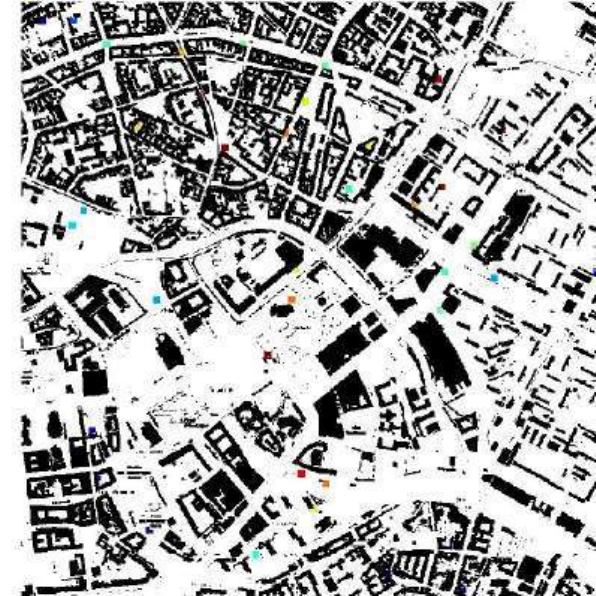} &
\includegraphics[width=0.101\textwidth]{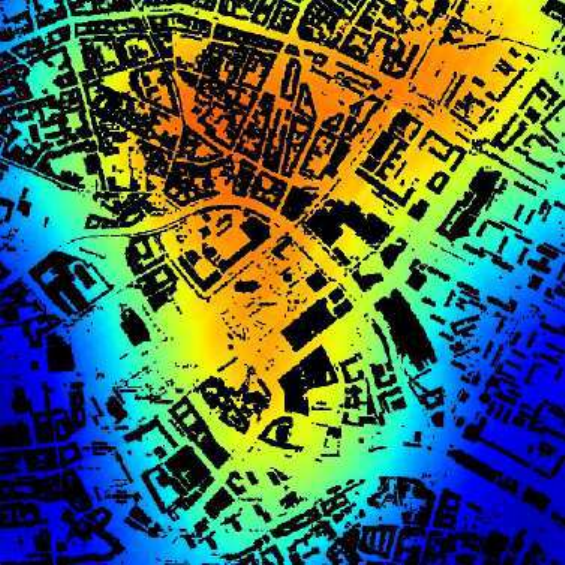} &
\includegraphics[width=0.101\textwidth]{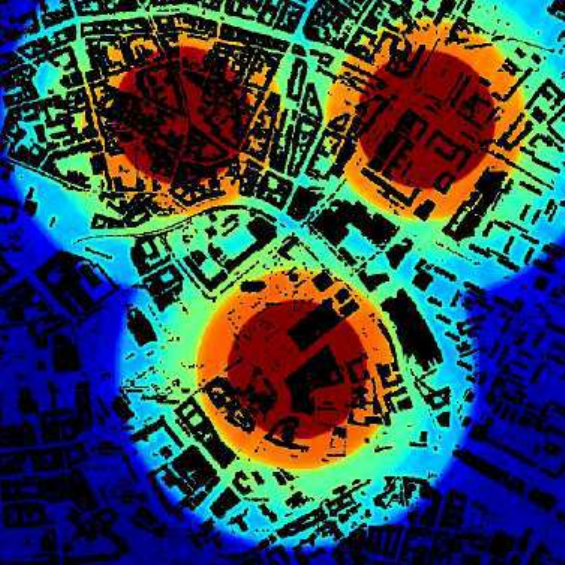} &
\includegraphics[width=0.101\textwidth]{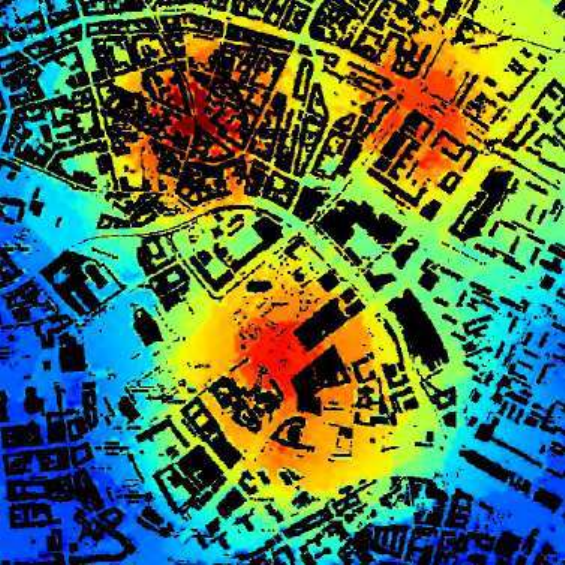} &
\includegraphics[width=0.101\textwidth]{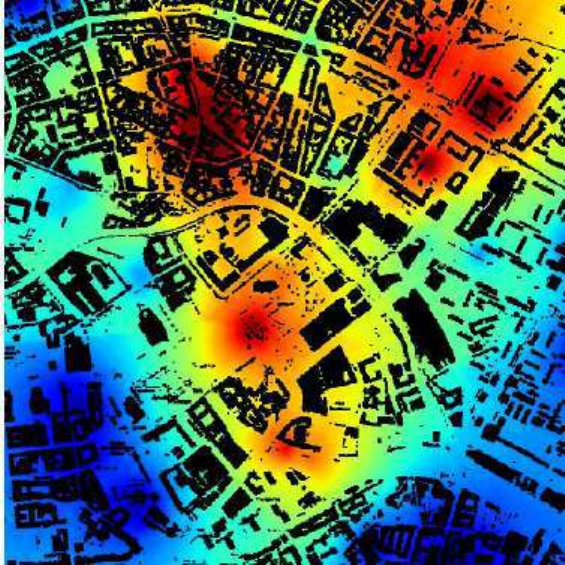} &
\includegraphics[width=0.101\textwidth]{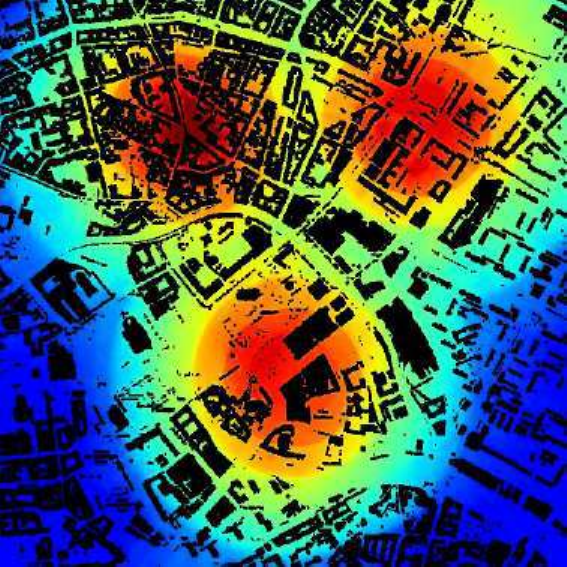} &
\includegraphics[width=0.101\textwidth]{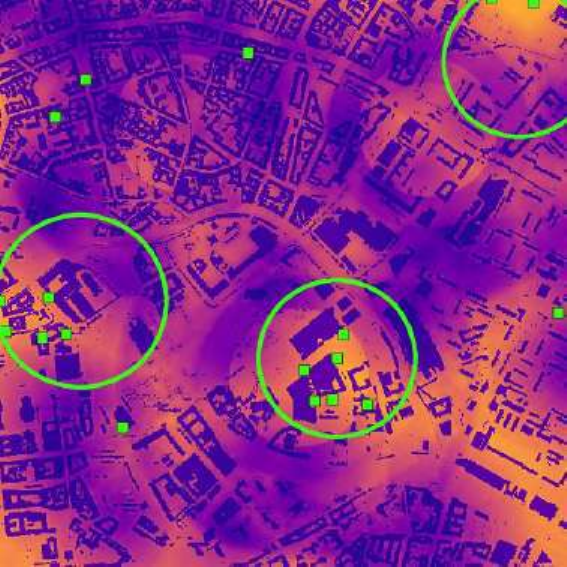} \\[0.10mm]
\resultrownoteletter{a}{Berlin Mitte}{2}{0}{1.6}
\includegraphics[width=0.101\textwidth]{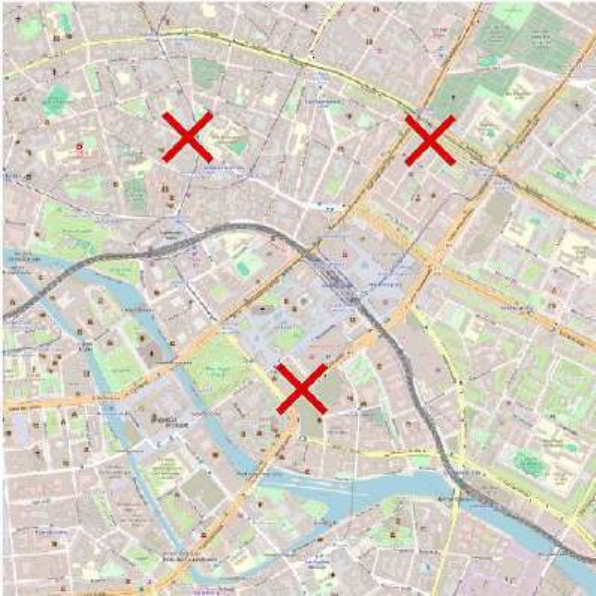} &
\includegraphics[width=0.101\textwidth]{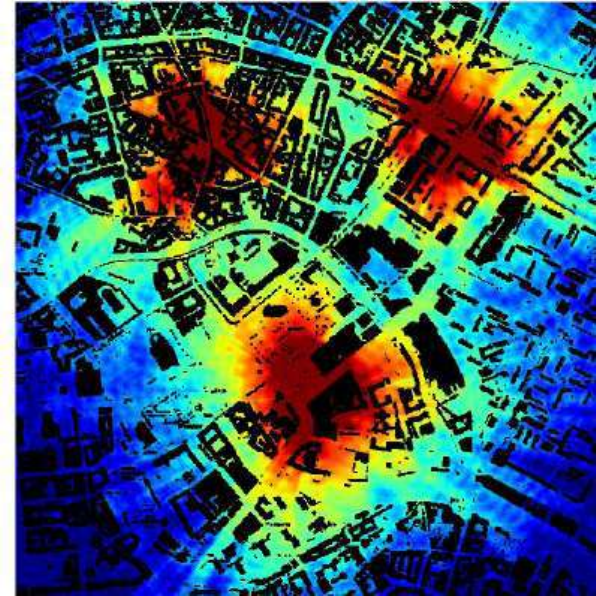} &
\includegraphics[width=0.101\textwidth]{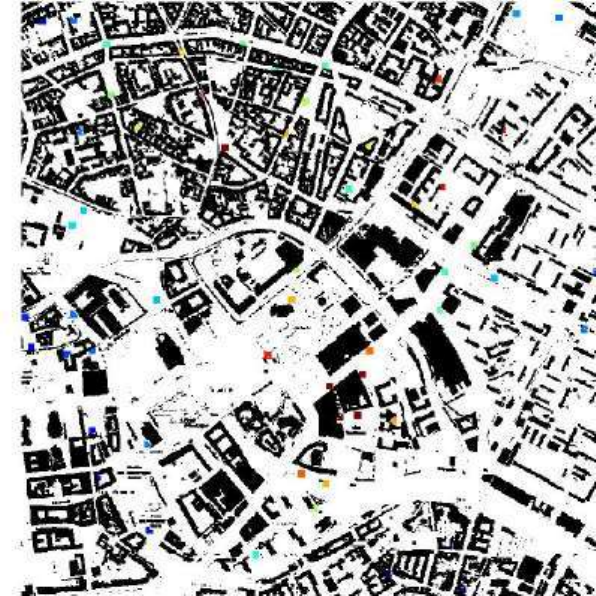} &
\includegraphics[width=0.101\textwidth]{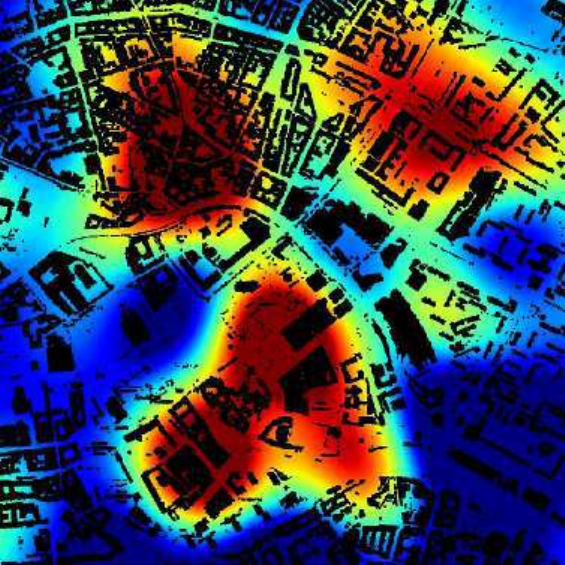} &
\includegraphics[width=0.101\textwidth]{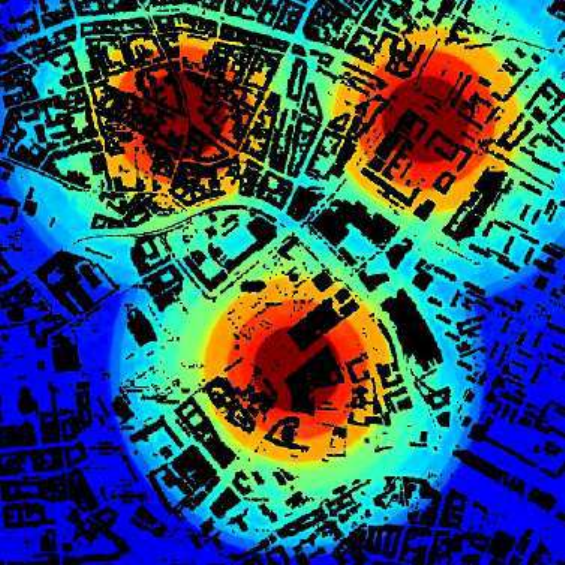} &
\includegraphics[width=0.101\textwidth]{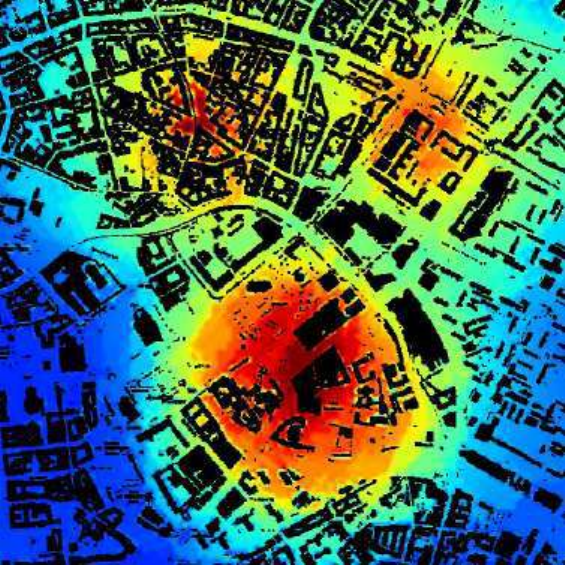} &
\includegraphics[width=0.101\textwidth]{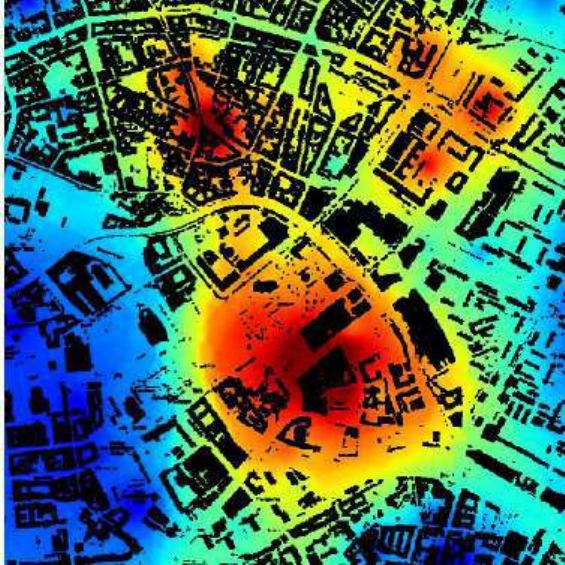} &
\includegraphics[width=0.101\textwidth]{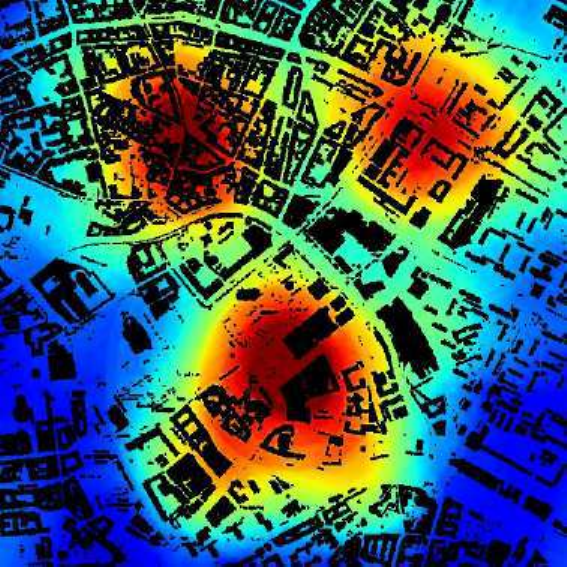} &
\includegraphics[width=0.101\textwidth]{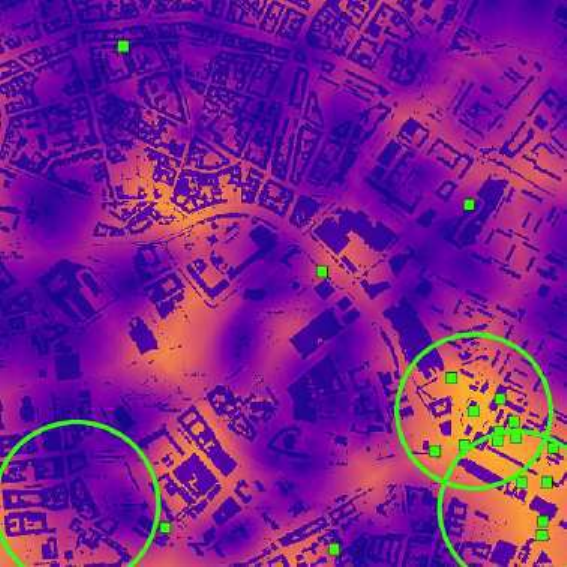} \\[0.10mm]
\resultrownoteletter{b}{Berlin Mitte}{2}{0}{2.4}
\includegraphics[width=0.101\textwidth]{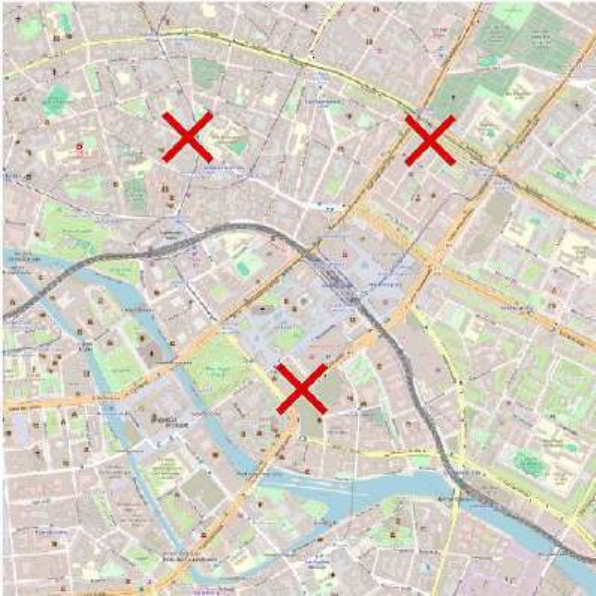} &
\includegraphics[width=0.101\textwidth]{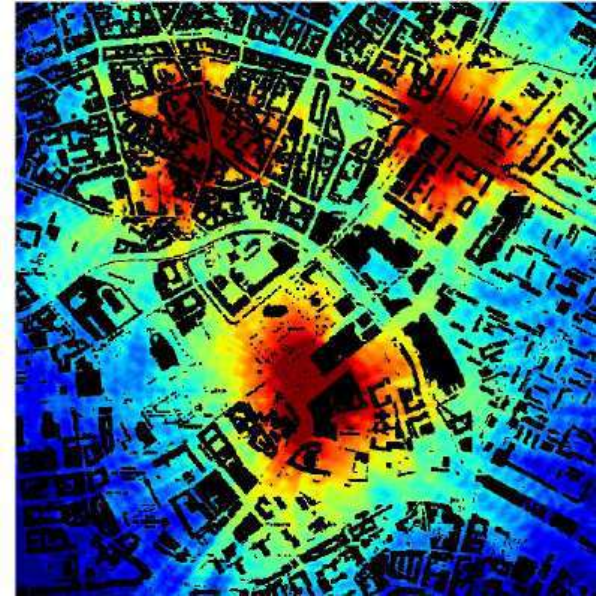} &
\includegraphics[width=0.101\textwidth]{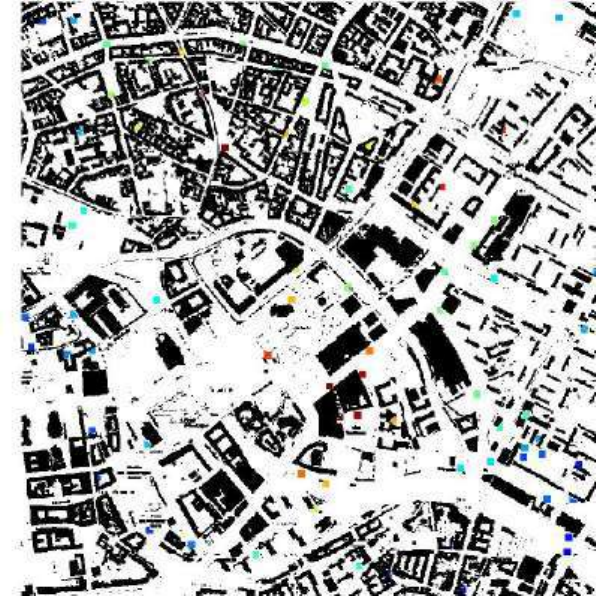} &
\includegraphics[width=0.101\textwidth]{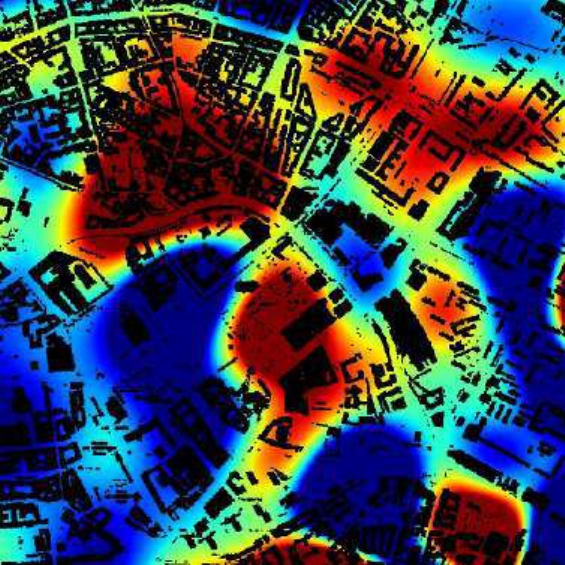} &
\includegraphics[width=0.101\textwidth]{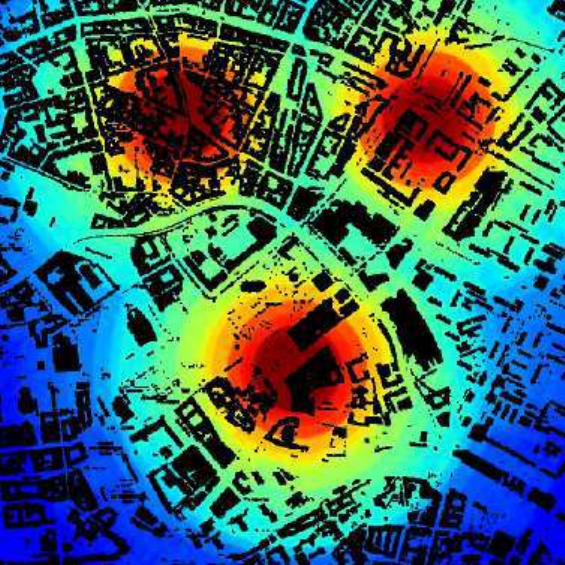} &
\includegraphics[width=0.101\textwidth]{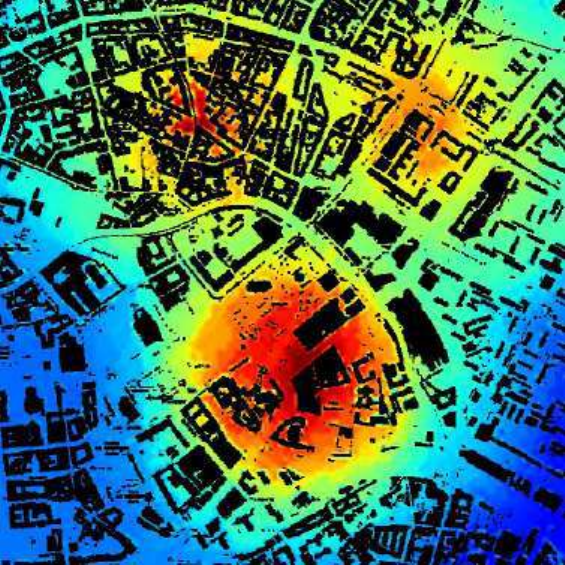} &
\includegraphics[width=0.101\textwidth]{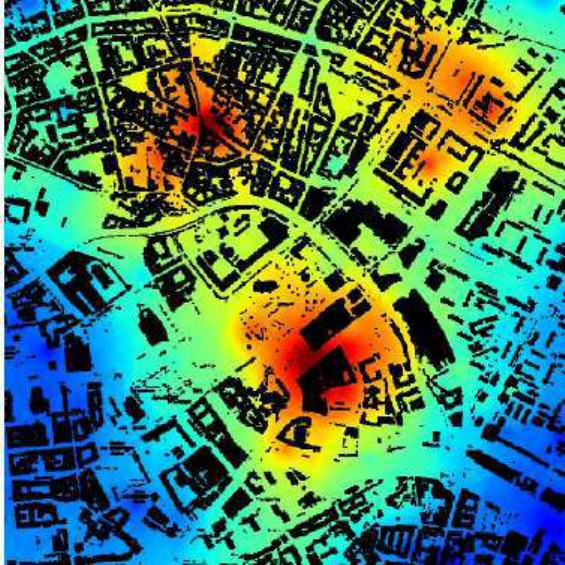} &
\includegraphics[width=0.101\textwidth]{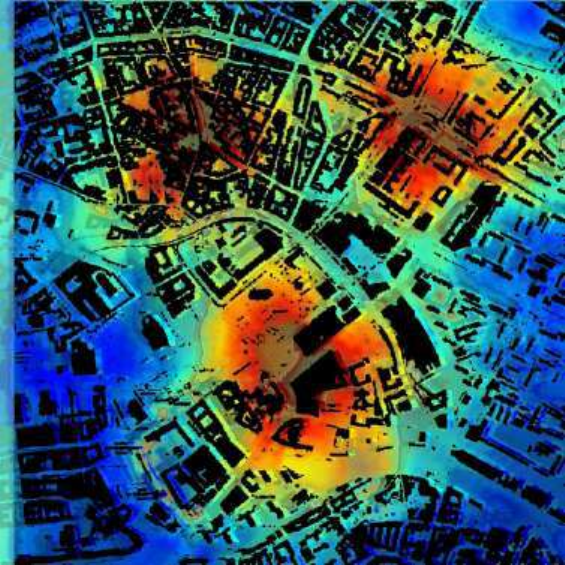} &
\includegraphics[width=0.101\textwidth]{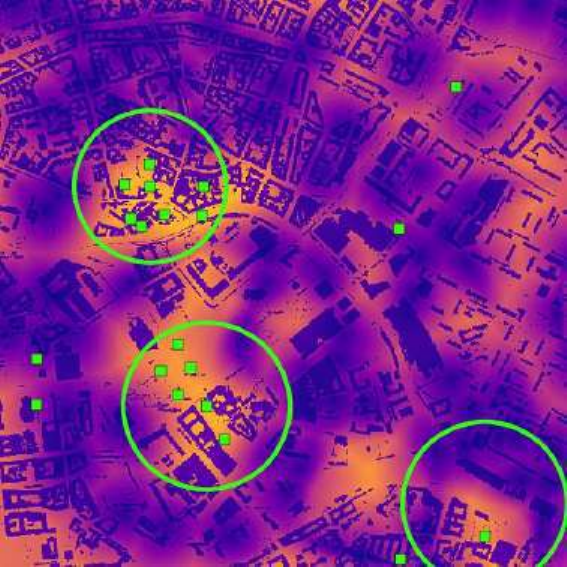} \\[0.10mm]
\resultrownoteletter{c}{Berlin Mitte}{2}{0}{3.2}
\includegraphics[width=0.101\textwidth]{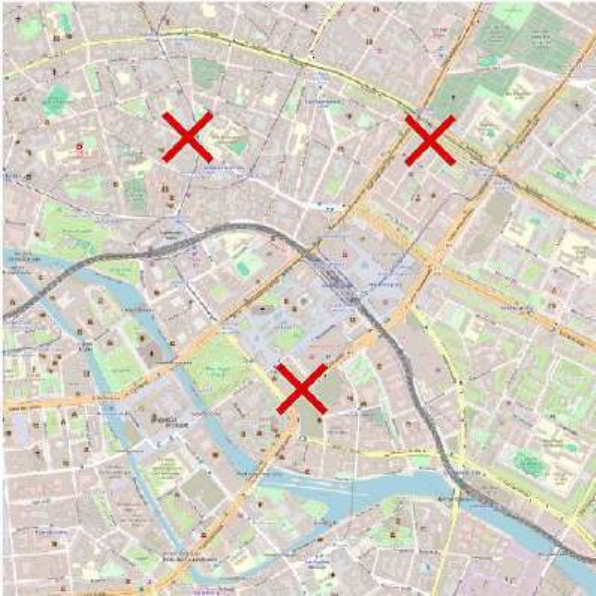} &
\includegraphics[width=0.101\textwidth]{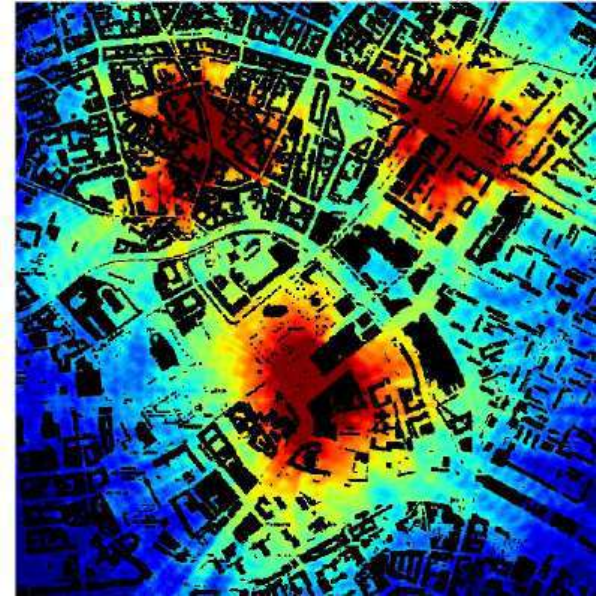} &
\includegraphics[width=0.101\textwidth]{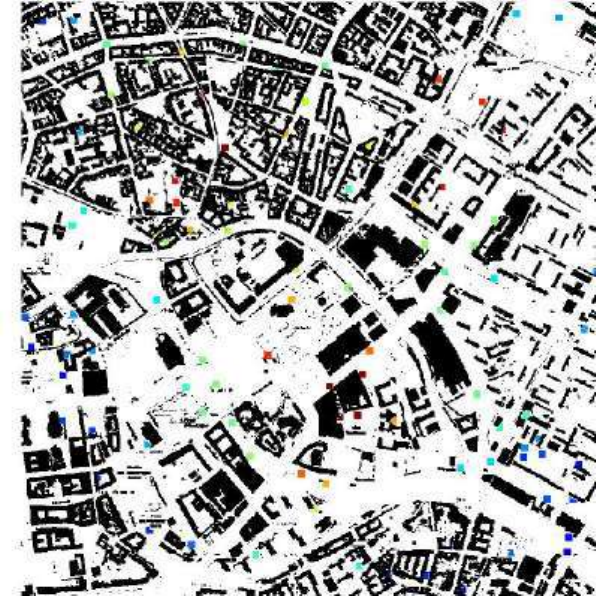} &
\includegraphics[width=0.101\textwidth]{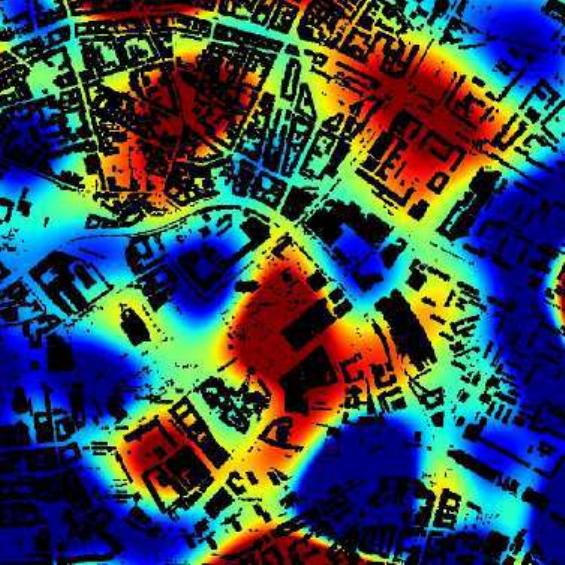} &
\includegraphics[width=0.101\textwidth]{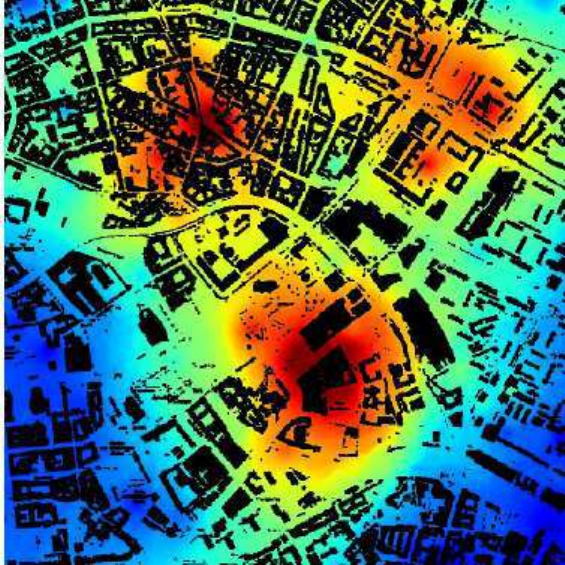} &
\includegraphics[width=0.101\textwidth]{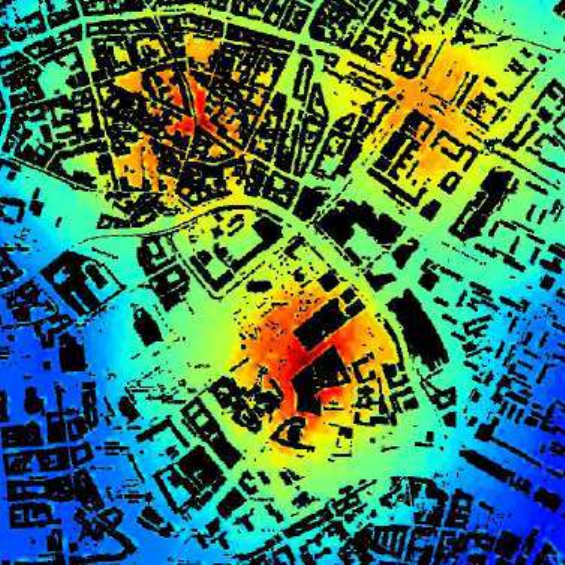} &
\includegraphics[width=0.101\textwidth]{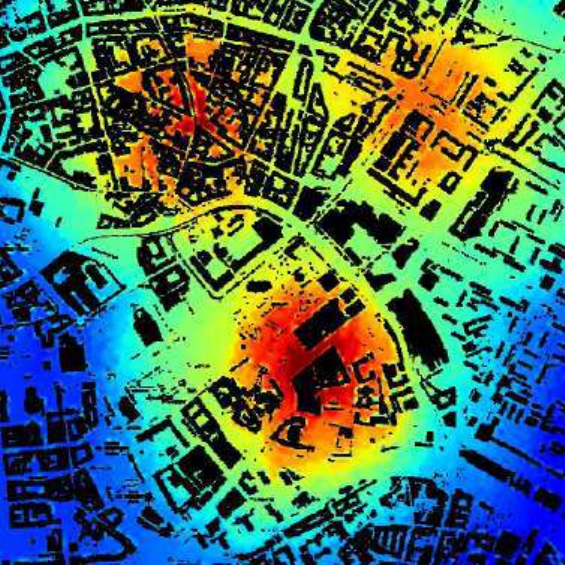} &
\includegraphics[width=0.101\textwidth]{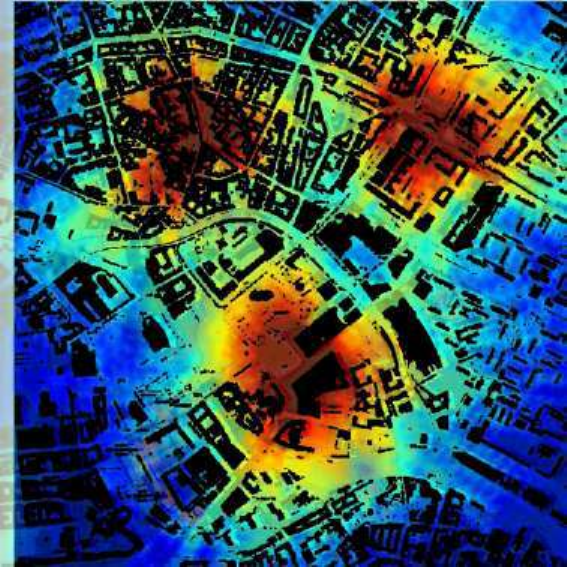} &
\includegraphics[width=0.101\textwidth]{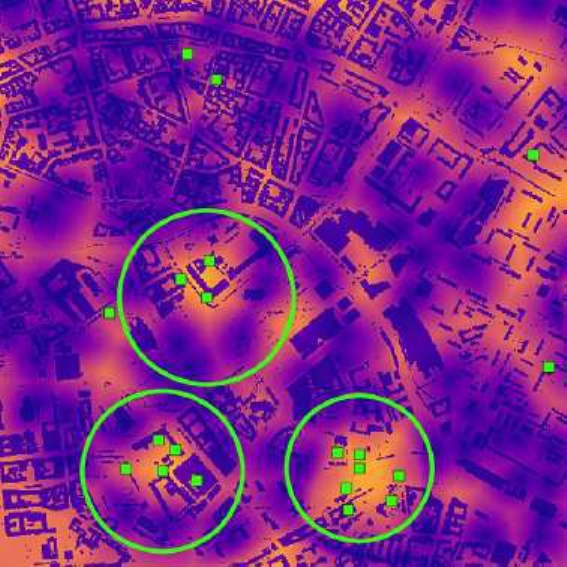} \\[0.10mm]
\resultrownoteletter{d}{Berlin Mitte}{2}{0}{4.0}
\includegraphics[width=0.101\textwidth]{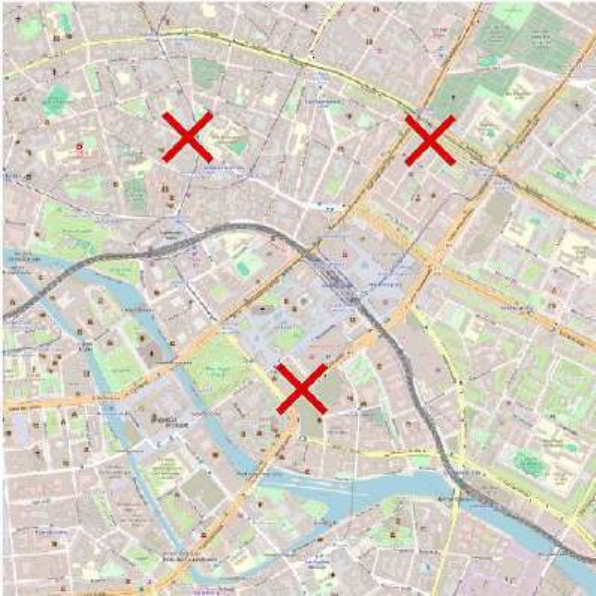} &
\includegraphics[width=0.101\textwidth]{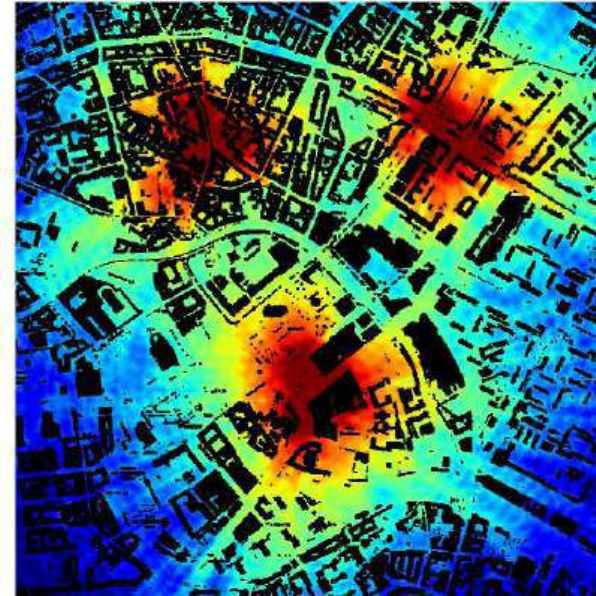} &
\includegraphics[width=0.101\textwidth]{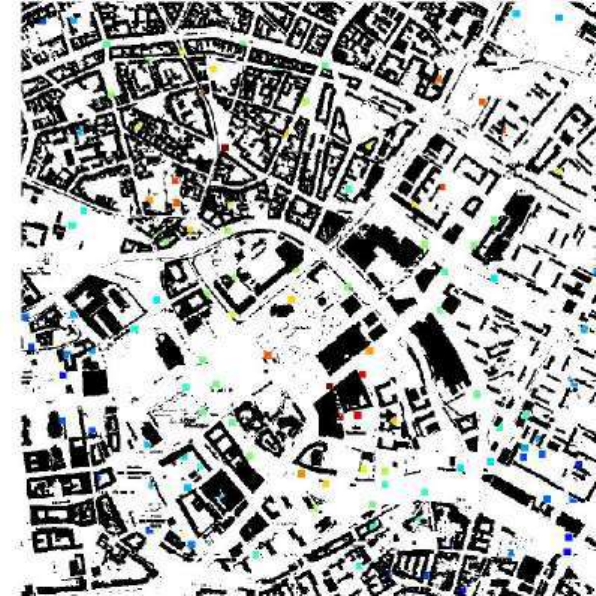} &
\includegraphics[width=0.101\textwidth]{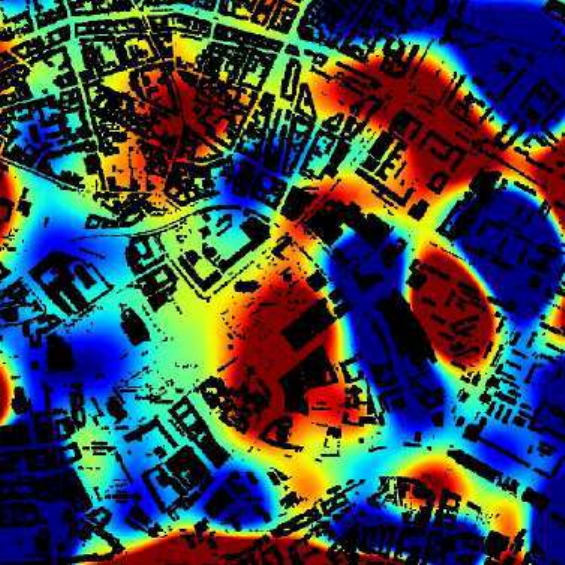} &
\includegraphics[width=0.101\textwidth]{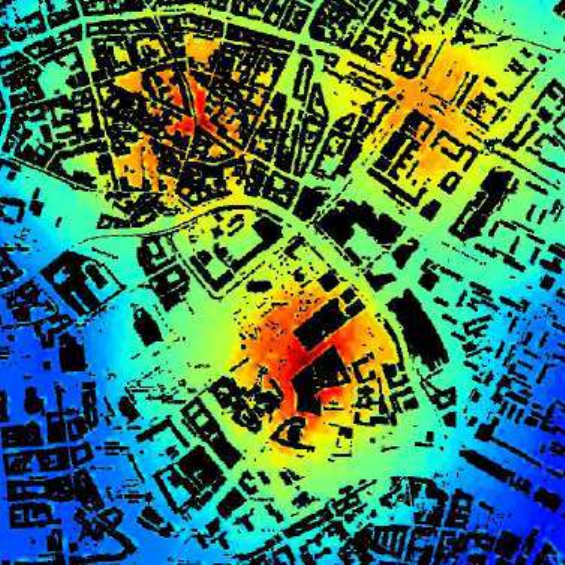} &
\includegraphics[width=0.101\textwidth]{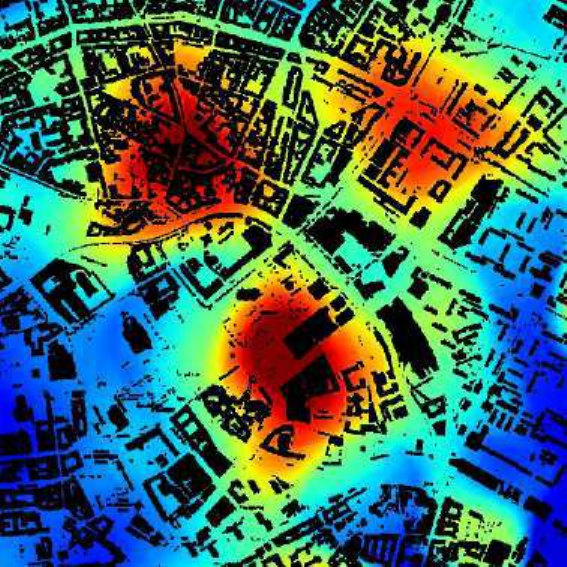} &
\includegraphics[width=0.101\textwidth]{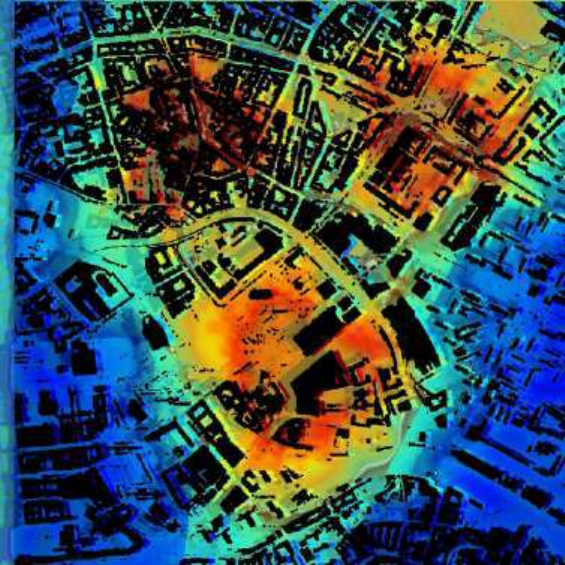} &
\includegraphics[width=0.101\textwidth]{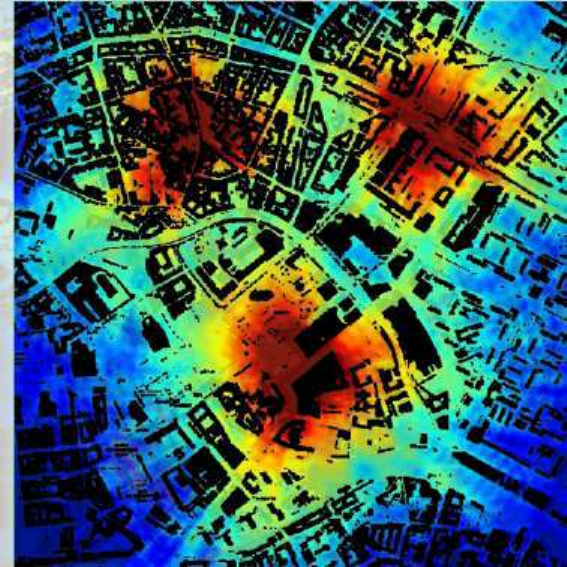} &
\includegraphics[width=0.101\textwidth]{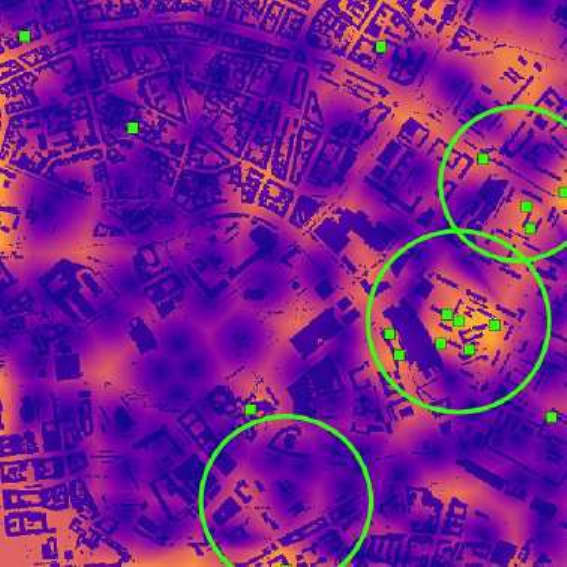} \\[0.10mm]
\resultrownoteletter{e}{Berlin Mitte}{2}{0}{4.8}
\end{tabular}
\caption{Closed-loop reconstruction examples for Berlin Mitte under transmitter location $\xi=2$ and receiver height $z=0~\mathrm{m}$. Panels (a)--(e) correspond to sensing ratios $\rho$ of 1.6\%, 2.4\%, 3.2\%, 4.0\%, and 4.8\%, respectively.}
\label{fig:berlin_tx2_result_grid}
\end{figure*}

%%% 新修改的解释！
% 这个图是berlin第三类tx位置的图（发射源比较集中），但是占据的空间比较大，而且和图5（第二个tx位置的图有类似性）所以我直接取消掉了。

%%%节省下来的空间我新画了更小尺度的结果图。

Second, we evaluate the sensing-efficiency gain. Unlike the previous experiment where the sensing sparsity changes by a relatively large interval, a finer-grained increment of $\Delta \rho=0.8\%$ is adopted. Moreover, unlike the previous experiment, we select a different TX location, namely $\xi=2$, to generate a different ground-truth base map, while the receiver height is fixed at $z=0~\mathrm{m}$. The corresponding intuitive results are illustrated in Fig.~\ref{fig:berlin_tx2_result_grid}. In this figure, consecutive rows can be viewed as successive construction stages, where newly acquired measurements are recursively incorporated into the inference process.  Among different IRM construction frameworks, Kriging remains sensitive to local observations, and therefore its performance cannot improve monotonically as the sensing sparsity increases. When the transmitter locations become closer and cause spectrum overlap, KNN exhibits relatively coarse intensity estimation in the central area. RadioDiff performs accurate inference around the transmitter, but exhibits weaker prediction capability near map boundaries. RadioLAM achieves high construction fidelity, yet its refinement process is relatively slow; for example, the IRM generated by RadioLAM at $\rho=4.8\%$ exhibits quality comparable to that of the proposed framework at $\rho=3.2\%$. Overall, the proposed closed-loop multi-stage framework exhibits the fastest and most stable construction improvement as the sensing sparsity gradually increases. Based on uncertainty-aware feedback, the framework continuously selects the most informative sensing candidates among stages, thereby accelerating IRM refinement. To further provide quantitative evidence, the sensing-efficiency gain is also computed according to \eqref{ele:gain} and recorded in Table~\ref{tab:gain_results}. 

\begin{table}[!t]
\centering
\caption{Quantitative comparison of sensing-efficiency gain ($\mathcal{M}_{\mathrm{eff}}\uparrow, \quad \Delta\rho:0.8\%)$.}
\label{tab:gain_results}
\resizebox{\columnwidth}{!}{
\begin{tabular}{lccccc}
\toprule
\textbf{Scenario}
& \textbf{Kriging}
& \textbf{KNN}
& \textbf{RadioDiff}
& \textbf{RadioLAM}
& \textbf{Proposed}
\\
\midrule

$\xi=1$, $z=30$ m, 
& 0.77
& 1.34
& 1.51
& 1.49
& \textbf{1.57}
\\
\midrule
$\xi=2$, $z=0$ m, 
& 0.73
& 1.29
& 1.54
& 1.46
& \textbf{1.65}
\\

\midrule

$\xi=3$, $z=15$ m, 
& 0.91
& 1.35
& 1.52
& 1.65
& \textbf{1.78}
\\

\bottomrule
\end{tabular}}
\end{table}

%%%鲁棒性的分析
Third, when it comes to robustness across heterogeneous conditions, 5 different urban morphologies are selected for computation and 
the corresponding results are summarized in Table~\ref{tab2:robustness_results}.
Specifically, KNN exhibits the largest inference accuracy variation across different cities, since it mainly relies on local feature similarity and lacks explicit modeling of radio propagation and blockage effects. Kriging demonstrates relatively stable performance, but at the cost of inferior construction quality. In contrast, learning-based methods, including RadioDiff, RadioLAM, and the proposed framework, exhibit consistently strong robustness since they are trained on the comprehensive dataset containing diverse building blockages and geographical contexts. The proposed framework further outperforms owing to its dedicated token design and multi-stage progressive refinement, which enables more effective utilization of environmental priors and sensing outcomes.

\begin{table}[!pb]
\centering
\caption{Quantitative comparison of robustness among different cities. ($\mathcal{M}_{\mathrm{rob}}\downarrow,\quad z=0 \mathrm{(m)}, \mathcal{C}=5 $).}
\label{tab2:robustness_results}
\resizebox{\columnwidth}{!}{
\begin{tabular}{lccccc}
\toprule
\textbf{Scenario}
& \textbf{Kriging}
& \textbf{KNN}
& \textbf{RadioDiff}
& \textbf{RadioLAM}
& \textbf{Proposed}
\\
\midrule

$\xi=1$ , $\rho=3.2\%$
& 2.84 & 5.72 & 1.03 & 1.02 & \textbf{0.87}
\\

$\xi=1$ , $\rho=4.8\%$
& 2.46 & 4.31 & 0.72 & 0.76 & \textbf{0.70}
\\

$\xi=2$ , $\rho=3.2\%$
& 2.52 & 5.34 & 1.43 & 1.38 & \textbf{0.97}
\\

$\xi=2$ , $\rho=4.8\%$
& 2.21 & 3.86 & 0.81 & 0.75 & \textbf{0.71}
\\

$\xi=3$ , $\rho=3.2\%$
& 3.88 & 6.62 & 1.22 & 1.25 & \textbf{0.88}
\\

$\xi=3$ , $\rho=4.8\%$
& 2.71 & 4.43 & 0.84 & 0.87 & \textbf{0.75}
\\

\bottomrule
\end{tabular}}
\end{table}

\begin{figure}[htbp!]
\centering
\includegraphics[width=0.95\columnwidth]{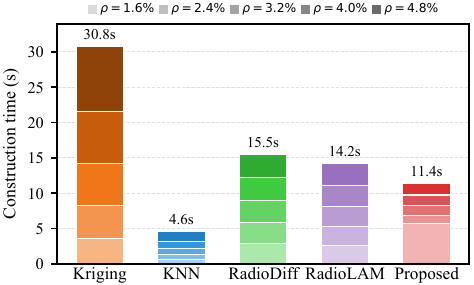}
\caption{Accumulated construction time of different methods over five sensing sparsity levels under $\xi=1$ and $z=0~\mathrm{m}$.}
\label{fig:latency_results}
\end{figure}

Last but not least, the practical deployment efficiency is evaluated by construction time, calculated by \eqref{metric-time}. Specifically, we select a specific setting of $\xi=1$ and $z=0~\mathrm{m}$, and accumulate the construction time over multiple sensing stages (i.e., five sensing sparsity levels). The corresponding results are summarized in Fig.~\ref{fig:latency_results}. It can be observed that KNN achieves the lowest computational latency, whereas Kriging incurs the highest overhead due to its repeated interpolation operations over large-scale spatial grids. Although the proposed framework requires a relatively longer execution time at the initial stage because of the complex teacher model, the subsequent stages are executed using the lightweight distilled student model with much lower computation overhead. Therefore, under continuously evolving urban environments, the proposed framework offers a more favorable tradeoff between construction quality and time.

\subsection{Multi-Scale Deployment}
\label{subsec:multiscale}
\begin{figure}[!t]
\centering
\begin{minipage}[t]{0.245\columnwidth}
\centering
\includegraphics[height=0.70in,keepaspectratio]{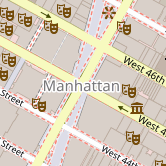}\\[-0.5mm]
{\fontsize{6}{6.5}\selectfont (a) OpenStreetMap}
\end{minipage}\hfill
\begin{minipage}[t]{0.245\columnwidth}
\centering
\includegraphics[height=0.70in,keepaspectratio]{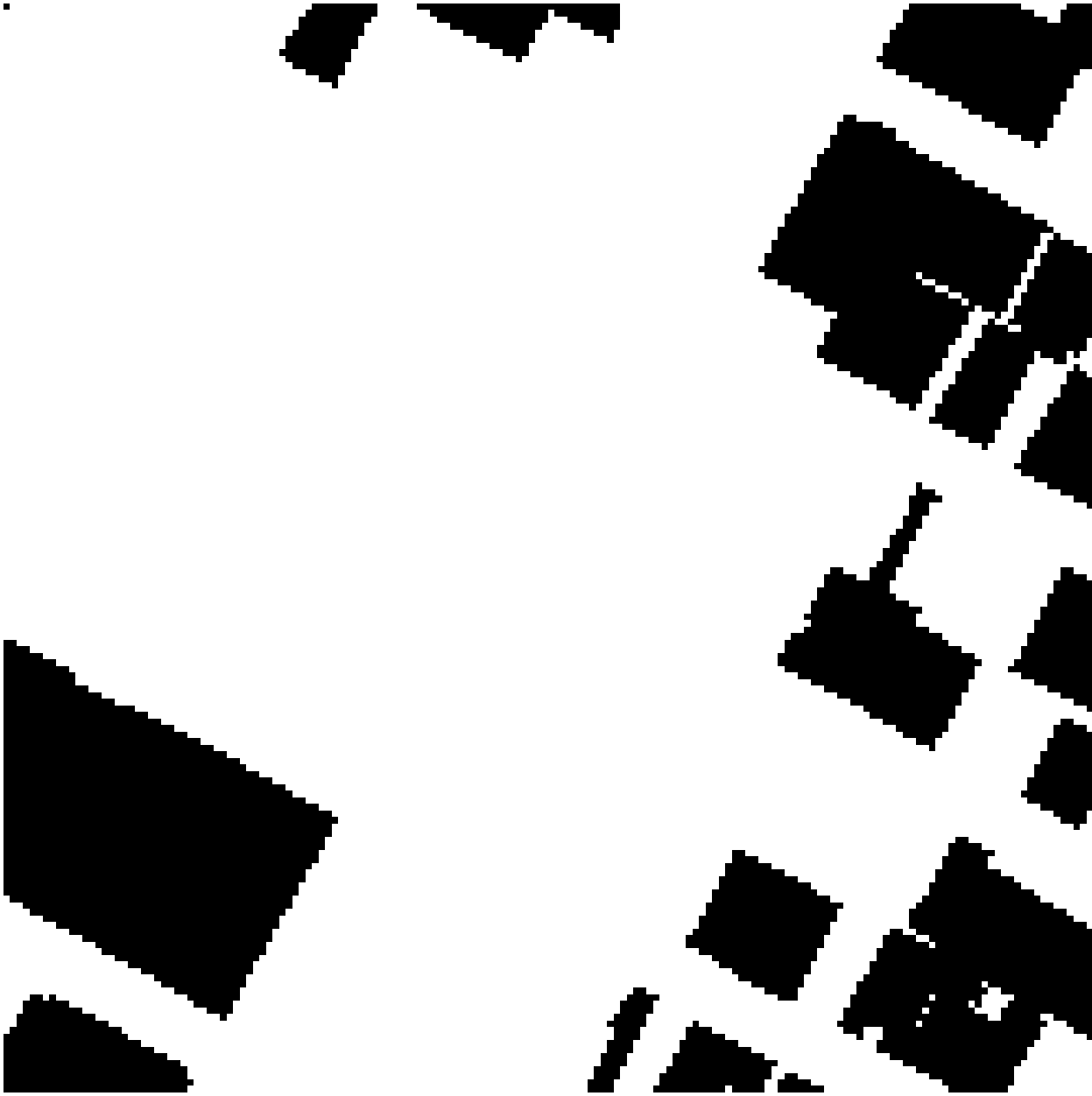}\\[-0.5mm]
{\fontsize{6}{6.5}\selectfont (b) Effective mask}
\end{minipage}\hfill
\begin{minipage}[t]{0.245\columnwidth}
\centering
\includegraphics[height=0.70in,keepaspectratio]{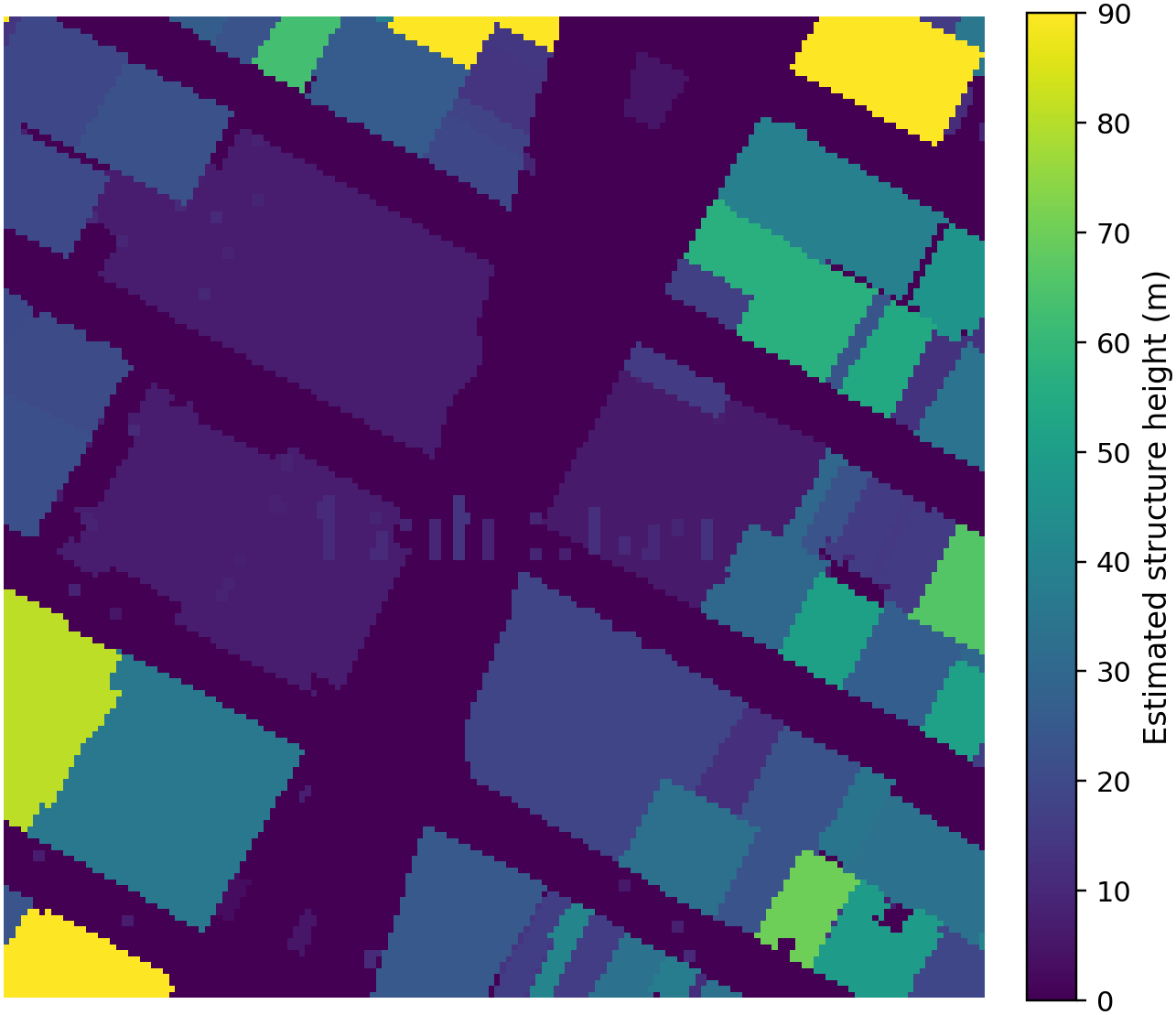}\\[-0.5mm]
{\fontsize{6}{6.5}\selectfont (c) Height map}
\end{minipage}\hfill
\begin{minipage}[t]{0.245\columnwidth}
\centering
\includegraphics[height=0.70in,keepaspectratio]{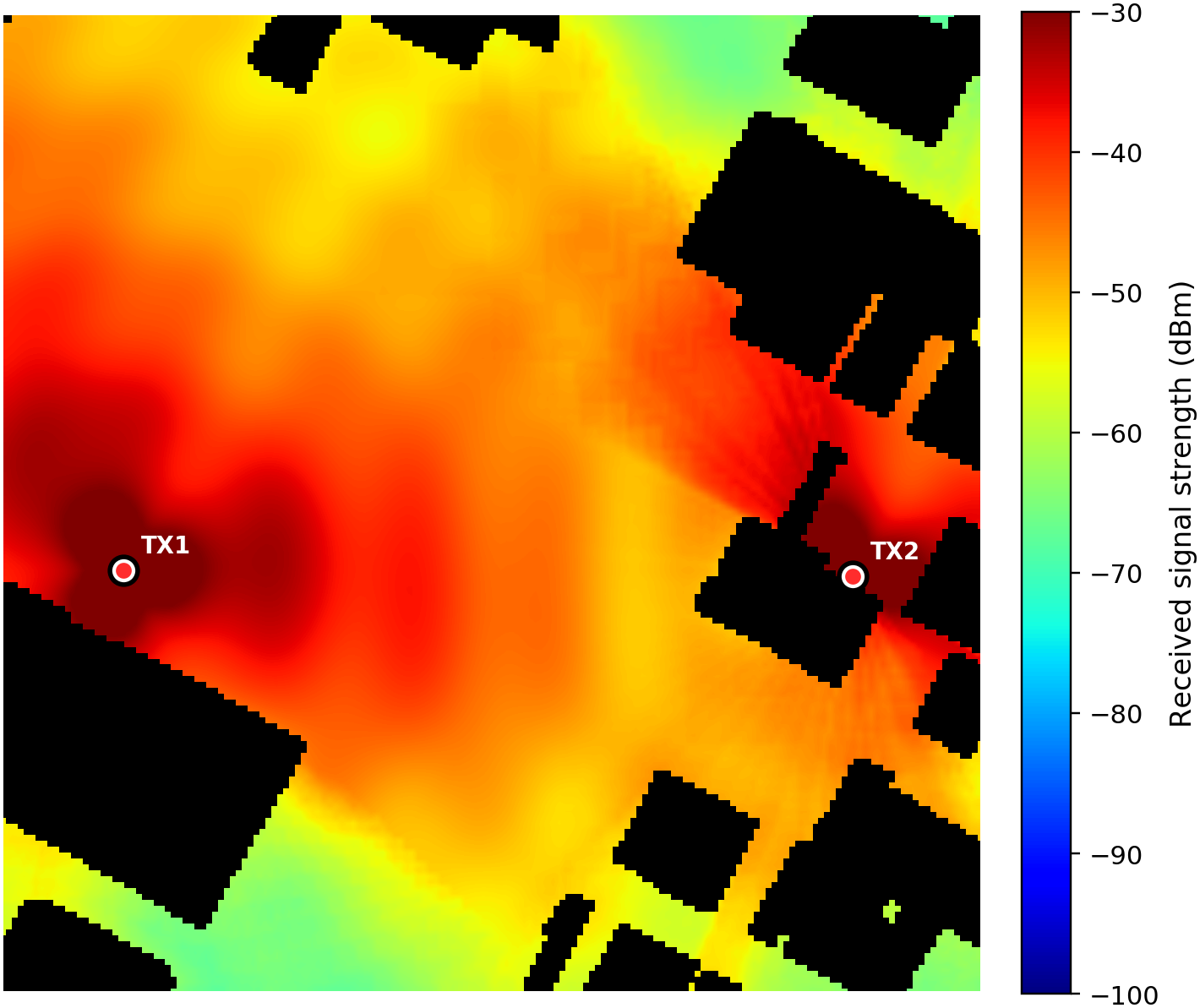}\\[-0.5mm]
{\fontsize{6}{6.5}\selectfont (d) Ground-truth IRM}
\end{minipage}

\caption{A small-scale example in Downtown Manhattan.}
\label{fig:small_scale_example}
\end{figure}

The previous experiments mainly focus on large-scale urban environments. However, practical wireless applications may also require fine-grained spectrum awareness within small local regions to enhance wireless communication. Therefore, we further investigate the scalability of the proposed framework. For example, a local region of size $300~\mathrm{m}\times300~\mathrm{m}$ in Manhattan is extracted from the large-scale map, with only two fixed transmitters. Moreover, the receiver height is fixed at $z=30~\mathrm{m}$, while the frequency is set as $3000~\mathrm{MHz}$. Using the customized simulation strategy, the corresponding geographical map, height map, blockage masks, and ground-truth IRM are illustrated in Fig.~\ref{fig:small_scale_example}(a)--(d), respectively.

\begin{figure}[!t]
\centering
\begin{minipage}[t]{0.245\columnwidth}
\centering
\includegraphics[height=0.68in,keepaspectratio]{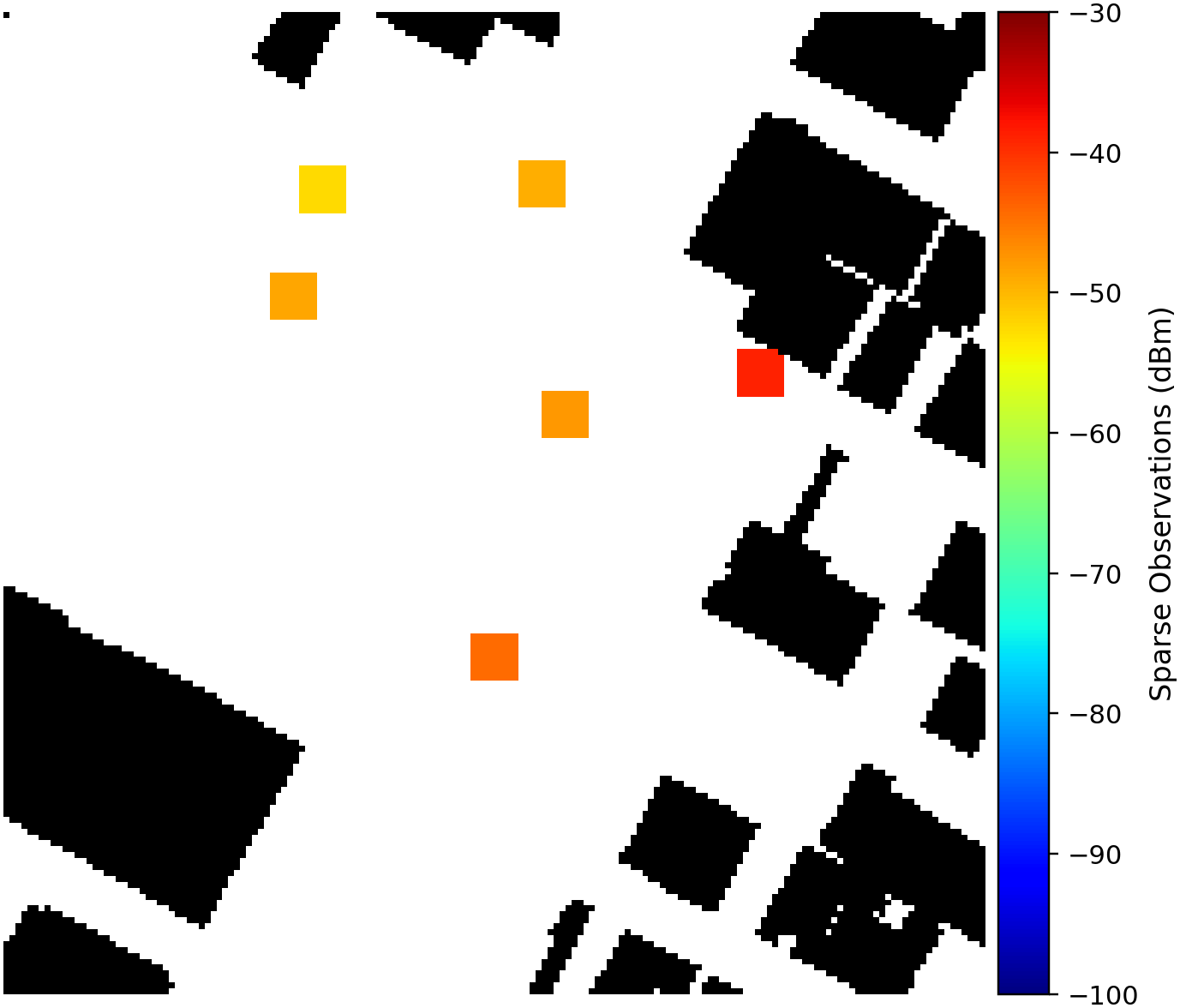}\\[-0.5mm]
{\fontsize{6}{6.5}\selectfont (a) Measurements (1.6\%)}
\end{minipage}\hfill
\begin{minipage}[t]{0.245\columnwidth}
\centering
\includegraphics[height=0.68in,keepaspectratio]{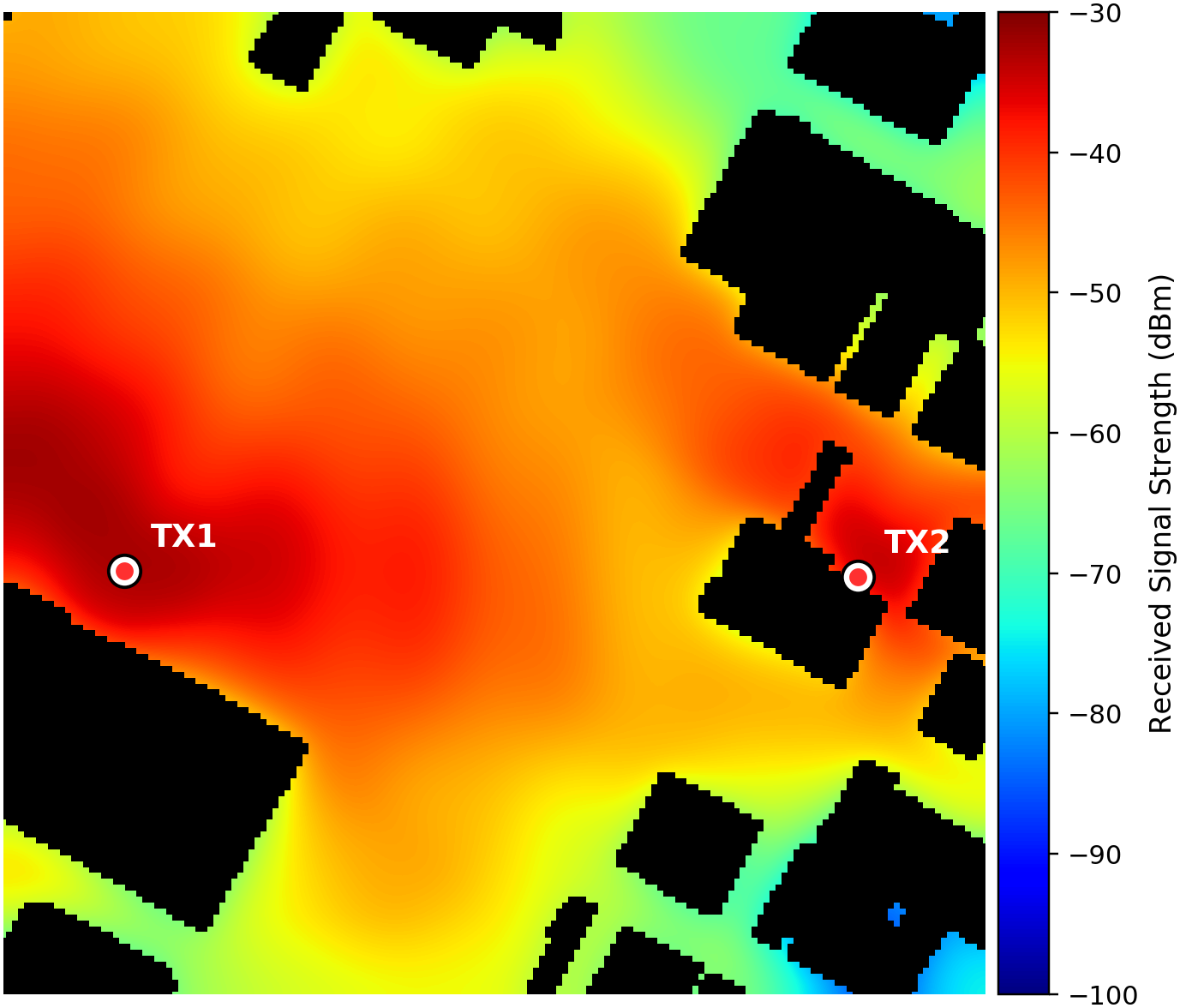}\\[-0.5mm]
{\fontsize{6}{6.5}\selectfont (b) Inference (1.6\%)}
\end{minipage}\hfill
\begin{minipage}[t]{0.245\columnwidth}
\centering
\includegraphics[height=0.68in,keepaspectratio]{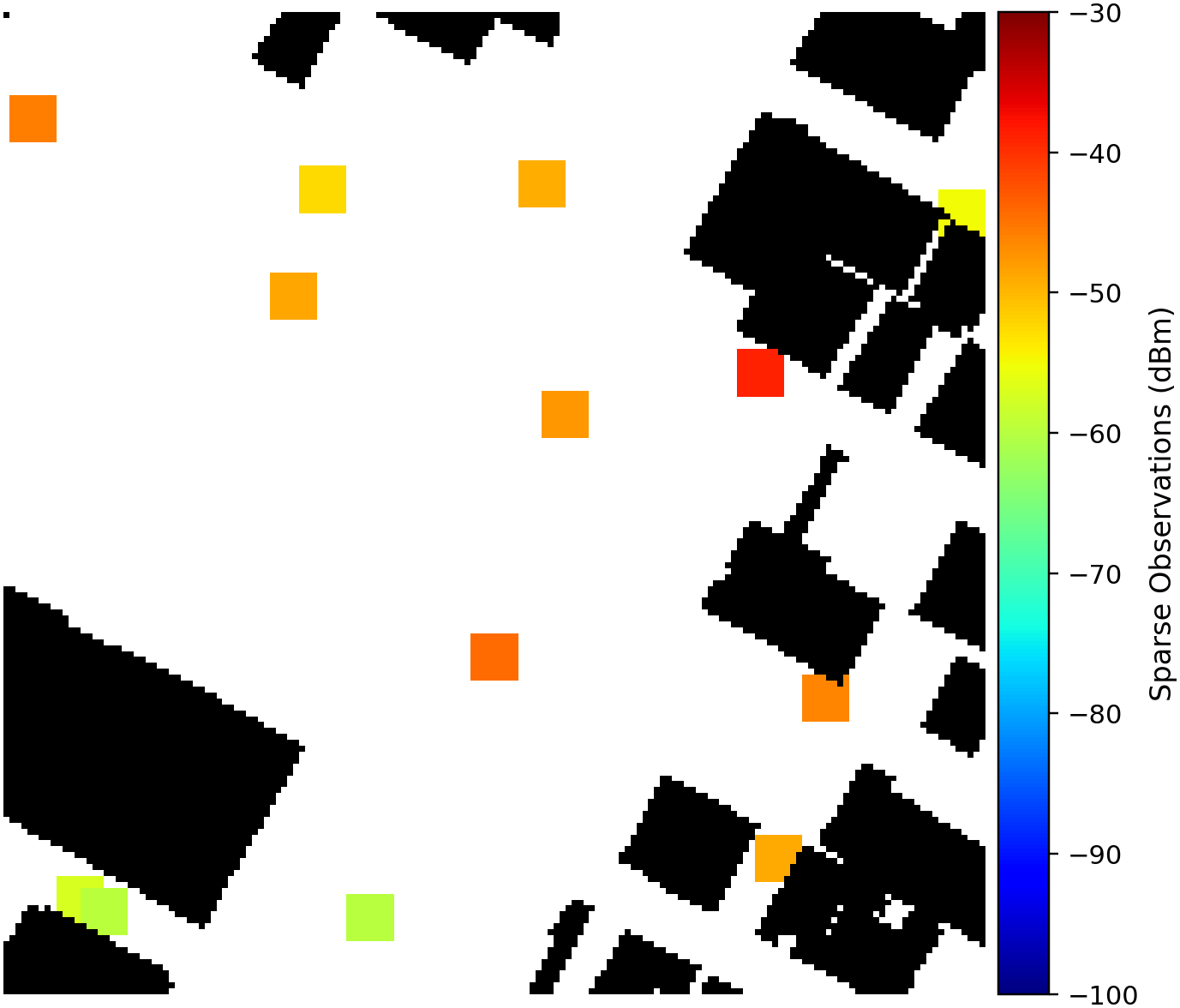}\\[-0.5mm]
{\fontsize{6}{6.5}\selectfont (c) Measurements (3.2\%)}
\end{minipage}\hfill
\begin{minipage}[t]{0.245\columnwidth}
\centering
\includegraphics[height=0.68in,keepaspectratio]{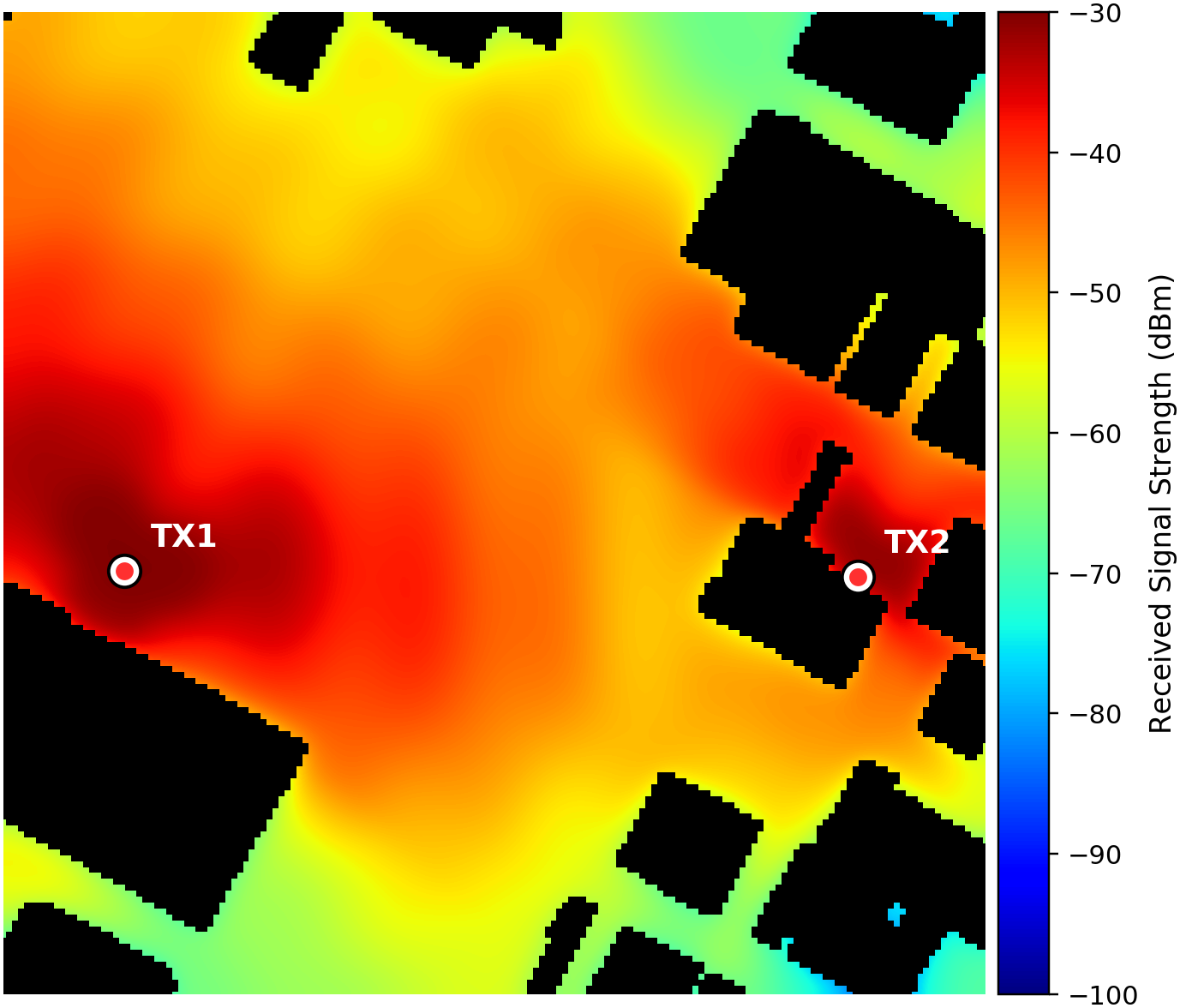}\\[-0.5mm]
{\fontsize{6}{6.5}\selectfont (d) Inference (3.2\%)}
\end{minipage}

\caption{Illustration of the proposed closed-loop multi-stage construction with different sparsity.}
\label{fig:scale-results}
\end{figure}

Although the spatial scale is significantly reduced, the selected region still contains heterogeneous building layouts. The radio fields generated by the two transmitters exhibit strong overlap and shadowing interactions, which makes accurate IRM construction considerably challenging. However, the proposed method can still accurately characterize local transmitter hotspots, building-induced shadowing regions, and sharp spectrum variations around geographical boundaries, benefiting from the physics-aware priors with powerful diffusion blocks, uncertainty-guided sensing strategy, and multi-stage inference refinements. The corresponding progressive construction results under different sensing sparsity levels are presented in Fig.~\ref{fig:scale-results}. Thus, the proposed framework shows excellent scalability, which can provide reliable IRM construction across multiple spatial scales, ranging from city-level environments to fine-grained local deployments.

\section{Conclusion}

\label{sec:conclusion}

This paper presents a closed-loop multi-stage framework for practical IRM construction. The proposed design couples physics-aware sparse sensing, hybrid-conditioned DiT inference, and an uncertainty-guided feedback mechanism in multiple stages. Compared to baseline methods, the proposed framework provides a progressive coarse-to-fine refinement for 3D spatiotemporal IRMs. Extensive experiments conducted on the customized spectrum dataset, which incorporates diverse urban morphologies, demonstrate that the proposed framework consistently achieves superior inference accuracy, sensing-efficiency gain, and robustness. 
As future work, the proposed framework can be extended from reactive sensing to predictive knowledge inference by explicitly modeling temporal dependencies across historical IRMs.
Moreover, the current few-transmitter setting will be generalized to multi-scale urban scenarios, allowing the IRM to elaborately characterize complex spectrum interactions among practical wireless networks.

\balance
\bibliographystyle{IEEEtran}
\bibliography{IEEEabrv,ref}

\end{document}